\documentclass[12pt]{article}
 \usepackage{graphicx,amsmath,amssymb}
\usepackage{indentfirst}
 \usepackage[usenames]{color}
 \usepackage[colorlinks=true, urlcolor=navyblue, linkcolor=navyblue, citecolor=navyblue]{hyperref}
\usepackage{epstopdf}
\usepackage{appendix}

\definecolor{navyblue}{rgb}{0,0.08,0.45}

\newcommand{\mbf}[1]{\mathbf{#1}}

\newcommand{\half}{{\frac{1}{2}}}
\newcommand{\threehalf}{{\frac{3}{2}}}
\newcommand{\fivehalf}{{\frac{5}{2}}}
\newcommand{\sevenhalf}{{\frac{7}{2}}}
\newcommand{\ninehalf}{{\frac{9}{2}}}
\newcommand{\elevenhalf}{{\frac{11}{2}}}
\newcommand{\thirteenhalf}{{\frac{13}{2}}}

\def\Dslash{\raise.15ex\hbox{/}\kern-.7em D}
\def\Pslash{\raise.15ex\hbox{/}\kern-.7em P}

\def\Journal#1#2#3#4{{#1} {{\bf #2},} {#4} {(#3)}}

\def\PRD{{Phys. Rev.} D}
\def\PRC{{Phys. Rev.} C}

\def\ZPC{{Z. Phys. C}}

\def\MPLA{{Mod. Phys. Lett.} A}

\begin{document}

\begin{flushright}
{
\small
SLAC--PUB--14860\\
\date{today}}
\end{flushright}

\vspace{80pt}

\centerline{\LARGE Hadronic Form Factor Models and Spectroscopy}

\vspace{10pt}

\centerline{\LARGE Within the Gauge/Gravity Correspondence}

\vspace{40pt}

\centerline{{
Guy F. de T\'eramond$^{a}$ 
\footnote{E-mail: \href{mailto:gdt@asterix.crnet.cr}{gdt@asterix.crnet.cr}}
and
Stanley J. Brodsky,$^{b}$ 
\footnote{E-mail: \href{mailto:sjbth@slac.stanford.edu}{sjbth@slac.stanford.edu}}
}}

\vspace{30pt}

{\centerline {$^{a}${\it Universidad de Costa Rica, San Jos\'e, Costa Rica}}

\vspace{4pt}

{\centerline {$^{b}${\it SLAC National Accelerator Laboratory, 
Stanford University, Stanford, CA 94309, USA}}

 \vspace{40pt}
 
\begin{abstract}
We show that the nonperturbative light-front  dynamics of relativistic hadronic bound states has a dual semiclassical gravity description on a higher dimensional warped AdS space in the limit of zero quark masses. This mapping of AdS gravity theory to the boundary quantum field theory, quantized at fixed light-front time, allows one to establish a precise relation between holographic wave functions in AdS space and the light-front wavefunctions describing the internal structure of hadrons. The resulting AdS/QCD model gives a remarkably good accounting of the spectrum, elastic and transition form factors of the light-quark hadrons in terms of one parameter, the QCD gap scale. The light-front holographic approach described here thus provides a frame-independent first approximation to the light-front Hamiltonian problem for QCD.   This article is based on lectures at the Niccol\`o Cabeo International School of Hadronic Physics, Ferrara, Italy, May 2011.
\end{abstract}

\date{today}

\vspace{20pt}

\newpage

\tableofcontents

\newpage

\section{Introduction}
\label{intro}

One of the most challenging problems in particle physics is to understand hadron dynamics and spectroscopy in terms of the confined quark and gluon quanta of quantum chromodynamics, the fundamental theory of the strong interactions.   A central goal is to compute detailed hadronic properties, such as moments, structure functions, distribution amplitudes, transversity distributions,  elastic and transition form factors, and the excitation dynamics of hadron resonances from first principles; {\it i.e.}, directly from the QCD Lagrangian. The most successful theoretical approach thus far has been to quantize QCD on discrete lattices in Euclidean space-time~\cite{Wilson:1974sk}.  Lattice numerical results follow from the computation of frame-dependent moments of distributions in Euclidean space; however,  dynamical observables in Minkowski space-time, such as the time-like hadronic form factors, are not  obtained directly from Euclidean-space lattice computations.   Dyson-Schwinger methods have led to many important insights, such as the infrared fixed-point behavior of the strong coupling constant~\cite{Cornwall:1981zr}; however,
in practice, these analyses are limited to ladder approximation in Landau gauge.

In principle, one could  calculate hadronic spectroscopy and wavefunctions by solving for the eigenvalues and eigenfunctions of the QCD Hamiltonian: 
$H \vert  \Psi \rangle = E \vert \Psi \rangle$
 at fixed time $t.$ However, this traditional method -- called the ``instant form" by Dirac~\cite{Dirac:1949cp}, is  plagued by complex vacuum and relativistic effects. 
In contrast,  quantization at fixed light-front (LF) time $\tau = t + z/c$ -- the ``front-form'' of Dirac~\cite{Dirac:1949cp} -- provides a powerful boost-invariant nonperturbative method for solving QCD and constitutes the ideal framework to describe the
structure of hadrons in terms of their quark and gluon degrees of
freedom. 
The simple structure of the light-front  vacuum allows an unambiguous
definition of the partonic content of a hadron in QCD and of hadronic light-front wavefunctions (LFWFs),
the underlying link between large distance hadronic states and the
constituent degrees of freedom at short distances.  Thus, one can also solve QCD by diagonalizing the light-front QCD Hamiltonian $H_{LF}$. The spectrum and light-front Fock-state wavefunctions are obtained from the eigenvalues and eigensolutions of the Heisenberg problem $H_{LF} \vert \psi \rangle = M^2 \vert \psi \rangle$, which becomes an infinite
set of coupled integral equations for the light-front components $\psi_n = \langle n \vert \psi \rangle$ in the Fock expansion~\cite{Brodsky:1997de, Burkardt:1995ct}.
This nonperturbative method has the advantage that it is frame-independent, operates in physical Minkowski space-time, and has no fermion-doubling problem~\cite{Brodsky:1997de}.  It has been applied successfully in lower space-time dimensions. In practice, however, the resulting large matrix diagonalization problem in $3+1$ space-time has proven to be a daunting task, so alternative methods and approximations  are necessary.

The AdS/CFT correspondence between gravity or string theory on a higher-dimensional anti--de Sitter (AdS) space and conformal field theories (CFT) in physical space-time~\cite{Maldacena:1997re},
has led to a semiclassical approximation for strongly-coupled quantum field theories
which provides physical insights into its nonperturbative dynamics. The correspondence is holographic in the sense that it determines a duality between  theories in different number of space-time dimensions.  In practice, the duality provides an effective gravity description in a ($d+1$)-dimensional AdS
space-time in terms of a flat $d$-dimensional conformally-invariant quantum field theory defined at the AdS
asymptotic boundary~\cite{Gubser:1998bc, Witten:1998qj}.
Thus, in principle, one can compute physical observables in a strongly coupled gauge theory  in terms of a classical gravity theory.

Anti-de Sitter $AdS_{5}$ space is the maximally symmetric space-time
with negative curvature and a four-dimensional space-time boundary.
The most general group of transformations that leave the $AdS_{d+1}$ differential line element 
\begin{equation} \label{eq:AdSz}
ds^2 = \frac{R^2}{z^2} \left(\eta_{\mu \nu} dx^\mu dx^\nu - dz^2\right),
\end{equation}
invariant, the isometry group, has dimensions $(d+1)(d+2)/2$ ($R$ is the AdS radius). Five-dimensional anti-de Sitter space AdS$_5$ has 15 isometries, in agreement with the number of generators of the conformal group in four dimensions.
Since the AdS metric (\ref{eq:AdSz}) is invariant under a dilatation of all coordinates $x^\mu  \to \lambda x^\mu$  and $z \to \lambda z$, it follows that the additional dimension, the holographic variable $z$, acts as a scaling variable in Minkowski space: different values of $z$ correspond to
different energy scales at which the hadron is examined.  As a result,
a short space-like or time-like invariant interval
near the light-cone, $x_\mu x^\mu \to 0$ maps to the conformal  AdS boundary near $z \to 0$. This also corresponds to the
$Q \to \infty$ ultraviolet (UV) zero separation distance. On the other hand, a large invariant four-dimensional  interval of
confinement dimensions $x_\mu x^\mu \sim 1/\Lambda_{\rm QCD}^2$ maps to the large infrared (IR) region
of AdS space $z \sim 1 / \Lambda_{\rm QCD}$.

QCD is fundamentally different from conformal theories since its scale invariance is broken  by quantum effects.
A gravity dual to QCD is not known, but the mechanisms of
confinement can be incorporated in the gauge/gravity
correspondence by modifying the AdS geometry in the  large IR domain  $z \sim 1/\Lambda_{\rm QCD}$, which also
sets the scale of the strong
interactions~\cite{Polchinski:2001tt}.  In this simplified
approach, we consider the propagation of hadronic modes in a fixed
effective gravitational background asymptotic to AdS space, which
encodes salient properties of the QCD dual theory, such as the
UV conformal limit at the AdS boundary, as
well as modifications of the background geometry in the large-$z$
IR region to describe confinement.

The physical  states in AdS space are
represented by normalizable modes $\Phi_P(x, z) = e^{-iP \cdot x} \Phi(z)$,
with plane waves along Minkowski coordinates $x^\mu$ and a profile function $\Phi(z)$
along the holographic coordinate $z$. 
The hadronic invariant mass states
$P_\mu P^\mu = \mathcal{M}^2$ are found by solving the eigenvalue problem for the
AdS wave equation. The modified theory generates the point-like hard behavior expected from QCD, instead of the soft
behavior characteristic of extended objects~\cite{Polchinski:2001tt}.  It is rather remarkable that the QCD dimensional counting rules~\cite{Brodsky:1973kr, Matveev:ra} are also a key feature of nonperturbative models~\cite{Polchinski:2001tt} based on the gauge/gravity duality.  Although the mechanisms are different, both the perturbative QCD and the  AdS/QCD approaches depend on the leading-twist (dimension minus spin) interpolating operators of the hadrons and their structure at short distances.

The gauge/gravity duality leads  to a simple analytical
 and phenomenologically compelling nonperturbative  frame-independent first approximation to the light-front Hamiltonian problem for QCD -- ``Light-Front Holography"~\cite{deTeramond:2008ht}.
Incorporating the AdS/CFT correspondence~\cite{Maldacena:1997re} as a useful guide, light-front  holographic methods  were originally
introduced~\cite{Brodsky:2006uqa, Brodsky:2007hb} by mapping the Polchinski-Strassler formula for the electromagnetic (EM) form factors in AdS space~\cite{Polchinski:2002jw} to the corresponding Drell-Yan-West expression at fixed light-front time in
physical space-time~\cite{Drell:1969km, West:1970av}.   It was also shown that one obtains  identical light-front holographic mapping for the gravitational form factor~\cite{Brodsky:2008pf} -- the matrix elements of the energy-momentum tensor, by perturbing the AdS
metric (\ref{eq:AdSz}) around its static solution~\cite{Abidin:2008ku}.
In the usual ``bottom-up" approach to  the gauge/gravity duality~\cite{Erlich:2005qh, DaRold:2005zs},  fields in the bulk geometry are introduced to match the
chiral symmetries of QCD and axial and vector currents become the
primary entities as in effective chiral theory. In contrast, in light-front holography a direct connection with the internal constituent structure of hadrons is established using light-front quantization~\cite{deTeramond:2008ht, Brodsky:2006uqa, Brodsky:2007hb, Brodsky:2008pf, Brodsky:2003px}.

The identification of higher dimensional AdS space with partonic physics in physical space-time is specific to the light front: the Polchinski-Strassler formula for computing transition matrix elements is a simple overlap of AdS amplitudes, which maps to a convolution of frame-independent light-front wavefunctions. ÊThis AdS Êconvolution formula cannot be mapped to current matrix elements at ordinary fixed time $t$, since the instant-time wavefunctions must be boosted away from the hadron's rest frame -- an intractable dynamical problem. 
In fact, the boost of a composite system at fixed time $t$ is only known at weak binding. Moreover, the form factors in instant time also require computing the contributions of currents which arise from the vacuum in the initial state and which connect to the hadron in the final state. ÊThus instant form wavefunctions alone are not sufficient to compute covariant current matrix elements in the instant form. There is no analog of such contributions in AdS. In contrast, there are no vacuum contributions in the light-front formulae for current matrix elements -- in agreement with the AdS formulae.

Unlike ordinary instant-time quantization, the Hamiltonian equation of motion in the light-front is frame independent and has a structure similar to  eigenmode equations in AdS space.
This makes a direct connection of QCD with AdS/CFT methods possible.  In fact, one can also study the AdS/CFT duality and its modifications starting from the LF Hamiltonian equation of motion for a relativistic bound-state system
 $H_{LF} \vert  \psi \rangle  = \mathcal{M}^2 \vert  \psi \rangle$
in physical space-time~\cite{deTeramond:2008ht}, where 
the QCD light-front Hamiltonian $H_{LF} \equiv  P_\mu P^\mu = P^+ P^-  \! - \mbf{P}^2_\perp$, $P^\pm = P^0 \pm P^3$, is constructed from the QCD Lagrangian using the  standard methods of quantum field theory~\cite{Brodsky:1997de}.
To a first semiclassical approximation, where quantum loops and quark masses
are not included, LF holography leads to a LF Hamiltonian equation which
describes the bound-state dynamics of light hadrons in terms of
an invariant impact kinematical variable $\zeta$ which
measures the
separation of the partons within the hadron at equal light-front
time $\tau = x^+= x^0 + x^3$. The transverse coordinate $\zeta$ is closely related to the invariant mass squared  of the constituents in the LFWF  and its off-shellness  in  the LF kinetic energy,  and it is thus the natural variable to characterize the hadronic wavefunction.  In fact $\zeta$ is the only variable to appear 
in the relativistic light-front Schr\"odinger equations predicted from 
holographic QCD  in the limit of zero quark masses. 
The coordinate $z$ in AdS space is thus uniquely identified with  a Lorentz-invariant  coordinate $\zeta$ which measures the separation of the constituents within a hadron at equal light-front time.  The ÊAdS/CFT correspondence shows that the holographic coordinate $z$ in AdS Êspace is  related inversely to the internal relative momentum. In fact, light-front holography makes this identification precise.

Remarkably, the unmodified AdS equations
correspond to the kinetic energy terms of  the partons inside a
hadron, whereas the interaction terms in the QCD Lagrangian build confinement and
correspond to the truncation of AdS space in an effective dual
gravity  approximation~\cite{deTeramond:2008ht}.  Thus, all the complexities of the strong interaction dynamics are hidden in an effective potential $U(\zeta)$, and the central question -- how to derive the effective color-confining  potential $U(\zeta)$ directly from QCD, remains open. To circumvent this obstacle, the effective confinement potential can be introduced either with a sharp cut-off in the infrared region of AdS space, as in the ``hard-wall'' model~\cite{Polchinski:2001tt}, or, more successfully,  using a ``dilaton" background in the holographic coordinate to produce a smooth cutoff at large distances  as  in the ``soft-wall'' model~\cite{Karch:2006pv}. Furthermore, one can impose from the onset a correct phenomenological confining structure to determine the effective IR warping of AdS space, for example, by adjusting the dilaton background to reproduce the observed linear Regge behavior of the hadronic mass spectrum $\mathcal{M}^2$ as a function of the excitation quantum numbers~\cite{Karch:2006pv, Shifman:2005zn}~\footnote{Using a mean-field mechanism, an effective harmonic confinement interaction was obtained in Ref.~\cite{Glazek:1987ic} in a constituent quark model.}.
By using light-front holographic mapping techniques,
one also obtains a connection between the mass parameter $\mu R$ of the AdS theory with the orbital angular momentum of the constituents in the light-front bound-state Hamiltonian equation~\cite{deTeramond:2008ht}.
The identification of orbital angular momentum of the constituents is a key element in our description of the internal structure of hadrons using holographic principles, 
since hadrons with the same quark content, but different orbital angular momenta, have different masses.

In our approach, the holographic mapping is carried out in the  strongly coupled regime where QCD is almost conformal, 
corresponding to an infrared fixed-point.  A QCD infrared fixed point arises since the propagators of the confined quarks and gluons in the  loop integrals contributing to the $\beta$-function have a maximal 
wavelength~\cite{Brodsky:2007hb, Brodsky:2008be}; thus, an infrared fixed point appears as a natural consequence of confinement. The decoupling of quantum loops in the infrared is analogous to QED dynamics where vacuum polarization corrections to the photon propagator decouple at $Q^2 \to 0$. Since there is a  window where the QCD coupling is
large and approximately constant,
QCD resembles a conformal theory for massless quarks. One then uses the isometries of AdS$_5$ to represent scale transformations within the conformal window. We thus begin with a conformal approximation to QCD to model an effective dual gravity description in AdS space. The large-distance non-conformal effects are taken into account with the introduction of an effective confinement potential as described above.

Early attempts to derive effective one-body equations in light-front QCD are described in reference~\cite{Pauli:2002tj}.
We should also mention previous work by 't Hooft, who obtained
the spectrum of two-dimensional QCD in the large $N_C$ limit
in terms of a Schr\"odinger equation as a function of the parton $x$-variable~\cite{'tHooft:1974hx}.
In the scale-invariant limit, this equation is equivalent
to the equation of motion for a scalar field in AdS$_3$ space~\cite{Katz:2007br}. In this case,
there is a mapping between  the variable $x$ and the radial coordinate in AdS$_3$.

\section{Light-front bound-state Hamiltonian equation of motion and light-front holography \label{LF Hamiltonian}}

A key step in the analysis of an atomic system, such as positronium,
is the introduction of the spherical coordinates $r, \theta, \phi$
which  separates the dynamics of Coulomb binding from the
kinematical effects of the quantized orbital angular momentum $L$.
The essential dynamics of the atom is specified by the radial
Schr\"odinger equation whose eigensolutions $\psi_{n,L}(r)$
determine the bound-state wavefunction and eigenspectrum. In our recent 
work, we have shown that there is an analogous invariant
light-front coordinate $\zeta$ which allows one to separate the
essential dynamics of quark and gluon binding from the kinematical
physics of constituent spin and internal orbital angular momentum.
The result is a single-variable light-front Schr\"odinger equation for QCD
which determines the eigenspectrum and the light-front wavefunctions
of hadrons for general spin and orbital angular momentum~\cite{deTeramond:2008ht}, thus
providing a description of the 
internal dynamics of hadronic states in terms of their massless constituents  
at the same LF time  $\tau = x^+= x^0 + x^3$, the time marked by the
front of a light wave~\cite{Dirac:1949cp}, instead of the ordinary instant time $t = x^0$.

\subsection{Light-front quantization of QCD\label{LFquant}}

Our starting point is the $SU(3)_C$ invariant Lagrangian of QCD
\begin{equation} \label{LFH}
\mathcal{L}_{\rm QCD} = \bar \psi \left( i \gamma^\mu D_\mu - m\right) \psi 
- \tfrac{1}{4} G^a_{\mu \nu} G^{a \, \mu \nu} ,
\end{equation}
where $D_\mu = \partial_\mu - i g_s A^a_\mu T^a$ and 
$G^a_{\mu \nu} = \partial_\mu A^a_\nu - \partial_\nu A^a_\mu + 
g_s c^{abc} A_\mu^b A_\nu^c$, with $\left[T^a, T^b\right] = i c^{abc} T^c$ and
$a, b ,c$ are $SU(3)_C$ color indices.

One can express the  hadron four-momentum  generator $P =  (P^+, P^-, \mbf{P}_{\!\perp})$,
$P^\pm = P^0 \pm P^3$,  in terms of the
dynamical fields, the Dirac field $\psi_+$, where $\psi_\pm = \Lambda_\pm
\psi$, $\Lambda_\pm = \gamma^0 \gamma^\pm$, and the transverse field
$\mbf{A}_\perp$ in the $A^+ = 0$ gauge~\cite{Brodsky:1997de}
quantized on the light-front at fixed light-cone time $x^+ $, $x^\pm = x^0 \pm x^3$
\begin{eqnarray} \label{eq:Pm}
P^-  &\!\!=\!&  \half \int \! dx^- d^2 \mbf{x}_\perp \bar \psi_+ \, \gamma^+
\frac{ \left( i \mbf{\nabla}_{\! \perp} \right)^2 + m^2 }{ i \partial^+}  \psi_+
 + {\rm(interactions)} ,\\ \label{eq:Pp}
P^+ &\!\!=\!& \int \! dx^- d^2 \mbf{x}_\perp
 \bar \psi_+ \gamma^+   i \partial^+ \psi_+ ,  \\ \label{eq:Pperp}
 \mbf{P}_{\! \perp}  &\!\!=\!&  \half \int \! dx^- d^2 \mbf{x}_\perp
 \bar \psi_+ \gamma^+   i \mbf{\nabla}_{\! \perp} \psi_+   ,
\end{eqnarray}
where the integrals are over the null plane $\tau = x^+ = 0$, the hyper-plane tangent to the light cone. This is the initial-value surface  for the fields where the
commutation relations are fixed.
The LF Hamiltonian $P^-$ generates LF time translations
\begin{equation}
\left[\psi_+(x), P^-\right] = i \frac{\partial}{\partial x^+} \psi_+(x),
\end{equation}
 to evolve the initial conditions to all space-time,
whereas the LF longitudinal  $P^+$ and  transverse momentum $\mbf{P}_\perp$ are kinematical generators. For simplicity we have omitted from (\ref{eq:Pm}-\ref{eq:Pperp}) the contributions from the gluon field $\mbf{A}_\perp$.

According to Dirac's classification of the forms of relativistic dynamics~\cite{Dirac:1949cp}, the fundamental generators of the Poincar\'e group
can be separated into kinematical and dynamical generators. The kinematical generators   act along the initial surface and leave the light-front plane invariant: they are thus independent of the dynamics and therefore contain no interactions. The dynamical generators change the light-front position and depend consequently
on  the interactions.   In addition to $P^+$ and $\mbf{P}_\perp$, the kinematical generators in the light-front frame are the $z$-component of the angular momentum $J^z$ and the boost operator $\mbf{K}$.  In addition to the Hamiltonian $P^-$,  $J^z$ and $J^y$ are also  dynamical generators. The light-front frame has the maximal number of kinematical generators.

\subsection{A semiclassical approximation to QCD \label{LFholog}}

A physical hadron in four-dimensional Minkowski space has four-momentum $P_\mu$ and invariant
hadronic mass states $P_\mu P^\mu = \mathcal{M}^2$ determined by the
Lorentz-invariant Hamiltonian equation for the relativistic bound-state system
\begin{equation} \label{LFH}
H_{LF} \vert  \psi(P) \rangle =  \mathcal{M}^2 \vert  \psi(P) \rangle,
\end{equation}
with  $H_{LF} \equiv P_\mu P^\mu  =  P^- P^+ -  \mbf{P}_\perp^2$,
where  the hadronic state $\vert\psi\rangle$ is an expansion in multiparticle Fock eigenstates
$\vert n \rangle$ of the free light-front  Hamiltonian:
$\vert \psi \rangle = \sum_n \psi_n \vert \psi \rangle$.
The  state $\vert \psi(P^+,\mbf{P}_\perp,J^z) \bigr\rangle$
is an eigenstate of the total momentum $P^+$
and $\mbf{P}_{\! \perp}$ and the  total spin  $J^z$. Quark and gluons appear from the light-front quantization
of the excitations of the dynamical fields $\psi_+$ and $\mbf{A}_\perp$, expanded in terms of creation and
annihilation operators at fixed LF time $\tau$. The Fock components $\psi_n(x_i, {\mathbf{k}_{\perp i}}, \lambda_i)$
are independent of  $P^+$ and $\mbf{P}_{\! \perp}$
and depend only on relative partonic coordinates:
the momentum fraction
 $x_i = k^+_i/P^+$, the transverse momentum  ${\mathbf{k}_{\perp i}}$ and spin
 component $\lambda_i^z$. Momentum conservation requires
 $\sum_{i=1}^n x_i = 1$ and
 $\sum_{i=1}^n \mathbf{k}_{\perp i}=0$.
The LFWFs $\psi_n$ provide a
{\it frame-independent } representation of a hadron which relates its quark
and gluon degrees of freedom to their asymptotic hadronic state.
Since for each constituent $k^+_i = \sqrt{\mbf{k}_i^2 + m_i^2} + k^z_i > 0$ there are no contributions from the vacuum. Thus, apart from possible zero modes, the light-front QCD vacuum  is the trivial vacuum.
The constituent spin and orbital angular momentum properties of the hadrons are also encoded in the LFWFs.  Actually, the definition of quark and gluon angular momentum is unambiguous in Dirac's
front form in light-cone gauge $A^+=0$, and the gluons  have physical polarization $S^z_g= \pm 1$.

One can also derive light-front holography using a first semiclassical approximation  to transform the fixed 
light-front time bound-state Hamiltonian equation of motion in QCD (\ref{LFH})
to  a corresponding wave equation in AdS 
space~\cite{deTeramond:2008ht}.
To this end we
expand the initial and final hadronic states in terms of its Fock components. The computation is  simplified in the
frame $P = \big(P^+, \mathcal{M}^2/P^+, \vec{0}_\perp \big)$ where $P^2 =  P^+ P^-$.
We find
 \begin{equation} \label{eq:Mk}
\mathcal{M}^2  =  \sum_n  \! \int \! \big[d x_i\big]  \! \left[d^2 \mbf{k}_{\perp i}\right]
 \sum_q \Big(\frac{ \mbf{k}_{\perp q}^2 + m_q^2}{x_q} \Big)
 \left\vert \psi_n (x_i, \mbf{k}_{\perp i}) \right \vert^2  + {\rm (interactions)} ,
 \end{equation}
plus similar terms for antiquarks and gluons ($m_g = 0)$. The integrals in (\ref{eq:Mk}) are over
the internal coordinates of the $n$ constituents for each Fock state
\begin{equation} \label{phases}
\int \big[d x_i\big] \equiv
\prod_{i=1}^n \int dx_i \,\delta \Bigl(1 - \sum_{j=1}^n x_j\Bigr) , ~~~
\int \left[d^2 \mbf{k}_{\perp i}\right] \equiv \prod_{i=1}^n \int
\frac{d^2 \mbf{k}_{\perp i}}{2 (2\pi)^3} \, 16 \pi^3 \,
\delta^{(2)} \negthinspace\Bigl(\sum_{j=1}^n\mbf{k}_{\perp j}\Bigr),
\end{equation}
with phase space normalization
\begin{equation}
\sum_n  \int \big[d x_i\big] \left[d^2 \mbf{k}_{\perp i}\right]
\,\left\vert \psi_n(x_i, \mbf{k}_{\perp i}) \right\vert^2 = 1.
\end{equation}

 Each constituent of the light-front wavefunction  $\psi_{n}(x_i, \mbf{k}_{\perp i}, \lambda_i)$  of a hadron is on its respective mass shell 
 $k^2_i= k^+_i k^-_i - \mbf{k}^2_{\perp i} = m^2_i$, $i = 1, 2 \cdots n.$   Thus $k^-_i= \left({\mbf k}^2_{\perp i}+  m^2_i\right)/ x_i P^+$.
However,  the light-front wavefunction represents a state which is off the light-front energy shell: $P^-  - \sum_i^n k^-_i < 0$, for a stable hadron.  Scaling out $P^+ = \sum^n_i k^+_i$, 
the invariant mass  of the constituents $\mathcal{M}_n$ is
\begin{equation}
\mathcal{M}_n^2  = \Big( \sum_{i=1}^n k_i^\mu\Big)^2 = \sum_i \frac{\mbf{k}_{\perp i}^2 +  m_i^2}{x_i}.
 \end{equation}
 The functional dependence  for a given Fock state is expressed in terms of the invariant mass,  the measure of the off-energy shell of the bound state of the $n$-parton LFWF:  $\mathcal{M}^2 \! - \mathcal{M}_n^2$.

The LFWF $\psi_n(x_i, \mathbf{k}_{\perp i}, \lambda_i)$ can be expanded in terms of  $n-1$ independent
position coordinates $\mathbf{b}_{\perp j}$,  $j = 1,2,\dots,n-1$,
conjugate to the relative coordinates $\mbf{k}_{\perp i}$, with $\sum_{i = 1}^n \mbf{b}_{\perp i} = 0$.
We can also express Eq.  (\ref{eq:Mk})
in terms of the internal impact coordinates $\mbf{b}_{\perp j}$ with the result
\begin{equation}
 \mathcal{M}^2  =  \sum_n  \prod_{j=1}^{n-1} \int d x_j \, d^2 \mbf{b}_{\perp j} \,
\psi_n^*(x_j, \mbf{b}_{\perp j})  \\
 \sum_q   \left(\frac{ \mbf{- \nabla}_{ \mbf{b}_{\perp q}}^2  \! + m_q^2 }{x_q} \right)
 \psi_n(x_j, \mbf{b}_{\perp j}) \\
  + {\rm (interactions)} . \label{eq:Mb}
 \end{equation}
The normalization is defined by
\begin{equation}
\sum_n  \prod_{j=1}^{n-1} \int d x_j d^2 \mathbf{b}_{\perp j}
\left \vert \psi_n(x_j, \mathbf{b}_{\perp j})\right\vert^2 = 1.
\end{equation}

If we want to simplify further the description of the multiple parton system and reduce its dynamics to a single variable problem, we must take the limit of quark masses to zero. 
Indeed, the underlying classical QCD Lagrangian with massless quarks is scale and conformal invariant~\cite{Parisi:1972zy}, and consequently only in this limit it is possible to map the equations of motion and transition matrix elements to their correspondent conformal AdS  expressions. 

To simplify the discussion we will consider a two-parton hadronic bound state.  In the limit
of zero quark mass
$m_q \to 0$
\begin{equation}  \label{eq:Mbpion}
\mathcal{M}^2  =  \int_0^1 \! \frac{d x}{x(1-x)} \int  \! d^2 \mbf{b}_\perp  \,
  \psi^*(x, \mbf{b}_\perp)
  \left( - \mbf{\nabla}_{ {\mbf{b}}_{\perp}}^2\right)
  \psi(x, \mbf{b}_\perp) +   {\rm (interactions)}.
 \end{equation}
 For $n=2$, the invariant mass is $\mathcal{M}_{n=2}^2 = \frac{\mbf{k}_\perp^2}{x(1-x)}$. 
 Similarly, in impact space the relevant variable for a two-parton state is  $\zeta^2= x(1-x)\mbf{b}_\perp^2$.
Thus, to first approximation  LF dynamics  depend only on the boost invariant variable
$\mathcal{M}_n$ or $\zeta$,
and hadronic properties are encoded in the hadronic mode $\phi(\zeta)$ from the relation
\begin{equation} \label{eq:psiphi}
\psi(x,\zeta, \varphi) = e^{i L \varphi} X(x) \frac{\phi(\zeta)}{\sqrt{2 \pi \zeta}} ,
\end{equation}
thus factoring out the angular dependence $\varphi$ and the longitudinal, $X(x)$, and transverse mode $\phi(\zeta)$.
This is a natural factorization in the light front since the
corresponding canonical generators, the longitudinal and transverse generators $P^+$ and $\mbf{P}_\perp$ and the $z$-component of the orbital angular momentum
$J^z$ are kinematical generators which commute with the LF Hamiltonian generator $P^-$.
We choose the normalization
 $ \langle\phi\vert\phi\rangle = \int \! d \zeta \,
 \vert \langle \zeta \vert \phi\rangle\vert^2 = P_{q \bar q}$,  where $P_{q \bar q}$ is the probability of finding the $q \bar q$ component in the pion light-front wavefunction. The longitudinal mode is thus normalized as
$ \int_0^1 \frac{X^2(x)}{x(1-x)}=1$.

We can write the Laplacian operator in (\ref{eq:Mbpion}) in circular cylindrical coordinates $(\zeta, \varphi)$
\begin{equation} \label{eq:Lzeta}
\nabla_\zeta^2 = \frac{1}{\zeta} \frac{d}{d\zeta} \left( \zeta \frac{d}{d\zeta} \right)
+ \frac{1}{\zeta^2} \frac{\partial^2}{\partial \varphi^2},
\end{equation}
and factor out the angular dependence of the
modes in terms of the $SO(2)$ Casimir representation $L^2$ of orbital angular momentum in the
transverse plane. Using  (\ref{eq:psiphi}) we find~\cite{deTeramond:2008ht}
\begin{equation} \label{eq:KV}
\mathcal{M}^2   =  \int \! d\zeta \, \phi^*(\zeta) \sqrt{\zeta}
\left( -\frac{d^2}{d\zeta^2} -\frac{1}{\zeta} \frac{d}{d\zeta}
+ \frac{L^2}{\zeta^2}\right)
\frac{\phi(\zeta)}{\sqrt{\zeta}}   \\
+ \int \! d\zeta \, \phi^*(\zeta) U(\zeta) \phi(\zeta) ,
\end{equation}
where $L = \vert L^z \vert $.  In writing the above equation we have summed the complexity of the interaction terms in the QCD Lagrangian  by the introduction of the effective
potential $U(\zeta)$, which is  modeled to enforce confinement at some IR scale.
The LF eigenvalue equation $P_\mu P^\mu \vert \phi \rangle  =  \mathcal{M}^2 \vert \phi \rangle$
is thus a light-front  wave equation for $\phi$
\begin{equation} \label{LFWE}
\left(-\frac{d^2}{d\zeta^2}
- \frac{1 - 4L^2}{4\zeta^2} + U(\zeta) \right)
\phi(\zeta) = \mathcal{M}^2 \phi(\zeta),
\end{equation}
a relativistic single-variable LF Schr\"odinger equation.   Its eigenmodes $\phi(\zeta)$
determine the hadronic mass spectrum and represent the probability
amplitude to find $n$-partons at transverse impact separation $\zeta$,
the invariant separation between pointlike constituents within the hadron~\cite{Brodsky:2006uqa} at equal LF time.  Thus the effective interaction potential is instantaneous in LF time $\tau$, not instantaneous in ordinary time $t$.
The LF potential thus satisfies causality, unlike the instantaneous Coulomb interaction.
Extension of the results to arbitrary $n$ follows from the $x$-weighted definition of the
transverse impact variable of the $n-1$ spectator system~\cite{Brodsky:2006uqa}
\begin{equation} \label{zetan}
\zeta = \sqrt{\frac{x}{1-x}} \Big\vert \sum_{j=1}^{n-1} x_j \mbf{b}_{\perp j} \Big\vert,
\end{equation}
 where $x = x_n$ is the longitudinal
momentum fraction of the active quark. One can also
generalize the equations to allow for the kinetic energy of massive
quarks using Eqs. (\ref{eq:Mk}) or (\ref{eq:Mb})~\cite{Brodsky:2008pg}. In this case, however,
the longitudinal mode $X(x)$ does not decouple from the effective LF bound-state equations.

\subsection{Higher spin hadronic modes in AdS space \label{HigherSpin} }

We now turn to the formulation of bound-state equations for mesons of arbitrary spin $J$ in AdS space~\footnote{This section is based on our collaboration with Hans Guenter Dosch. A detailed discussion of higher integer and half-integer spin wave equations  in modified AdS spaces
will be given in Ref.~\cite{BDdT:2012}. See also the discussion in Ref.~\cite{Gutsche:2011vb}.}.  As we shall show in the next section, there is a remarkable correspondence between the equations of motion in AdS space and the Hamiltonian equation for the relativistic bound-state system for the corresponding angular momentum in light-front theory.

The description of higher spin modes in AdS space is a notoriously difficult problem~\cite{Fronsdal:1978vb, Fradkin:1986qy, Metsaev:2008fs}.
 A spin-$J$ field in AdS$_{d+1}$ is represented by a rank $J$ tensor field $\Phi(x^A)_{M_1 \cdots M_J}$, which is totally symmetric in all its indices. Such a tensor contains also lower spins, which can be eliminated by imposing gauge conditions. The action for a spin-$J$ field in AdS$_{d+1}$ space-time in presence of a dilaton background field $\varphi(z)$ (the string frame) is given by
\begin{multline} \label{SJ}
S = \half \int \! d^d x \, dz  \,\sqrt{g} \,e^{\varphi(z)}
  \Big( g^{N N'} g^{M_1 M'_1} \cdots g^{M_J M'_J} D_N \Phi_{M_1 \cdots M_J} D_{N'}  \Phi_{M'_1 \cdots M'_J}    \\
 - \mu^2  g^{M_1 M'_1} \cdots g^{M_J M'_J} \Phi_{M_1 \cdots M_J} \Phi_{M'_1 \cdots M'_J}  + \cdots \Big)  ,
\end{multline}
where $M, N = 1, \cdots , d+1$, $\sqrt{g} = (R/z)^{d+1}$ and $D_M$ is the covariant derivative which includes parallel transport. The omitted terms in (\ref{SJ}) refer to
terms with different contractions. The coordinates of AdS are the Minkowski coordinates $x^\mu$ and the holographic variable $z$ labeled $x^M = \left(x^\mu, z\right)$. The  d + 1 dimensional mass $\mu$ is not a physical observable and is {\it a priory} an arbitrary
parameter. The dilaton background field $\varphi(z)$ in  (\ref{SJ})   introduces an energy scale in the five-dimensional AdS action, thus breaking its conformal invariance. It is  a function of the holographic coordinate $z$ which vanishes
 in the conformal ultraviolet limit $z \to 0$.   In the hard wall model $\varphi = 0$ and the conformality is broken by the IR boundary conditions at $z = z_0 \sim 1 /\Lambda_{\rm QCD}$.
 
 A physical hadron has plane-wave solutions and polarization indices $M$ along the 3 + 1 physical coordinates
 \begin{equation}
 \Phi_P(x,z)_{\mu_1 \cdots \mu_J} = e^{- i P \cdot x} \Phi(z)_{\mu_1 \cdots \mu_J},
 \end{equation}
 with four-momentum $P_\mu$ and  invariant hadronic mass  $P_\mu P^\mu \! = \! \mathcal{M}^2$. All other components vanish identically:  
 $\Phi_{z \mu_2 \cdots \mu_J} = \Phi_{\mu_1 z \cdots \mu_J} = \cdots = \Phi_{\mu_ 1 \mu_2 \cdots z} = 0$. One can then construct an effective action in terms
 of high spin modes $\Phi_J = \Phi_{\mu_1 \mu_2 \cdots \mu_J}$, with only  physical degrees of 
 freedom~\cite{Karch:2006pv}. In this case the system of coupled differential equations which follow from (\ref{SJ}) reduce to a homogeneous equation in terms of the physical field $\Phi_J$.

In terms of  fields with tangent indices
\begin{equation} \label{PhiT}
 {\hat \Phi}_{A_1 A_2 \cdots A_J}
 = e_{A_1}^{M_1} e_{A_2}^{M_2} \cdots e_{A_J}^{M_J} \,
 {\Phi}_{M_1 M_2 \cdots M_J}
 =  \left(\frac{z}{R}\right)^J  \negthinspace {\Phi}_{A_1 A_2 \cdots A_J}  ,
 \end{equation}
we find the effective action~\cite{BDdT:2012} ($\hat \Phi_J \equiv \hat \Phi_{\mu_1 \cdots \mu_J}$)
\begin{equation} \label{ShatPhiJ}
S = \half \int \! d^d x \, dz  \,\sqrt{g} \,e^{\varphi(z)}
  \Big( g^{N N'}  \partial_N \hat \Phi_J \partial_{N'} \hat  \Phi_J 
 - \mu^2 \hat \Phi_J^2  \Big)  ,
\end{equation}
containing only the physical degrees of freedom and usual derivatives. Thus, the effect of the covariant derivatives in the effective action for spin-$J$ fields with polarization components along the physical coordinates is a shift in the AdS mass $\mu$.  The vielbein $e^A_M$
 is defined by $g_{M N} = e^A_M e^B_N\eta_{A B}$,
 where  $A, B = 1, \cdots , d+1$ are tangent AdS space indices and $\eta_{A B}$ has diagonal components  $(1, -1, \cdots, -1)$. In AdS the vielbein is $e^A_M = (R/z) \delta^A_M$. 

In terms of the AdS field $\Phi_J \equiv  \Phi_{\mu_1 \cdots \mu_J}$ we can express  the effective action (\ref{ShatPhiJ}) 
\begin{equation} \label{SPhiJ}
S = \half \int \! d^d x \, dz  \,\sqrt{g_J} \,e^{\varphi(z)}
  \Big( g^{N N'}  \partial_N  \Phi_J \partial_{N'}  \Phi_J 
 - \mu^2  \Phi_J^2  \Big)  ,
\end{equation}
where we have defined an effective metric determinant  
\begin{equation} \label{gJ}
\sqrt{g_J} = (R/z)^{d+1 - 2 J},
\end{equation}
 and rescaled the AdS mass $\mu$ in  $(\ref{ShatPhiJ})$.
Variation of the higher-dimensional action (\ref{SPhiJ}) gives the AdS wave equation  for the spin-$J$ mode $\Phi_J$
\begin{equation} \label{AdSWEJ}
\left[-\frac{ z^{d-1 -2 J}}{e^{\varphi(z)}}   \partial_z \left(\frac{e^\varphi(z)}{z^{d-1 - 2 J}} \partial_z\right)
+ \left(\frac{\mu R}{z}\right)^2\right] \Phi(z)_J = \mathcal{M}^2 \Phi(z)_J,
 \end{equation}
where the eigenmode  $ \Phi_J$  is normalized according to
\begin{equation}  \label{Phinorm}
R^{d - 1 - 2 J} \int_0^{\infty} \! \frac{dz}{z^{d -1 - 2 J}} \, e^{\varphi(z)} \Phi_J^2 (z) = 1.
\end{equation}
The AdS mass is $\mu$ obeys the relation
 \begin{equation} \label{muR}
 (\mu R)^2 = (\tau - J)(\tau -  d + J) ,
 \end{equation}
 which follows from the scaling  behavior of the tangent AdS field near $z \to 0$, $\hat \Phi_J \sim z^\tau$.

We can also derive (\ref{AdSWEJ}) by shifting dimensions for a $J$-spin mode~\cite{deTeramond:2008ht, deTeramond:2010ge}}. To this end, we start with the scalar wave equation which follows from the variation of (\ref{SJ}) for $J = 0$. This case is particularly simple as the covariant derivative of a scalar field is the usual derivative. We  obtain the eigenvalue equation
\begin{equation} \label{WeS}
\left[-\frac{ z^{d-1}}{e^{\varphi(z)}}   \partial_z \left(\frac{e^{\varphi(z)}}{z^{d-1}} \partial_z\right) 
+ \left(\frac{\mu R}{z}\right)^2\right] \Phi = \mathcal{M}^2 \Phi.
\end{equation}
A physical  spin-$J$ mode $\Phi_{\mu_1 \cdots \mu_J}$  with all  indices 
along 3+1 is constructed by shifting dimensions
$\Phi_J(z) = ( z/R)^{-J}  \Phi(z)$. It is simple to show that the shifted field $\Phi_{\mu_1 \mu_2 \cdots \mu_J}$ obeys the wave equation
(\ref{AdSWEJ}) which follows from (\ref{WeS})
upon mass rescaling $(\mu R)^2 \to (\mu R)^2 - J(d-J)  + J z \,  \varphi'(z)$.

\subsubsection{Non-conformal warped metrics}

In the Einstein frame conformal invariance is broken by the introduction of an additional warp factor 
in the AdS metric in order to include confinement forces
\begin{eqnarray}  \label{gE}
ds^2 &=& (g_E)_{M N} dx^M dx^N \\   \nonumber
&=& \frac{R^2}{z^2} e^{\lambda(z)} \left( \eta_{\mu \nu} dx^\mu dx^\nu - dz^2\right).
\end{eqnarray}
The action is
\begin{multline} \label{SJE}
S = \half \int \! d^d x \, dz  \,\sqrt{g_E} \,
  \Big( g_E^{N N'} g_E^{M_1 M'_1} \cdots g_E^{M_J M'_J} D_N \Phi_{M_1 \cdots M_J} D_{N'}  \Phi_{M'_1 \cdots M'_J}    \\
 - \mu^2  g_E^{M_1 M'_1} \cdots g_E^{M_J M'_J} \Phi_{M_1 \cdots M_J} \Phi_{M'_1 \cdots M'_J}  + \cdots \Big)  ,
\end{multline}
 where $g_E^{MN} \equiv (g_E)^{MN}$ and  $(g_E)_{MN} = \frac{R^2}{z^2} e^{\lambda(z)}\eta_{MN}$. The flat metric  $\eta_{M N}$ has diagonal components  $(1, -1, \cdots, -1)$.  
 
The use of warped metrics is useful to visualize the overall confinement behavior
as we follow an object in warped AdS space as it falls to the infrared region by the effects of gravity. The gravitational potential energy for an object of mass $m$ in general relativity is
given in terms of  the time-time component of the metric tensor $g_{00}$
\begin{equation} \label{V}
V = mc^2 \sqrt{(g_E)_{00}} = mc^2 R \, \frac{e^{\lambda(z)/2}}{z},
\end{equation}
thus we may expect a potential that has a minimum at the hadronic scale $z_0$ and grows fast for larger values of $z$ to confine effectively a particle in a hadron within distances $z \sim z_0$. In fact, 
according to Sonnenscheim~\cite{Sonnenschein:2000qm} a background
dual to a confining theory should satisfy the conditions for the
metric component $g_{00}$ 
\begin{equation}\label{Scond}
\partial_z (g_{00}) \vert_{z=z_0} = 0 , ~~~~ g_{00} \vert_{z = z_0} \ne 0,
\end{equation}
to display the Wilson loop area law for confinement of strings.

 To relate the results in the Einstein frame  where hadronic modes  propagate  in the non-conformal warped metrics
 (\ref{gE}) to the results in the String-Jordan frame (\ref{SJ}), we scale away the dilaton profile by a redefinition of the fields in the action.
 This corresponds to the multiplication of  
  the metric determinant $\sqrt{g_E} = \left(\frac{R}{z}\right)^{d+1} e^{(d+1)\lambda(z)/2}$ by the
 contravariant tensor $(g_E)^{MN}$. Thus the result~\cite{BDdT:2012}
 $\varphi(z)  \to \frac{d-1}{2} \lambda(z)$, or $\varphi \to \frac{3}{2} \lambda$ for AdS$_5$.

 \subsubsection{Effective confining potentials in AdS \label{Vz}}

For some applications  it is convenient to scale away the dilaton factor in the action by a field redefinition~\cite{Afonin:2010hn}. 
For example, for a scalar field we can shift   $\Phi \to e^{- \varphi/2} \Phi$, and the bilinear component in the action is transformed into the equivalent problem of a free kinetic part plus an effective confining potential $V(z)$ which breaks the conformal invariance.~\footnote{In fact, for fermions the conformality cannot be broken by the introduction of a dilaton background or by explicitly deforming the AdS metric as discussed above, since the additional warp factor is scaled away by a field redefinition. In this case the breaking of the conformal invariance and the generation of the fermion spectrum can only be accomplished by the introduction of an effective potential. This is further discussed in Sec. \ref{SWB}.} For the spin-$J$ effective action (\ref{SPhiJ})  we find upon the field redefinition $\Phi_J \to e^{- \varphi/2} \Phi_J$ 
\begin{multline} \label{SPhiJV}
S = \half \int \! d^d x \, dz  \,\sqrt{g_J} 
  \Big( g^{N N'}  \partial_N  \Phi_J \partial_{N'}  \Phi_J 
 - \mu^2  \Phi_J^2  -  V(z)  \Phi_J^2  \Big)  \\
 -\frac{1}{4} \lim_{\epsilon \to 0} \int d^dx  \left(\frac{R}{z}\right)^{d - 1 - 2 J} \! \varphi'(z) \Phi_J^2 \Big|_\epsilon^\infty,
\end{multline}
with effective metric determinant (\ref{gJ}) $\sqrt{g_J} = (R/z)^{d + 1 - 2 J}$ and  effective potential $V(z) = \frac{z^2}{R^2} U(z)$, where
\begin{equation} \label{Uz}
U(z) = \half \varphi''(z) + \frac{1}{4} \varphi'(z)^2  + \frac{2J - d + 1}{2 z} \varphi'(z) .
\end{equation}

The action  (\ref{SPhiJ}) is thus equivalent, modulo  a surface term,  to the action  (\ref{SPhiJV})  written in terms of the rotated fields $\Phi_J \to e^{- \varphi/2} \Phi_J$.
The result (\ref{Uz}) is identical to the result obtained in Ref.  \cite{Gutsche:2011vb}.  As we will show in the following section, the effective potential (\ref{Uz}), for $d=4$, is precisely the effective light-front potential which appears in Eq. (\ref{LFWE}), where the LF transverse impact variable  $\zeta$ is identified with the holographic variable $z$. 
A different approach is discussed in Ref.~\cite{Csaki:2006ji} where the infrared physics is introduced by a back-reaction model to the AdS metric. See also Refs.~\cite{Gursoy:2007er, dePaula:2008fp, Gherghetta:2009ac, Abidin:2011ht}.

\subsection{Light-front holographic mapping \label{LFmapping}}

The structure of the QCD light-front Hamiltonian equation  (\ref{LFH}) for the state $\vert \psi(P) \rangle$ is similar to the structure of the wave equation (\ref{AdSWEJ}) for the
$J$-mode $\Phi_{\mu_1 \cdots \mu_J}$ in AdS space; they are both frame-independent and have identical eigenvalues $\mathcal{M}^2$, the mass spectrum of the color-singlet states of QCD. This provides the basis for a profound connection between physical QCD formulated in the light-front  and the physics of hadronic modes in AdS space. However, important differences are also apparent:  Eq. (\ref{LFH}) is a linear quantum-mechanical equation of states in Hilbert space, whereas Eq. (\ref{AdSWEJ}) is a classical gravity equation; its solutions describe spin-$J$ modes propagating in a higher dimensional
warped space. Physical hadrons are composite, and thus inexorably endowed of orbital angular momentum. Thus, the identification
of orbital angular momentum is of primary interest in establishing a connection between both approaches. In fact, to a first semiclassical approximation,
light-front QCD  is formally equivalent to the equations of motion on a fixed gravitational background~\cite{deTeramond:2008ht} asymptotic to AdS$_5$, where the prominent properties of confinement are encoded in a dilaton profile $\varphi(z)$. 

As shown in  Sect. \ref{LFholog}, one can indeed systematically reduce  the LF  Hamiltonian eigenvalue Eq.  (\ref{LFH}) to an effective relativistic wave equation (\ref{LFWE}), analogous to the AdS equations, by observing that each $n$-particle Fock state has an essential dependence on the invariant mass of the system  and
thus, to a first approximation, LF dynamics depend only on $\mathcal{M}_n^2$.
In  impact space the relevant variable is the boost invariant  variable $\zeta$,
 which measures the separation of quarks and gluons, and which also allows one to separate the bound state dynamics
of the constituents from the kinematics of their
internal angular momentum.

Upon the substitution $z \! \to\! \zeta$  and
\begin{equation}
\phi_J(\zeta)   = \left(\zeta/R\right)^{-3/2 + J} e^{\varphi(z)/2} \, \Phi_J(\zeta) ,
\end{equation}
in (\ref{AdSWEJ}), we find for $d=4$ the QCD light-front wave equation (\ref{LFWE}) with effective potential~\cite{deTeramond:2010ge}
\begin{equation} \label{U}
U(\zeta) = \half \varphi''(\zeta) +\frac{1}{4} \varphi'(\zeta)^2  + \frac{2J - 3}{2 z} \varphi'(\zeta) ,
\end{equation}
provided that the fifth dimensional mass $\mu$ is related to the internal orbital angular momentum $L = max \vert L^z \vert$ and the total angular momentum $J^z = L^z + S^z$ of the hadron.  Light-front holographic mapping thus implies that the fifth dimensional AdS mass $\mu$ is not a free parameter  but scales as 
\begin{equation} \label{muRJL}
(\mu R)^2 = - (2-J)^2 + L^2.
\end{equation}
The angular momentum projections in the light-front $\hat z$ direction $L^z, S^z$ and $J^z$ are kinematical generators in the front form, so they are the natural quantum numbers to label the eigenstates of light-front  physics.  In general, a hadronic eigenstate with spin $J^z$ in the front form corresponds to an eigenstate of $J^2 = j(j+1)$ in the rest frame in the conventional instant form.
It thus has   $2 j + 1 $ degenerate states with $J^z= - j, -j+1, \cdots j-1, +j$~\cite{Brodsky:1997de}, thus $J$ represents the maximum value of $|J^z|$, $J = max \vert J^z \vert$.

If $L^2 < 0$, the LF Hamiltonian  
defined in Eq.  (\ref{LFH}) is unbounded from below
$\langle \phi \vert H_{LF} \vert \phi \rangle <0$  and the spectrum contains an
infinite number of unphysical negative values of $\mathcal{M}^2 $ which can be arbitrarily large.
As $\mathcal{M}^2$ increases in absolute value, the particle becomes
localized within a very small region near $\zeta = 0$, since  the effective potential is conformal at small  $\zeta$.
For $\mathcal{M}^2 \to - \infty$ the particle is localized at $\zeta = 0$, the
particle ``falls towards the center''~\cite{LL:1958}.
The critical value  $L=0$  corresponds to the lowest possible stable solution, the ground state of the light-front Hamiltonian.
For $J = 0$ the five dimensional mass $\mu$
 is related to the orbital  momentum of the hadronic bound state by
 $(\mu R)^2 = - 4 + L^2$ and thus  $(\mu R)^2 \ge - 4$. The quantum mechanical stability condition $L^2 \ge 0$ is thus equivalent to the Breitenlohner-Freedman stability bound in AdS~\cite{Breitenlohner:1982jf}.
The scaling dimensions are $2 + L$ independent of $J$, in agreement with the
twist-scaling dimension of a two-parton bound state in QCD.
It is important to notice that in the light-front the $SO(2)$ Casimir for orbital angular momentum $L^2$
is a kinematical quantity, in contrast to the usual $SO(3)$ Casimir $L(L+1)$ from non-relativistic physics which is
rotational, but not boost invariant. 
The  $SO(2)$ Casimir form $L^2$  corresponds to the group of rotations in the transverse LF plane.
Indeed, the Casimir operator for $SO(N) \sim S^{N-1}$ is $L(L+N-2)$.

\section{Mesons in light-front holography}

Considerable progress has recently been achieved in the study of the meson excitation spectrum in QCD from discrete lattices which is a first-principles method~\cite{Dudek:2011zz}. In practice, lattice gauge theory computations of eigenvalues beyond the ground-state are very challenging. Furthermore, states at rest are not classified according to total angular momentum  $J$ and $J_z$, but according to the irreducible representation of the lattice, and thus a large basis of interpolating operators is required for the extraction of meaningful data~\cite{Dudek:2011zz}. In contrast, the semiclassical light-front holographic wave equation (\ref{LFWE}) obtained in the previous section describes relativistic bound states at equal light-front time  with a simplicity comparable to the Schr\"odinger equation of atomic physics at equal instant time. It thus provides a framework for a first-order analytical exploration of the spectrum of mesons.
In the limit of zero-quark masses, the light-front wave equation has a geometrical equivalent to the equation of  motion in a warped AdS space-time.

\subsection{A hard-wall model for mesons\label{hardwallmesons}}

As the simplest example we consider a truncated model where quarks propagate freely in the
hadronic interior up to the confinement scale  $1/\Lambda_{\rm QCD}$. The interaction terms in the QCD Lagrangian  effectively build confinement, here depicted by 
a hard wall potential
\begin{equation}
U(\zeta) =
\left\lbrace
\begin{array}{l}
0 \quad\text{  if} \quad \zeta \le \frac{1}{\Lambda_{\rm QCD}} , \\
\infty \quad\text{if}    \quad  \zeta > \frac{1}{\Lambda_{\rm QCD}}  .
\end{array}
\right. 
\label{hardwallU}
\end{equation}
This provides an  analog of the MIT bag model~\cite{Chodos:1974je}
where quarks are permanently confined inside a finite region of space.
In contrast to bag models, boundary conditions are imposed on the 
boost-invariant variable $\zeta$, not on the bag radius at fixed time.  
The  wave functions
have support for longitudinal momentum fraction $0 < x < 1$.
The resulting model is a manifestly Lorentz invariant model
with confinement at large distances, while incorporating conformal
 behavior at small physical separation.

The eigenvalues of the  LF wave equation (\ref{LFWE}) for the potential (\ref{hardwallU}) are determined by the boundary conditions $\phi(z = 1/\Lambda_{\rm QCD}) = 0$, and are given in terms of the roots of the Bessel functions: $\mathcal{M}_{L,k} = \beta_{L,k} \Lambda_{\rm QCD}$. Light-front eigenmodes $\phi(\zeta)$ are normalized according to
\begin{equation} \label{eq:normphi}
\int_0^{\Lambda^{-1}_{\rm QCD}} d\zeta \, \phi^2(\zeta) = 1,
\end{equation}
and are given by
\begin{equation} \label{eq:phiL}
\phi_{L,k}(\zeta) = \frac{\sqrt{2} \Lambda_{QCD}}{
J_{1+L}(\beta_{L,k})} \sqrt{\zeta} J_L \! \left(\zeta \beta_{L,k} \Lambda_{QCD}\right) .
\end{equation}

Individual hadron states can be identified by their interpolating operators, which are defined at the $z \to 0$ asymptotic boundary of AdS space, and couple to the AdS field $\hat \Phi(x,z)$ (\ref{PhiT}) at the boundary limit (See Appendix \ref{interop}).  The short-distance behavior of a hadronic state is characterized by its twist (canonical dimension minus spin) $\tau = \Delta - \sigma$, where $\sigma$ is the sum over the constituent's spin $\sigma = \sum_{i =1}^n \sigma_i$. The twist of a hadron is also equal to the number of its constituent partons $n$.~\footnote{To include orbital $L$-dependence we make the substitution $\tau \to n + L$.}  

Pion interpolating operators are constructed by examining the behavior of
bilinear covariants $\bar \psi \Gamma \psi$ under charge conjugation and parity transformation.
Thus, for example, a pion interpolating operator $\bar q \gamma^+ \gamma_5 q$ creates
a state with quantum numbers $J^{PC} = 0^{- +}$, and a vector meson
interpolating operator $\bar q \gamma_\mu q$ a state $1^{- -}$. Likewise the operator $\bar q \gamma_\mu \gamma_5 q$ creates a state with
$1^{++}$ quantum numbers, for example the $a_1(1260)$ positive parity meson. 
If we include  orbital excitations, 
the pion interpolating operator is
$\mathcal{O}_{2+L} = \bar q \gamma^+ \gamma_5  D_{\{\ell_1} \cdots D_{\ell_m\}} q$. This is an operator  with total internal  orbital
momentum $L = \sum_{i=1}^m \ell_i$, twist $\tau = 2 + L$ and canonical dimension $\Delta = 3 + L$.  Similarly the vector-meson interpolating operator is given by
$\mathcal{O}_{2+L}^\mu = \bar q \gamma^\mu D_{\{\ell_1} \cdots D_{\ell_m\}} q$.   
The scaling of  the AdS field $\hat \Phi$ (\ref{PhiT}) near $z \to 0$, 
$\hat \Phi(z) \sim z^\tau$,  is precisely the scaling required to match the scaling dimension of the local meson interpolating operators.

 \begin{table}[htdp]
 \caption{\small Confirmed $I=1$ mesons listed by PDG~\cite{Nakamura:2010zzi}. The labels $L$,  $S$ and  $n$ refer to assigned internal orbital angular momentum, internal spin and radial quantum number respectively.  For a $q \bar q$ state $P = (-1)^{L+1}$, $C = (-1)^{L+S}$. }
  \begin{center} 
 {\begin{tabular}{@{}cccccc@{}}
 \hline\hline \vspace{0pt}
 $L$ &    $S$ & $ n$ &  $J^{PC}$ &   Meson State & 
 \\[0.2ex]
 \hline
 \multicolumn{6}{c}{}\\[-3.0ex]
 0 & 0 & 0  &  $0^{-+}$   & $\pi(140)~$    \\[0.0ex]
 0 & 0 & 1  &  $0^{-+}$   & $\pi(1300)$    \\[0.0ex]
 0 & 0 & 2  &  $0^{-+}$   & $\pi(1800)$    \\[0.0ex]
 0 & 1 & 0  &  $1^{- -}$  &  $\rho(770)~$   \\[0.0ex] 
 0 & 1 & 1  &  $1^{- -}$  &  $\rho(1450)~$   \\[0.0ex] 
 0 & 1 & 2  &  $1^{- -}$  &  $\rho(1700)~$   \\[0.5ex]   \hline \\  [-3.0ex] 
 1 & 0 & 0  &  $ 1^{+-}$  & $b_1(1235)$  \\[0.0ex]
 1  & 1 & 0 &  $0^{++}$  & $a_0(980)~$ \\[0.0ex]
 1  & 1 & 1 &  $0^{++}$  & $a_0(1450)~$ \\[0.0ex]
 1  & 1 & 0 &  $1^{++}$  & $a_1(1260)$ \\[0.0ex]
 1  & 1 & 0 &  $2^{++}$  & $a_2(1320)$   \\[0.5ex]  \hline \\  [-3.0ex] 
 2  & 0 & 0 &  $2^{-+}$   & $\pi_2(1670)$ \\[0.0ex]
 2  & 0 & 1 &  $2^{-+}$   & $\pi_2(1880)$ \\[0.0ex]
 2  & 1 & 0 &  $3^{--}$    &$\rho_3(1690)$ \\[0.5ex] \hline \\  [-3.0ex] 
 3  & 1 & 0 &  $4^{++}$  & $ a_4(2040)$ \\[0.5ex] 
 \hline\hline
 \end{tabular}}
 \end{center}
 \label{mesons}
 \end{table}

We list in Table \ref{mesons} the confirmed (4-star and 3-star) isospin $I =1$ mesons states from the updated Particle Data Group (PDG)~\cite{Nakamura:2010zzi},
with their assigned internal spin, orbital angular momentum and radial quantum numbers. The $I = 1$ mesons have quark content $\vert u \bar d \rangle$,
$\frac{1}{\sqrt{2}} \vert u \bar u - d \bar d \rangle$ and  $\vert d \bar u \rangle$. The $I=1$ mesons are the $\pi$, $b$, $\rho$ and $a$ mesons.  We have not listed in Table  \ref{mesons} the $I=0$ mesons which are a mix of $u \bar u$, $d \bar d$ and $ s \bar s$, thus more complex entities. The light $I = 0$ mesons  are $\eta$, $\eta'$, $h$, $h'$ $\omega$, $\phi$, $f$ and $f'$.
This list comprises the puzzling $I=0$ scalar $f$-mesons, which may be interpreted as a superposition of tetra-quark  states with a $q \bar q$,  $L=1$, $S=1$,  configuration which couple to a $J=0$ state~\cite{Klempt:2007cp}.~\footnote{The interpretation of the $\pi_1(1400)$ is not very clear~\cite{Klempt:2007cp} and is not included in Table  \ref{mesons}. Likewise we do not include the $\pi_1(1600)$ in the present analysis.}

\begin{figure}[h]
\centering
\includegraphics[width=8.0cm]{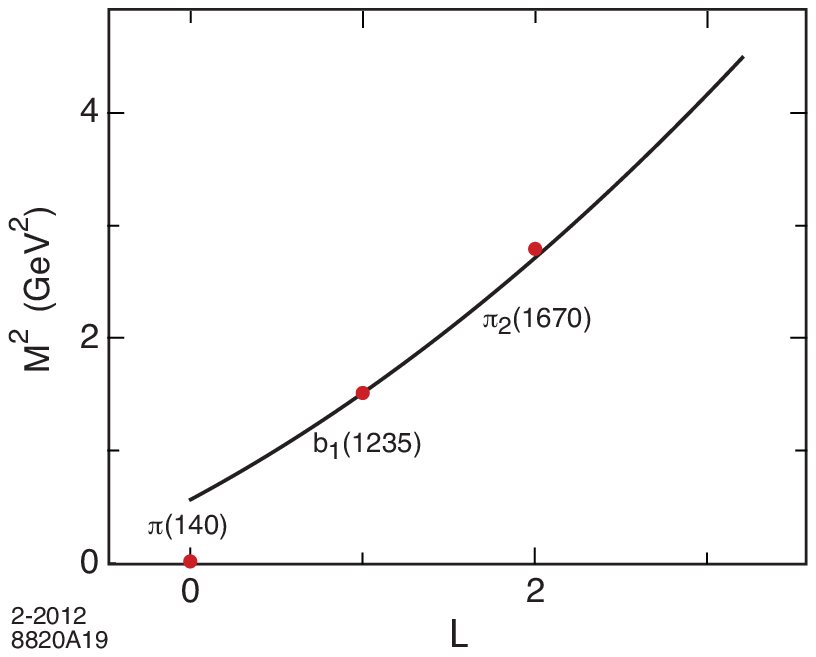}  \hspace{0pt}
\includegraphics[width=8.0cm]{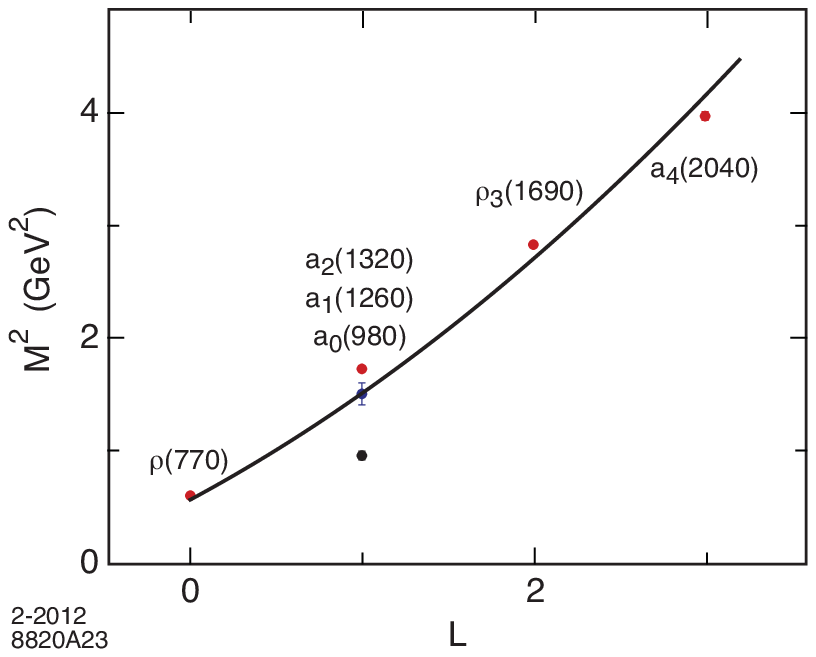}
 \caption{\small $I=1$  light-meson orbital states in the hard wall model for $\Lambda_{\rm QCD}$ = 0.32 GeV: pseudoscalar mesons  (left) and  vector mesons (right).}
\label{pionspechw}
\end{figure} 

The light $I=1$  orbital meson spectrum is compared in Fig. \ref{pionspechw} with the truncated-space model for $n=0$. The data is from PDG~\cite{Nakamura:2010zzi}.
The predictions   for the lower mass mesons are in better agreement with data as compared with Ref. \cite{deTeramond:2005su}, where naive conformal dimensions were used instead. However the model fails to account for the pion as a chiral $\mathcal{M}_\pi = 0$ state.  The hard-wall model for mesons is degenerate with respect to the orbital quantum number $L$, and thus fails to account for the important $L = \vert L^z \vert = 1$ triplet splitting shown in Fig.  \ref{pionspechw} (right);  the $a_0(980)$, $a_1(1260)$ and $a_2(1320)$ states, which corresponds to $J = \vert J^z \vert = 0, 1, 2$ respectively.  Using the asymptotic expansion  of the Bessel function for large arguments we find that $\mathcal{M} \sim 2n + L$, in contrast to the usual Regge dependence $\mathcal{M}^2 \sim n + L$ found experimentally~\cite{Klempt:2007cp}. As a consequence, the radial modes are not well described in the truncated-space model. For example the first radial AdS eigenvalue has a mass 1.77 GeV, which is too high compared to the mass of the observed first  radial excitation of the meson, the $\pi(1300)$.  The  shortcomings of the hard-wall model described in this section are evaded in the soft wall model discussed below, where the sharp cutoff is modified.

\subsection{A soft-wall model for mesons \label{softwallmesons}}

As we discussed in Sec. \ref{LFmapping}, the conformal metric of AdS space can be modified  within the gauge/gravity framework  to include  confinement by  the introduction of an additional warp factor or, equivalently, with a dilaton
background $\varphi(z)$, which breaks the conformal invariance of the theory.
A particularly interesting case is a dilaton profile $\exp{\left(\pm \kappa^2 z^2\right)}$ of either sign, since it 
leads to linear Regge trajectories~\cite{Karch:2006pv} and avoids the ambiguities in the choice of boundary conditions at the infrared wall.  
The corresponding modified metric  can be interpreted in the higher dimensional warped AdS space as a gravitational potential in the fifth dimension
\begin{equation}
V(z) = mc^2 \sqrt{g_{00}} = mc^2 R  \frac{e^{\pm 3 \kappa^2 z^2/4}}{z} .
\end{equation}
In the case of the negative solution,
the potential decreases monotonically, and thus an object located in the boundary of AdS space will fall to infinitely large 
values of $z$.  This is illustrated in detail by Klebanov and Maldacena in Ref.  \cite{Klebanov:2009zz}.
For the positive solution, the potential is nonmonotonic and has an absolute minimum at $z_0 \sim 1/\kappa$.  
Furthermore, for large values of $z$ the gravitational potential increases exponentially, thus confining any object  to distances 
$\langle z \rangle \sim 1/\kappa$~\cite {Andreev:2006vy, deTeramond:2009xk}.

From (\ref{U}) we obtain for the positive sign confining solution $\varphi = \exp{\left(\kappa^2 z^2\right)}$ the effective potential~\cite{deTeramond:2009xk}
\begin{equation} \label{Ukappa}
U(\zeta) =   \kappa^4 \zeta^2 + 2 \kappa^2(J - 1),
\end{equation}
which  corresponds  to a transverse oscillator in the light-front.
For the effective potential (\ref{Ukappa}) equation  (\ref{LFWE}) has eigenfunctions
\begin{equation} \label{phi}
\phi_{n, L}(\zeta) = \kappa^{1+L} \sqrt{\frac{2 n!}{(n\!+\!L\!)!}} \, \zeta^{1/2+L}
e^{- \kappa^2 \zeta^2/2} L^L_n(\kappa^2 \zeta^2) ,
\end{equation}
and eigenvalues~\footnote{Similar results are found in Ref. \cite{Gutsche:2011vb}.}
\begin{equation} \label{M2SFM}
\mathcal{M}_{n, J, L}^2 = 4 \kappa^2 \left(n + \frac{J+L}{2} \right).
\end{equation}

\begin{figure}[!]
\centering
\includegraphics[angle=0,width=7.2cm]{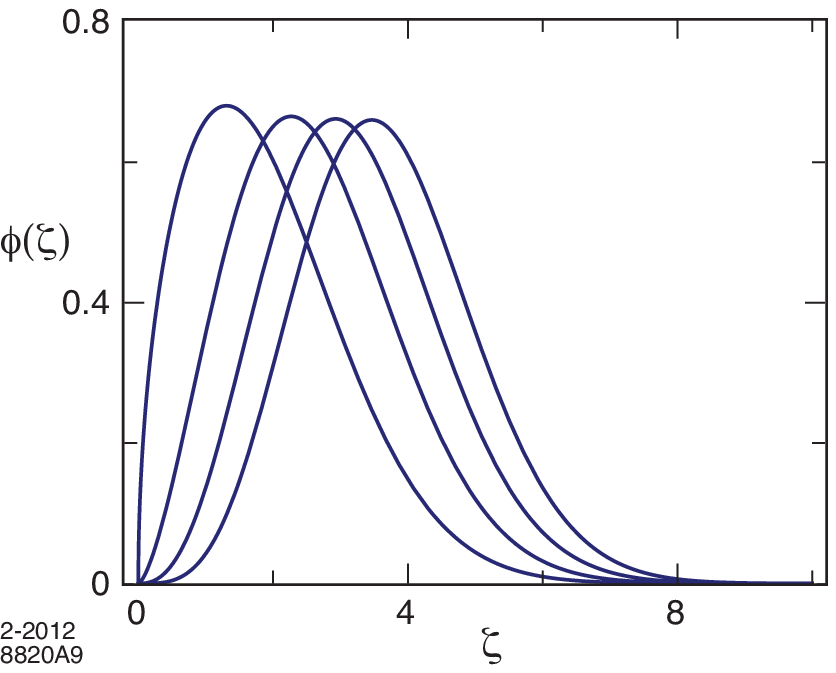} \hspace{10pt}
\includegraphics[angle=0,width=7.2cm]{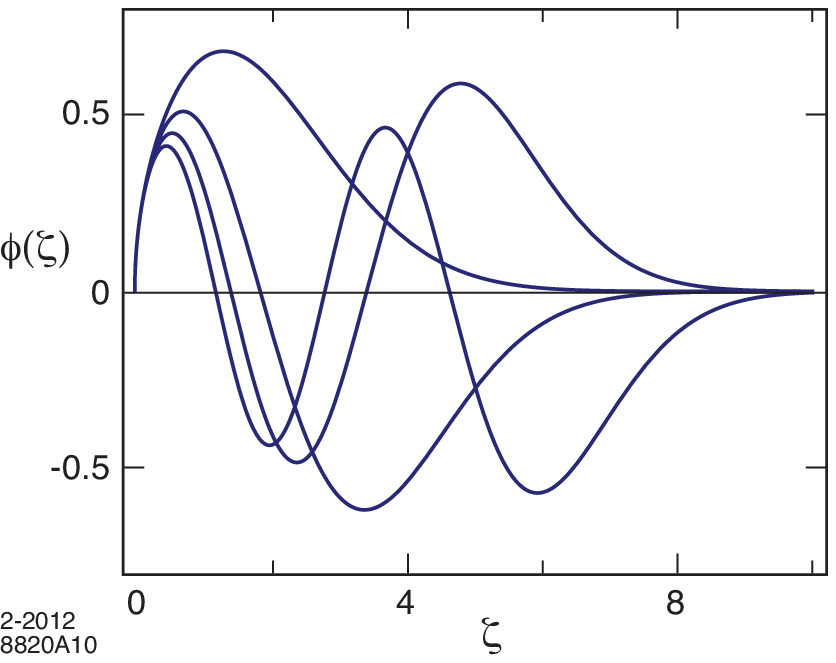}
\caption{\small Light-front wavefunctions $\phi_{n,L}(\zeta)$  in physical space-time corresponding to a dilaton profile $\exp(\kappa^2 z^2)$: (left) orbital modes ($n=0$) and (right) radial modes ($L=0$).}
\label{LFWFs}
\end{figure}

The meson spectrum (\ref{M2SFM}) has a string-theory Regge form $\mathcal{M}^2 \sim n + L$: the square of the eigenmasses are linear in both the angular momentum $L$ and radial quantum number $n$, where $n$ counts the number of nodes  of the wavefunction in the radial variable $\zeta$. 
The LFWFs (\ref{phi}) for different orbital and radial excitations are depicted in Fig. \ref{LFWFs}. Constituent quark and antiquark separate from each other as the orbital and
radial quantum numbers increase. The number of nodes in the light-front wave function depicted in Fig. \ref{LFWFs} (right) correspond to the radial excitation quantum number $n$.

\begin{figure}[h]
\centering
\includegraphics[width=8.0cm]{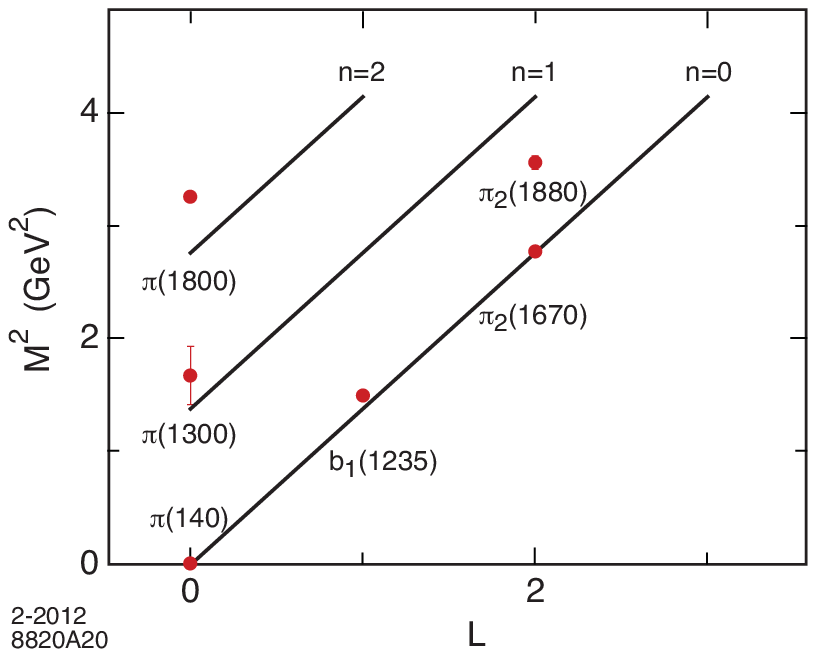}  \hspace{0pt}
\includegraphics[width=8.0cm]{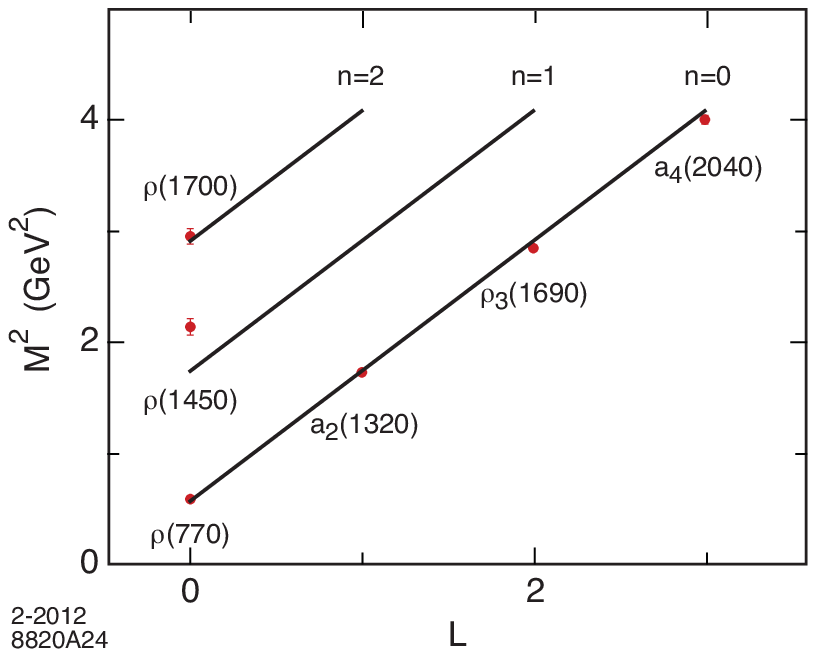}
 \caption{\small $I\!=\!1$ parent and daughter Regge trajectories for the $\pi$-meson family (left) with
$\kappa= 0.59$ GeV; and  the   $\rho$-meson
 family (right) with $\kappa= 0.54$ GeV.}
\label{pionspec}
\end{figure} 

 For the $J = L + S$ meson families Eq. (\ref{M2SFM}) becomes
 \begin{equation} \label{M2SFMLS}
 \mathcal{M}_{n,L,S}^2 = 4 \kappa^2  \left(n + L + \frac{S}{2}\right).
 \end{equation}
 The lowest possible solution for $n = J = 0$ has eigenvalue $\mathcal{M}^2 = 0$.
This is a chiral symmetric bound state of two massless quarks with scaling dimension 2 and size 
 $\langle \zeta^2 \rangle \sim 1/\kappa^2$, which we identify with the lowest state, the pion.
Thus one can compute the corresponding Regge families by simply adding  $4 \kappa^2$ for a unit change in the radial quantum number, $4 \kappa^2$ for a change in one unit in the orbital quantum number and $2 \kappa^2$ for a change of one unit of spin to the ground state value of $\mathcal{M}^2$. 
The spectral predictions  for the $J = L + S$ light pseudoscalar and vector meson  states,  listed in  Table. \ref{mesons}, are  compared with experimental data 
in Fig. \ref{pionspec} for the positive sign dilaton model discussed here.  The data is from PDG~\cite{Nakamura:2010zzi}.

 It is important to notice that in contrast to the hard-wall model, the soft-wall model with positive dilaton accounts for the mass pattern observed in radial excitations, as well as for the triplet splitting for the $L=1$, $J = 0,1,2$, vector meson $a$-states. Using the spectral formula (\ref{M2SFM}) we find 
 \begin{equation}
 \mathcal{M}_{a_2(1320)} >  \mathcal{M}_{a_1(1260)} >   \mathcal{M}_{a_0(980)}.
 \end{equation}
 The predicted values are 0.76, 1.08 and 1.32 GeV for the masses of the  $a_0(980)$, $a_1(1260)$ and   $a_2(1320)$ vector mesons, compared with the experimental values 0.98, 1.23 and 1.32  GeV respectively. The prediction for the mass of the $L=1$, $n=1$  state $a_0(1450)$ is  1.53 GeV, compared with the observed value 1.47 GeV.
 For other calculations of the hadronic spectrum in the framework of AdS/QCD, see Refs.~\cite{BoschiFilho:2005yh, Evans:2006ea, Hong:2006ta, Colangelo:2007pt, Forkel:2007cm, Forkel:2007ru, Vega:2008af, Nawa:2008xr,   Colangelo:2008us, Forkel:2008un, Ahn:2009px, Sui:2009xe, Kapusta:2010mf, Zhang:2010bn, Iatrakis:2010zf,  Branz:2010ub, Kirchbach:2010dm, Sui:2010ay}.~\footnote{For recent reviews see, for example, Refs.~\cite{Erlich:2009me, Kim:2011ey}.}

\section{Meson form factors  \label{lEMFF}}

A form factor in QCD is defined by the transition matrix element of a local quark current between hadronic states.  
The great advantage of the front form  -- as emphasized by Dirac --  is that boost operators are kinematic.  Unlike the instant form,  the boost operators in the front form have no interaction terms.  The calculation of a current matrix element $\langle P + q \vert J^\mu \vert P \rangle$ requires boosting the hadronic  eigenstate from $\vert P \rangle $ to $\vert P + q \rangle $, a task which becomes hopelessly complicated in the instant form.
In addition, the virtual photon couples to connected currents which arise from the instant form vacuum.

In AdS space form factors are computed from the overlap integral of normalizable modes with boundary currents which propagate in AdS space. The AdS/CFT duality incorporates the connection between the twist scaling dimension of the  QCD boundary interpolating operators to the falloff of the normalizable modes in AdS near its conformal boundary. If both quantities represent the same physical observable for any value of the transferred momentum squared $q^2$, 
a precise correspondence can be established between the string modes $\Phi$ in AdS space and the light front wavefunctions of hadrons $\psi_n$ in physical four dimensional space-time~\cite{Brodsky:2006uqa}. 
In fact, Light-Front Holography was originally  derived by observing the correspondence between matrix elements obtained in AdS/CFT with the corresponding formula using the light-front
representation~\cite{Brodsky:2006uqa}.
The same results follow from comparing the relativistic light-front Hamiltonian equation describing bound states in QCD with the wave equations describing the propagation of modes in a warped AdS 
space, as shown in the previous section~\cite{deTeramond:2008ht}.

\subsection{Meson electromagnetic form factor \label{sec:MFF}}

In the higher dimensional gravity theory, the hadronic transition matrix element  corresponds to
the  coupling of an external electromagnetic field $A^M(x,z)$,  for a photon propagating in AdS space, with the extended field $\Phi_P(x,z)$ describing a meson in AdS is~\cite{Polchinski:2002jw}
 \begin{equation} \label{MFF}
 \int \! d^4x \, dz  \sqrt{g} \, A^M(x,z)
 \Phi^*_{P'}(x,z) \overleftrightarrow\partial_M \Phi_P(x,z)
  \sim
 (2 \pi)^4 \delta^4 \left( P'  \! - P - q\right) \epsilon_\mu  (P + P')^\mu F_M(q^2) .
 \end{equation}
To simplify the discussion we will first describe a  model with a wall at  $z \sim 1/\Lambda_{\rm QCD}$ -- the hard wall model -- which limits the propagation of the string modes in AdS space beyond  the IR separation $z \sim 1/\Lambda_{\rm QCD}$ and also
sets the gap scale~\cite{Polchinski:2001tt}. 
 We recall from Sec. \ref{HigherSpin} that the coordinates of AdS$_5$ are the Minkowski coordinates $x^\mu$ and $z$ labeled $x^M = (x^\mu, z)$,
 with $M, N = 1, \cdots 5$,  and $g$ is the determinant of the metric tensor. 
The pion has initial and final four momentum $P$ and $P'$ respectively and $q$ is the four-momentum transferred to the pion by the photon with polarization $\epsilon_\mu$.
The expression on the right-hand side
of (\ref{MFF}) represents the space-like QCD electromagnetic transition amplitude in physical space-time
\begin{equation}
\langle P' \vert J^\mu(0) \vert P \rangle = \left(P + P' \right)^\mu F_M(q^2).
\end{equation}
It is the EM matrix element of the quark current  $J^\mu = e_q \bar q \gamma^\mu q$, and represents a local coupling to pointlike constituents. Although the expressions for the transition amplitudes look very different, one can show  that a precise mapping of the matrix elements  can be carried out at fixed light-front time~\cite{Brodsky:2006uqa, Brodsky:2007hb}.

The form factor is computed in the light front from the matrix elements of the plus-component of the current $J^+$
 in order to avoid coupling to Fock states with different numbers of constituents.
Expanding the  initial and final meson states $\vert \psi_M(P^+ \! , \mbf{P}_\perp)\rangle$ in terms of Fock components, $\vert \psi_M \rangle = \sum_n \psi_{n/M} \vert n \rangle$, we obtain the
 DYW expression~\cite{Drell:1969km, West:1970av} upon the phase space integration over the intermediate variables in the $q^+=0$ frame:
 \begin{equation} \label{eq:DYW}
F_M(q^2) = \sum_n  \int \big[d x_i\big] \left[d^2 \mbf{k}_{\perp i}\right]
\sum_j e_j \psi^*_{n/M} (x_i, \mbf{k}'_{\perp i},\lambda_i)
\psi_{n/M} (x_i, \mbf{k}_{\perp i},\lambda_i),
\end{equation}
where the phase space factor $[d x_i\big] \left[d^2 \mbf{k}_{\perp i}\right]$ is given by (\ref{phases}) and the variables of the light cone Fock components in the
final-state are given by $\mbf{k}'_{\perp i} = \mbf{k}_{\perp i}
+ (1 - x_i)\, \mbf{q}_\perp $ for a struck  constituent quark and
$\mbf{k}'_{\perp i} = \mbf{k}_{\perp i} - x_i \, \mbf{q}_\perp$ for each
spectator. The formula is exact if the sum is over all Fock states $n$.
The form factor can also be conveniently written in impact space
as a sum of overlap of LFWFs of the $j = 1,2, \cdots, n-1$ spectator constituents~\cite{Soper:1976jc} 
\begin{equation} \label{eq:FFb}
F_M(q^2) =  \sum_n  \prod_{j=1}^{n-1}\int d x_j d^2 \mbf{b}_{\perp j}
\exp \! {\Bigl(i \mbf{q}_\perp \! \cdot \sum_{j=1}^{n-1} x_j \mbf{b}_{\perp j}\Bigr)}
\left\vert  \psi_{n/M}(x_j, \mbf{b}_{\perp j})\right\vert^2,
\end{equation}
corresponding to a change of transverse momentum $x_j \mbf{q}_\perp$ for each
of the $n-1$ spectators with  $\sum_{i = 1}^n \mbf{b}_{\perp i} = 0$.

For definiteness we shall consider 
the $\pi^+$  valence Fock state 
$\vert u \bar d\rangle$ with charges $e_u = \frac{2}{3}$ and $e_{\bar d} = \frac{1}{3}$.
For $n=2$, there are two terms which contribute to Eq. (\ref{eq:FFb}). 
Exchanging $x \leftrightarrow 1 \! - \! x$ in the second integral  we find 
\begin{equation}  \label{eq:PiFFb}
 F_{\pi^+}(q^2)  =  2 \pi \int_0^1 \! \frac{dx}{x(1-x)}  \int \zeta d \zeta \,
J_0 \! \left(\! \zeta q \sqrt{\frac{1-x}{x}}\right) 
\left\vert \psi_{u \bar d/ \pi}\!(x,\zeta)\right\vert^2,
\end{equation}
where $\zeta^2 =  x(1  -  x) \mathbf{b}_\perp^2$ and $F_{\pi^+}(q\!=\!0)=1$.

We now compare this result with the electromagnetic  form factor 
in  AdS  space-time. The incoming electromagnetic field propagates in AdS according to
$A_\mu(x^\mu ,z) = \epsilon_\mu(q) e^{-i q \cdot x} V(q^2, z)$ in the gauge $A_z = 0$ (no physical polarizations along the AdS variable $z$).  The bulk-to-boundary propagator
 $V(q^2,z)$ is the solution of the AdS wave equation for $A_\mu(x^\mu ,z)$
given by ($Q^2 = - q^2 > 0$)
\begin{equation} \label{eq:V}
V(Q^2, z) = z Q K_1(z Q),
\end{equation}
with boundary conditions~\cite{Polchinski:2002jw} 
\begin{equation} \label{BCV}
V(Q^2 = 0, z ) = V(Q^2, z = 0) = 1.
\end{equation}
The propagation of the pion in AdS space is described by a normalizable mode
$\Phi_P(x^\mu, z) = e^{-i P  \cdot x} \Phi(z)$ with invariant  mass $P_\mu P^\mu = \mathcal{M}^2$ and plane waves along Minkowski coordinates $x^\mu$.  
Extracting the overall factor  $(2 \pi)^4 \delta^4 \left( P'  \! - P - q\right)$ from momentum conservation at the vertex 
which arises from integration over Minkowski variables in (\ref{MFF}), we find
\cite{Polchinski:2002jw} 
\begin{equation}
F(Q^2) = R^3 \int \frac{dz}{z^3} \, V(Q^2, z)  \, \Phi^2(z),
\label{eq:FFAdS}
\end{equation}
where $F(Q^2\! = 0) = 1$. 
Using the integral representation of $V(Q^2,z)$
\begin{equation} \label{eq:intJ}
V(Q^2, z) = \int_0^1 \! dx \, J_0 \! \left(\! z  Q \sqrt{\frac{1-x}{x}}\right) ,
\end{equation} we write the AdS electromagnetic form-factor as
\begin{equation} 
F(Q^2)  =    R^3 \! \int_0^1 \! dx  \! \int \frac{dz}{z^3} \, 
J_0\!\left(\!z Q\sqrt{\frac{1-x}{x}}\right)  \Phi^2(z) .
\label{eq:AdSFx}
\end{equation}

To compare with  the light-front QCD  form factor expression (\ref{eq:PiFFb})  we 
use the expression of the LFWF  (\ref{eq:psiphi}) in the transverse LF plane, where we  factor out the longitudinal
 and transverse modes $\phi(\zeta)$ and $X(x)$ respectively. If both expressions for the form factor are to be
identical for arbitrary values of $Q$, we obtain $\phi(\zeta) = (\zeta/R)^{3/2} \Phi(\zeta)$ and $X(x) = \sqrt{x(1-x)}$~\cite{Brodsky:2006uqa},
where we identify the transverse impact LF variable $\zeta$ with the holographic variable $z$,
$z \to \zeta = \sqrt{x(1-x)} \vert \mbf b_\perp \vert$.~\footnote{Extension of the results to arbitrary $n$ follows from the $x$-weighted definition of the transverse impact variable of the $n-1$ spectator system given by Eq. (\ref{zetan}). In general the mapping relates the AdS density  $\Phi^2(z)$ to an effective LF single particle transverse density~\cite{Brodsky:2006uqa}. }
Thus, in addition of recovering the expression found in Sec. \ref{LFmapping} which relates the transverse mode
 $\phi(\zeta)$ in physical space-time to the field $\Phi$ in AdS space, we find a definite expression for the longitudinal LF mode $X(x)$.
Identical results follow from mapping the matrix elements of the energy-momentum tensor~\cite{Brodsky:2008pf}.

\subsection{Elastic form factor with a  dressed current \label{EFFDC}}

The results for the elastic form factor described above correspond to a ÒfreeÓ current propagating on AdS space. It is dual to the electromagnetic
point-like current in the Drell-Yan-West
light-front formula~\cite{Drell:1969km, West:1970av} for the pion form factor.  
The DYW formula is an exact expression for the form factor. It is written as an infinite sum of an overlap of LF Fock components with an arbitrary number of constituents.
This allows one to map state-by-state to the effective gravity theory in AdS space. 
However, this mapping has the shortcoming that the multiple pole structure of the time-like form factor does not appear in the time-like region unless an infinite number of Fock states is included. Furthermore, the moments of the form factor at  $Q^2 = 0$ diverge term-by-term; for example one obtains an infinite charge radius~\cite{deTeramond:2011yi}. This could have been expected, as we are dealing with a massless quark approximation. In fact, infinite slopes also occur in chiral theories when coupling to a massless pion.

\begin{figure}[h]
\centering
\includegraphics[angle=0,width=7.6cm]{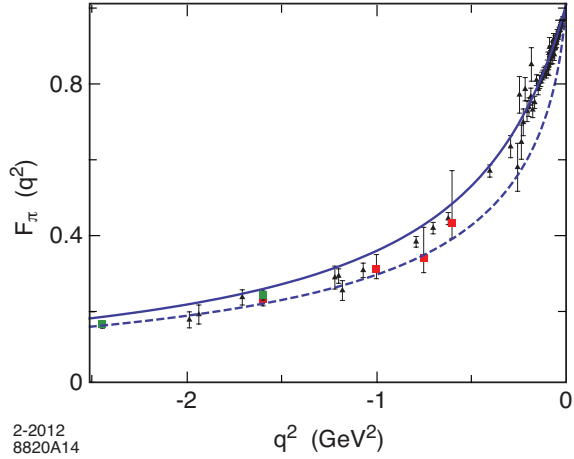} 
\caption{\small space-like electromagnetic pion form factor $F_\pi(q^2)$. Continuous line: confined current, dashed  line free current. Triangles are the data compilation  from Baldini~\cite{Baldini:1998qn}, boxes  are JLAB data~\cite{Tadevosyan:2007yd}. }
\label{PionFF}
\end{figure} 

Alternatively, one can use a truncated basis of states in the LF Fock expansion with a limited number of constituents and the nonperturbative pole structure can be generated with a  dressed EM current as in the Heisenberg picture, {\it i.e.},  the EM current becomes modified as it propagates in 
an IR deformed AdS space to simulate confinement.
The dressed current is dual to a hadronic EM current which includes any number of virtual $q \bar q$ components.  
The confined EM  current  also leads to finite moments at $Q^2=0$, as illustrated 
in  Fig. \ref{PionFF} for the EM pion form factor.

We describe briefly below how to compute a form factor for a confined current  in AdS space using a soft wall example. However, the actual computation of a form factor in AdS has several caveats which we will discuss in Sec. \ref{caveats}.

The effective potential corresponding to a dilaton profile $\exp(\pm \kappa^2 z^2)$ has the form of a harmonic oscillator confining potential $\kappa^4 z^2$. The normalizable solution for a meson of  twist $\tau$ (the number of constituents for a given Fock component) corresponding to
 the lowest radial $n = 0$ and orbital $L=0$ state is given by
\begin{equation}  \label{eq:Phitau}
\Phi^\tau(z) =   \sqrt{\frac{2 P_{\tau}}{\Gamma(\tau \! - \! 1)} } \, \kappa^{\tau -1} z ^{\tau} e^{- \kappa^2 z^2/2},
\end{equation} 
with normalization
\begin{equation} \label{eq:PhitauNorm}
\langle\Phi^\tau\vert\Phi^\tau\rangle = \int \frac{dz}{z^3} \, e^{- \kappa^2 z^2} \Phi^\tau(z)^2  = P_\tau,
\end{equation}
where $P_\tau$ is the probability for the twist $\tau$ mode (\ref{eq:Phitau}). 
This agrees with the fact that the field $\Phi^\tau$ couples to a local hadronic interpolating operator of twist $\tau$ 
defined at the asymptotic boundary of AdS space (See Appendix \ref{interop}), and thus
the scaling dimension of $\Phi^\tau$ is $\tau$. 

In the case of a soft-wall potential
the EM bulk-to-boundary propagator is~\cite{Brodsky:2007hb, Grigoryan:2007my}
\begin{equation} \label{eq:Vkappa}
V(Q^2,z) = \Gamma\left(1 + \frac{Q^2}{4 \kappa^2}\right) U\left(\frac{Q^2}{4 \kappa^2}, 0, \kappa^2 z^2\right),
\end{equation}
where $U(a,b,c)$ is the Tricomi confluent hypergeometric function
\begin{equation}
 \Gamma(a)  U(a,b,z) =  \int_0^\infty \! e^{- z t} t^{a-1} (1+t)^{b-a-1} dt.
  \end{equation} 
The modified current $V(Q^2,z)$, Eq. (\ref{eq:Vkappa}),  has the same boundary conditions (\ref{BCV}) as the free current (\ref{eq:V}),
and reduces to (\ref{eq:V}) in the  limit $Q^2 \to \infty$~\cite{Brodsky:2007hb}.
Eq.~(\ref{eq:Vkappa}) can be conveniently written in terms of the integral representation~\cite{Grigoryan:2007my}
\begin{equation}  \label{Vx}
V(Q^2,z) = \kappa^2 z^2 \int_0^1 \! \frac{dx}{(1-x)^2} \, x^{\frac{Q^2}{4 \kappa^2}} 
e^{-\kappa^2 z^2 x/(1-x)}.
\end{equation} 

Substituting in (\ref{eq:FFAdS}) the expression for the hadronic state (\ref{eq:Phitau}) with twist $\tau$  and the bulk-to-boundary propagator (\ref{Vx}), we find that the corresponding
elastic form factor for a twist $\tau$ Fock component  $F_\tau(Q^2)$
($Q^2 = - q^2 > 0$)~\cite{Brodsky:2007hb} 
\begin{equation} \label{Ftau}   
 F_\tau(Q^2) =  \frac{P_\tau}{{\Big(1 + \frac{Q^2}{\mathcal{M}^2_\rho} \Big) }
 \Big(1 + \frac{Q^2}{\mathcal{M}^2_{\rho'}}  \Big)  \cdots 
       \Big(1  + \frac{Q^2}{\mathcal{M}^2_{\rho^{\tau-2}}} \Big)} ,
\end{equation}
which is  expressed as a $\tau - 1$ product of poles along the vector meson Regge radial trajectory.
For a pion, for example, the lowest Fock state -- the valence state -- is a twist-2 state, and thus the form factor is the well known monopole form~\cite{Brodsky:2007hb}. Thus the mean-square charge radius of the pion $\langle r_\pi^2\rangle =  6/ \mathcal{M}_\rho^2$ in the valence  approximation. For $\mathcal{M}_\rho \simeq  770$ MeV we find $\langle r_\pi \rangle \simeq 0.63$ fm, compared with the experimental value  $\langle r_\pi \rangle = 0.672 \pm 0.008$ fm~\cite{Nakamura:2010zzi}.
In contrast, the computation with a free current gives the logarithmically divergent result~\footnote{The logarithmically divergent result does not appear in the hard-wall model if one uses Neumann boundary conditions. In this case the EM current is confined and $\langle r^2_\pi \rangle \sim 1 / \Lambda^2_{\rm QCD}$. A discussion of the pion form factor including chiral symmetry breaking effects in the hard-wall model is given in Refs. \cite{Kwee:2007dd} and  \cite{Grigoryan:2007wn}.}.
\begin{equation}
\langle r_\pi^2\rangle =  \frac{3}{2}  \ln\left(\frac {4 \kappa^2}{Q^2}\right)\Big\vert_{Q^2 \to 0}.
\end{equation}
The remarkable analytical form of (\ref{Ftau}),
expressed in terms of the $\rho$ vector meson mass and its radial excitations, incorporates the correct scaling behavior from the constituent's hard scattering with the photon and the mass gap from confinement.

\subsection{Effective wave function from holographic mapping of a confined current }

It is also possible to find a precise mapping of a confined EM current propagating in a warped AdS space to the light-front QCD Drell-Yan-West expression for the form factor. In this case we we find an effective LFWF, which corresponds to a superposition of an infinite number of Fock states generated by the ``dressed'' confined current. For the soft-wall model this mapping can be done analytically.   

The form factor in light-front  QCD can be expressed in terms of an effective single-particle density~\cite{Soper:1976jc}
\begin{equation} 
F(Q^2) =  \int_0^1 dx \, \rho(x,Q),
\end{equation}
where
\begin{equation} \label{rhoQCD}
\rho(x, Q) = 2 \pi \int_0^\infty \!  b \,  db \, J_0(b Q (1-x)) \vert \psi(x,b)\vert^2,
\end{equation}
for a two-parton state ($b = \vert \mbf{b}_\perp \vert$).

 We can also compute an effective density on the gravity side corresponding to a twist $\tau$ hadronic mode $\Phi_\tau$ in a modified AdS space.
 For the soft-wall model the expression is~\cite{Brodsky:2007hb}
 \begin{equation}  \label{rhoAdS}
\rho(x,Q) = (\tau \!-\!1) \, (1 - x)^{\tau-2} \, x^{\frac{Q^2}{4 \kappa^2}} .
\end{equation}
To compare (\ref{rhoAdS}) with the QCD expression (\ref{rhoQCD}) for twist-two we use the integral
\begin{equation}
\int_0^\infty \! u \, du  \, J_0(\alpha u) \,e^{- \beta u^2} = \frac{1}{2 \beta} \, e^{-\alpha^2/4\beta},
\end{equation}
and the relation $x^\gamma  = e^{\gamma \ln(x)}$. We find the effective two-parton  LFWF
\begin{equation} \label{ELFWF}
\psi(x, \mbf{b}_\perp) = \kappa \frac{ (1-x)}{\sqrt{\pi \ln(\frac{1}{x})}} \,
e^{- \half \kappa^2 \mbf{b}_\perp^2  (1-x)^2 / \ln(\frac{1}{x})},
\end{equation}
in impact space. The momentum space expression follows from the Fourier transform of  (\ref{ELFWF})
and it is given by  
\begin{eqnarray} 
\psi(x, \mbf{k}_\perp) &=& 4 \pi \, \frac{ \sqrt{\ln\left(\frac{1}{x}\right)}}{\kappa (1-x)} \,
x^{\mbf{k}_\perp^2/2 \kappa^2 (1-x)^2} \\
 &=& 4 \pi \, \frac{ \sqrt{\ln\left(\frac{1}{x}\right)}}{\kappa (1-x)} \,
e^{ - \mbf{k}_\perp^2/2 \kappa^2 (1-x)^2 \ln \left(\frac{1}{x}\right)}.
\end{eqnarray}
The effective LFWF  encodes  nonperturbative dynamical aspects that cannot be learned from a term-by-term holographic mapping, unless one includes an infinite number of terms.  Furthermore, it has the right analytical properties to reproduce the bound state vector meson pole in the time-like EM form factor. Unlike the ``true'' valence LFWF, the effective LFWF, which represents a sum of an infinite number of Fock components, is not symmetric in the longitudinal variables $x$ and $1-x$ for the active and spectator quarks,  respectively.

\subsection{Some caveats computing matrix elements in AdS/QCD \label{caveats}}

The positive dilaton background $\exp(+\kappa^2 z^2)$ used in Sec. \ref{softwallmesons} leads to a successful description  of the meson spectrum in terms of  the internal quantum numbers $n$, $L$ and $S$, and has been preferred for computations in the framework of light-front holography, where the internal structure of hadrons is encoded in the wave function.  The positive dilaton background has been discussed in the literature~\cite{Gutsche:2011vb,  Andreev:2006vy, deTeramond:2009xk,  Zuo:2009dz, Nicotri:2010at} since it has the expected behavior of a model dual to a confining theory~\cite{Sonnenschein:2000qm, Klebanov:2009zz}. 
This solution was studied in Ref.~\cite{Karch:2006pv} but discarded in the same paper, as it leads to a spurious massless scalar mode in the two-point correlation function for vector mesons~\cite{Karch:2010eg}, and a dilaton field with opposite sign, $\exp(- \kappa^2 z^2)$, was adopted instead~\cite{Karch:2006pv}.   However, using the results of  Sec. \ref{Vz},  one can readily show that the difference in the effective potential $U(z)$ corresponding to positive and negative dilaton factors $\exp(\pm\kappa^2 z^2)$
simply amounts to a $z$-independent shift in the light-front effective potential $U$, which in fact vanishes in the vector meson $J=1$ channel. From (\ref{Uz})
\begin{equation} \label{DelU}
\Delta U(z) = U_\varphi(z) - U_{-\varphi}(z) =  \varphi''(z)  + \frac{2J - d + 1}{z} \varphi'(z),
\end{equation}
in agreement with the results found in Ref.~\cite{Gutsche:2011vb}.

For the dilaton profile $\varphi= k^2 z^2$ we find for $d=4$
\begin{equation}
\Delta U =  4 (J-1) \kappa^2.
\end{equation}
Therefore, from the point of view of light-front physics, plus and minus dilaton soft-wall solutions are equivalent upon a redefinition of the eigenvalues for $J \ne 1$.  
For $J=1$ the effective potential is 
$U = \kappa^4 z^2$, identical for the plus and minus solutions~\cite{Afonin:2010hn}. Thus, the five-dimensional effective AdS action for a conserved EM current $V_M$ in presence of a confining potential $U = \kappa^4 z^2$~\cite{Afonin:2010hn}
\begin{equation} \label{SA}
S =  \int \! d^4 x \, dz  \sqrt{g}  
  \left( \frac{1}{4} F_{MN } F^{MN}
 -  \frac{ \kappa^4 z^4}{2 R^2} V_{M} V^{M}  \right) ,
\end{equation}
where $F_{MN} = \partial_M V_N - \partial_N V_M$, only differs by a surface term from the action corresponding to plus or minus dilaton profiles. Equivalently, one
can start from the five-dimensional action (\ref{SA}). Upon the field redefinition $V_M \to e^{ \pm \kappa^2 z^2/2} V_M$ one obtains the five-dimensional actions corresponding to plus or minus dilaton solutions, which differ from (\ref{SA}) only by a surface term. Consequently, essential physics cannot dependent on the particular choice of the dilaton sign.

Another difficulty found in the holographic approach to QCD   is that the vector meson masses obtained from the spin-1 equation of motion do not match the poles of the dressed current when computing a form factor. The discrepancy, in the case of the pion, is an overall  factor of $\sqrt 2$ between the value of the gap scale which follows from the spectrum or from the computation of the pion form factor in the valence state approximation.\footnote{This discrepancy is also present in the gap scale if one computes the spectrum and  form factors without recourse to holographic methods, for example using the semi-classical approximation of Ref. \cite{deTeramond:2008ht}. In this case a discrepancy of a factor factor $\sqrt 2$ is also found between the spectrum and the computation of space-like form factors.}  This is quite puzzling, since the same discrepancy  is also found, for example, when computing a space-like form factor using  the Drell-Yan-West expression, which is an exact expression if all Fock states are included.
In AdS conserved currents are not renormalized and correspond to five dimensional massless fields propagating in AdS according to the relation 
$(\mu R)^2 = (\Delta - p) (\Delta + p -  4)$  for a $p$ form. In the usual AdS/QCD framework~\cite{Erlich:2005qh, DaRold:2005zs} this  corresponds for $p=1$ to $\Delta = 3$ or 1,  the canonical dimensions of
an EM current and the massless gauge field respectively.  Normally, one uses a hadronic  interpolating operator  with minimum twist $\tau$ to identify a hadron and to predict the power-law fall-off behavior of its form factors and other hard 
scattering amplitudes~\cite{Polchinski:2001tt}; {\it e.g.},  for a two-parton bound state $\tau = 2$.   However, in the case of a current, one needs to  use  an effective field operator  with dimension $\Delta =3.$ The apparent inconsistency between twist (\ref{muR}) and canonical dimension is removed by noticing that in the light-front one chooses to calculate the  matrix element of the twist-3 plus  component of the ``good" current  $J^+$~\cite{Brodsky:2006uqa, Brodsky:2007hb}, in order to avoid coupling to Fock states with different numbers of constituents~\cite{Drell:1969km, West:1970av}.

As described in Sec. \ref{LFmapping}, light front holography provides a precise relation of the fifth-dimensional mass $\mu$ with the total and orbital angular momentum of a hadron in the  transverse LF plane   $(\mu R)^2 = - (2 - J)^2 + L^2$ (\ref{muRJL}). Thus the poles  computed from the AdS wave equations for a conserved  current $\mu R = 0$, correspond  to a $J = L = 1$ twist-3 state.  Following this, we can compute the mass of the radial excitations of the twist-3 vector family $J = L = 1$
using Eq. (\ref{M2SFM}). The result is  
\begin{equation}  \label{Mtwist3}
\mathcal{M}^2_{n, J=1, L=1} =  4 \kappa^2(n + 1),
\end{equation}
which is identical with the results obtained in Ref.~\cite{Karch:2006pv}, since, as explained above, the meson spectrum computed with positive or negative dilaton solutions is indistinguishable  for $J=1$. 

The twist-3 computation of the space-like form factor,  involves the current $J^+$, and the poles given by (\ref{Mtwist3}) do not correspond to the physical poles of the twist-2 transverse current $\mbf{J}_\perp$  present in the annihilation channel, namely the $J = 1, L = 0$ state. In this case Eq. (\ref{M2SFM}) gives for the twist-2, $J = 1$, $L=0$ vector family the result
\begin{equation}  \label{Mtwist2}
\mathcal{M}^2_{n, J=1, L=0} =  4 \kappa^2 \left(n + \half \right).
\end{equation}
Thus, to compare with physical data one must shift in (\ref{Ftau}) the twist-2 poles given by (\ref{Mtwist3}) to their physical positions (\ref{Mtwist2}). When the vector meson masses are shifted to their physical values the agreement of the predictions with observed data is very good~\cite{deTeramond:2011qp}. 
We presume that the problem arises because of the specific truncation used.

\subsection{Meson transition form factors}

 The photon-to-meson transition form factors~\footnote{This section is based on our collaboration with Fu-Guang Cao. Further details are given in \cite{Brodsky:2011yv, Brodsky:2011xx}. }  (TFFs) $F_{M \gamma}(Q^2)$ measured in $\gamma \gamma^* \to M$  reactions  have been of intense experimental and theoretical interest. The pion transition form factor between a photon and pion measured in the $e^- e^-\to e^- e^- \pi^0$  process, with one tagged electron, is the simplest bound-state process in QCD.
It can be predicted from first principles in the asymptotic $Q^2 \to \infty$  limit~\cite{Lepage:1980fj}.  More generally,
the pion TFF at large $Q^2$ can be calculated at leading twist as a convolution of a perturbative hard scattering amplitude $T_H(\gamma \gamma^* \to q \bar q)$
and a gauge-invariant meson distribution amplitude (DA), which incorporates the nonperturbative dynamics of the QCD bound-state~\cite{Lepage:1980fj}.

The BaBar\ Collaboration has reported measurements of the
transition form factors from $\gamma^* \gamma \to M$ process for the $\pi^0$~\cite{Aubert:2009mc},
$\eta$, and $\eta^\prime$~\cite{BaBar_eta, Druzhinin:2010bg} pseudoscalar mesons for a momentum   transfer  range much larger than
previous measurements~\cite{CELLO,CLEO}.  Surprisingly, the BaBar\ data for the $\pi^0$-$\gamma$ TFF
exhibit a rapid growth for $Q^2 > 15$ GeV$^2$, which is unexpected from QCD predictions. In contrast, the data for  the  $\eta$-$\gamma$ and  $\eta'$-$\gamma$
TFFs are in agreement with previous experiments and closer in agreement with theoretical predictions.
Many theoretical studies have been devoted to explaining BaBar's experimental results
\cite{BaBar_expln_LiM09, MikhailovS09, WuH10, RobertsRBGT10, BaBar_Expln_BroniowshiA10, Kroll10, GorchetinGS11, ADorokhov10, SAgaevBOP11, BakulevMPS11, Klopot:2011qq, Wu:2011gf, Noguera:2011fv, Balakireva:2011wp, McKeen:2011aa, Lih:2012yu, Czyz:2012nq}.

The pion transition form factor $F_{\pi \gamma}(Q^2)$  can be computed from first principles in QCD. To leading 
leading order in $\alpha_s(Q^2)$ and leading twist the result is~\cite{Lepage:1980fj} ($Q^2 = - q^2 >0$)
\begin{equation}
Q^2 F_{\pi \gamma}(Q^2)=\frac{4}{\sqrt{3}} \int_0^1  {\rm d} x \frac{\phi(x,{\bar x} Q)}{\bar x}
\left[ 1+ O \left(\alpha_s,\frac{m^2}{Q^2} \right) \right],
\label{eq:TFLB1}
\end{equation}
where $x$ is the longitudinal momentum fraction of the quark struck by the virtual photon in the hard scattering process
and ${\bar x}=1-x$ is the longitudinal momentum fraction of the spectator quark.  
The pion distribution amplitude $\phi(x,Q)$ in the light-front formalism~\cite{Lepage:1980fj} is the integral of the 
valence $q \bar q$ LFWF in light-cone gauge $A^+=0$
\begin{equation}
\phi(x,Q)=\int_0^{Q^2}\frac{ d^2 \mbf{k}_\perp}{16 \pi^3} \psi_{q \bar q/ \pi}(x, \mbf{k}_\perp),
\label{eq:DALC}
\end{equation}
and has the asymptotic form~\cite{Lepage:1980fj}  $\phi(x, Q \to \infty) = \sqrt{3} f_\pi x (1-x)$; thus the  leading order  QCD result  for the TFF at the asymptotic limit
is obtained~\cite{Lepage:1980fj},
\begin{equation} \label{TFFasy}
Q^2 F_{\pi \gamma}(Q^2 \rightarrow \infty)=2 f_\pi.
\end{equation}

To describe the two-photon processes $\gamma\gamma^* \rightarrow M$, using light-front holographic methods similar to those described in Sec. \ref{lEMFF}, we need to explore the mathematical structure of higher-dimensional forms in the five dimensional action, since the amplitude (\ref{MFF}) can only account for the elastic form factor $F_M(Q^2)$~\cite{Brodsky:2011xx}. For example, in the five-dimensional AdS action there is an additional Chern-Simons (CS) term  in addition to the usual Yang-Mills term $F^2$~\cite{Witten:1998qj}.
In the case of a $U(1)$ gauge theory the CS action is of the form $\epsilon^{L M N P Q} A_L \partial_M A_N \partial_P A_Q$.
The CS action is not gauge invariant: under a gauge transformation it changes by a total derivative which gives a surface term.   
The CS form is the product of three fields at the same point in five-dimensional space corresponding to a local interaction.  Indeed the five-dimensional CS action is responsible for the anomalous coupling of mesons to photons and has been used to describe, for example, the $\omega \to \pi \gamma$~\cite{Pomarol:2008aa} decay as well as the 
$\gamma  \gamma^* \to \pi^0$~\cite{Grigoryan:2008up, Grigoryan:2008cc}
 and  $\gamma^* \rho^0 \to \pi^0$~\cite{Zuo:2009hz} processes.~\footnote{The anomalous EM couplings to mesons in the Sakai and Sugimoto model is described in Ref. \cite{Sakai:2005yt}.}

The hadronic matrix element for the anomalous electromagnetic coupling to mesons in the higher gravity theory is
given by the five-dimensional CS  amplitude
\begin{multline} \label{eq:TFFAdS1}
\int d^4 x \int dz \, \epsilon^{L M N P Q} A_L \partial_M A_N \partial_P A_Q  \\ \sim
(2 \pi)^4 \delta^{(4)} \left(P + q - k\right) F_{\pi \gamma}(q^2) \epsilon^{\mu \nu \rho \sigma} \epsilon_\mu(q) P_\nu \epsilon_\rho(k) q_\sigma,
\end{multline}
which includes the pion field as well as the external photon fields by identifying the fifth component of $A$ with the meson mode in AdS space~\cite{Hill:2004uc}.
In the right-hand side of (\ref{eq:TFFAdS1})  $q$ and $k$ are the momenta of the virtual and on-shell incoming photons respectively  with  corresponding polarization vectors  $\epsilon_\mu(q)$ 
and $\epsilon_\mu(k)$ for  the amplitude   $\gamma \gamma^* \to \pi^0$.
The momentum of the outgoing  pion is $P$.

We now compare the QCD expression on the right-hand side  of  (\ref{eq:TFFAdS1}) with the AdS transition amplitude on the left-hand side.  As for the elastic form factor discussed in Sec. \ref{sec:MFF}, the incoming off-shell photon is represented by the propagation of the non-normalizable electromagnetic solution  in AdS space,
$A_\mu(x^\mu ,z) = \epsilon_\mu(q) e^{-i q \cdot x} V(q^2, z)$,
where $V(q^2,z)$ is the bulk-to-boundary propagator with boundary conditions (\ref{BCV})  $V(q^2 = 0, z ) = V(q^2, z = 0) = 1$.
Since the incoming photon with momentum $k$ is on its mass shell, $k^2 = 0$,  its wave function is $A_\mu(x^\mu, z) = \epsilon_\mu(k) e^{ i k \cdot x}$.   
Likewise, the propagation of the pion in AdS space is described by a normalizable mode
$\Phi_{P}(x^\mu, z) = e^{-i P  \cdot x} \Phi_\pi(z)$ with invariant  mass $P_\mu P^\mu = \mathcal{M}_\pi^2 =0$   
in the chiral limit for massless quarks. 
The normalizable mode $\Phi_\pi(z)$   scales as $\Phi_\pi(z) \to z^2$  in the limit $z \to 0$, since the leading interpolating operator for the pion has twist two.
 A simple dimensional analysis implies that  $A_z \sim  \Phi_\pi(z)/ z$, matching the twist scaling dimensions: two for the pion and one for the EM field.  Substituting in
(\ref{eq:TFFAdS1}) the expression given above for the 
 the pion and the EM fields  propagating in  AdS,   and extracting the overall factor  $(2 \pi)^4 \delta^4 \left( P'  \! - q - k\right)$ upon integration over Minkowski variables, we find $(Q^2\! = - q^2  \! > 0)$
\begin{equation}  \label{eq:TFFAdS2}
F_{\pi \gamma}(Q^2) = \frac{1}{2 \pi} \int_0^\infty   \frac{d z}{z} \,  \Phi_\pi(z)  V\!\left(Q^2, z\right) ,
\end{equation}
where  the normalization is fixed by the asymptotic QCD prediction (\ref{TFFasy}). We have defined our units such that the AdS radius $R=1$.

Since the LF mapping of (\ref{eq:TFFAdS2}) to the asymptotic QCD prediction (\ref{TFFasy}) only depends on the asymptotic behavior near the boundary of AdS space, the result is independent of the particular model used to 
modify the large $z$ IR region of AdS space.  At large enough $Q$, the important contribution to (\ref{TFFasy}) only comes from the region near $z \sim 1/Q$ where 
$\Phi(z) = 2 \pi f_\pi z^2 + \mathcal{O}(z^4)$.  Using the integral
$\int_0^\infty dx \, x^\alpha K_1(x) = 2^{\alpha -2} \alpha \, \left[\Gamma \! \left(\frac{\alpha}{2}\right)\right]^2$,
${\rm Re}(\alpha) >1$,
we recover the asymptotic result  (\ref{TFFasy})
\begin{equation}
Q^2 F_{\pi \gamma}(Q^2 \rightarrow \infty)=2 f_\pi + \mathcal{O}\left(\frac{1}{Q^2}\right),
\end{equation} 
with the pion decay constant $f_\pi$   \cite{Brodsky:2011xx}
\begin{equation} \label{eq:fpi}
f_\pi = \frac{1}{4 \pi} \frac{\partial_z\Phi^\pi(z)}{z} \Big\vert_{z=0}.
\end{equation}

A simple analytical expression for the pion TFF can be obtained from the ``soft-wall'' holographic model described in Sec. \ref{EFFDC}.
Using (\ref{eq:Phitau}) to describe the twist-two pion valence wave function in AdS space we find
\begin{equation} \label{eq:TFFAdSQCD}
Q^2 F_{\pi \gamma}(Q^2) = \frac{4}{\sqrt{3}} \int_0^1 dx  \frac{\phi(x)}{1-x}  \left[1 - \exp \left( - \frac{(1-x) P_{q \bar q} Q^2 }{4 \pi^2 f_\pi^2  x}\right) \right] ,
\end{equation}
where $\phi(x) = \sqrt{3} f_\pi x(1-x)$ is the asymptotic QCD distribution  with $f_\pi$ the pion decay constant and  $P_{q \bar q}$ is the probability for the valence state.
Remarkably, the holographic result for the pion TFF factor given by (\ref{eq:TFFAdSQCD}) 
for $P_{q \bar q} =1$
is identical to the 
results for the pion TFF obtained with the exponential light-front wave function model of 
Musatov and Radyushkin~\cite{Musatov:1997pu} consistent with the leading order  QCD result~\cite{Lepage:1980fj}.
Since the pion field is identified as the fifth component of $A_M$,
the CS form $\epsilon^{L M N P Q} A_L \partial_M A_N \partial_P A_Q$ is similar in form to an axial current;  this correspondence can explain why the resulting pion distribution amplitude has the asymptotic form.~\footnote{In Ref. \cite{Grigoryan:2008up} the pion TFF was studied in the framework of a CS extended hard-wall AdS/QCD model with  $A_z \sim \partial_z \Phi(z)$.  The  expression for the TFF
which follows from (\ref{eq:TFFAdS1}) then vanishes at $Q^2 =0$, and has to be corrected by the introduction of a surface term at the IR wall~\cite{Grigoryan:2008up}.
However, this procedure is only possible for a model with a sharp cutoff.}

\begin{figure}[htbp]
\begin{center}
\includegraphics[width=8.1cm]{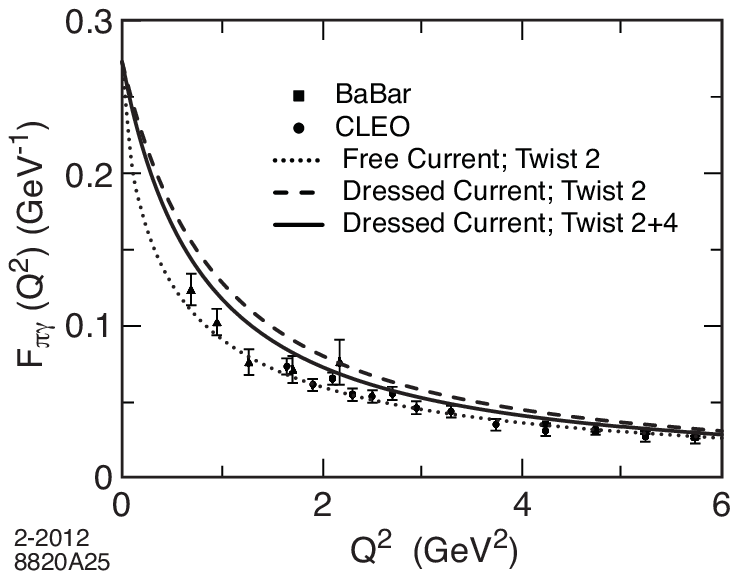}
\includegraphics[width=7.6cm]{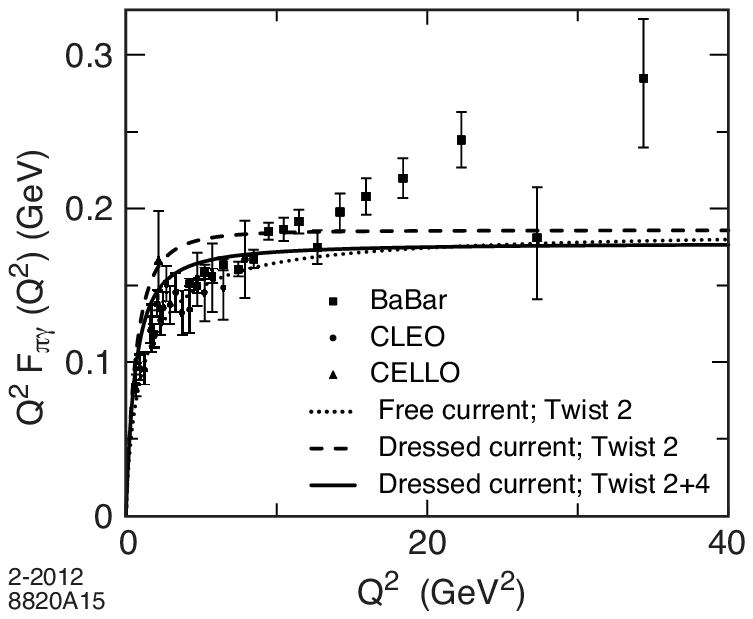}
\caption{\small The $\gamma \gamma^* \to \pi^0$ transition form factor $F_{\pi \gamma}(Q^2)$ (left) and $Q^2 F_{\pi \gamma}(Q^2)$ (right). The dotted curve is the asymptotic result.
The dashed and solid curves include the effects of using a confined EM current for twist-two and twist-two plus twist-four respectively. The data are from \cite{Aubert:2009mc, CELLO,  CLEO}. }
\label{Q2PiTFF}
\end{center}
\end{figure}

Taking $P_{q \bar q}=0.5$ in  (\ref{eq:TFFAdSQCD})  one obtains a  result in agreement with the Adler, Bell and Jackiw anomaly
result which agrees within a few percent with the observed value obtained from the
the decay $\pi^0 \to \gamma \gamma$. This suggests that the contribution from higher Fock states vanishes at $Q=0$ in this simple holographic confining model. Thus (\ref{eq:TFFAdSQCD}) represents
 a description of the pion TFF which encompasses the low-energy nonperturbative  and the high-energy hard domains, but includes only the asymptotic distribution amplitude of the $q \bar q$ component of the pion wave function at all scales.
The results from  (\ref{eq:TFFAdSQCD}) for $P_{q \bar q}=0.5$ are shown   in Fig.  \ref{Q2PiTFF}.  Also shown in Fig. \ref{Q2PiTFF} are the results for the free current approximation (which corresponds to the asymptotic result) with  $P_{q \bar q}=0.5$ and a twist-two plus twist-four model~\cite{Brodsky:2011xx} with  $P_{q \bar q} = 0.915$, and $P_{q \bar q q \bar q}= 0.085$.  The calculations~\cite{Brodsky:2011xx} agree reasonably well with the experimental data at low- and medium-$Q^2$ regions ($Q^2<10$ GeV$^2$), but disagree with BaBar's large $Q^2$ data.

\begin{figure}[htbp]
\begin{center}
\includegraphics[width=7.8cm]{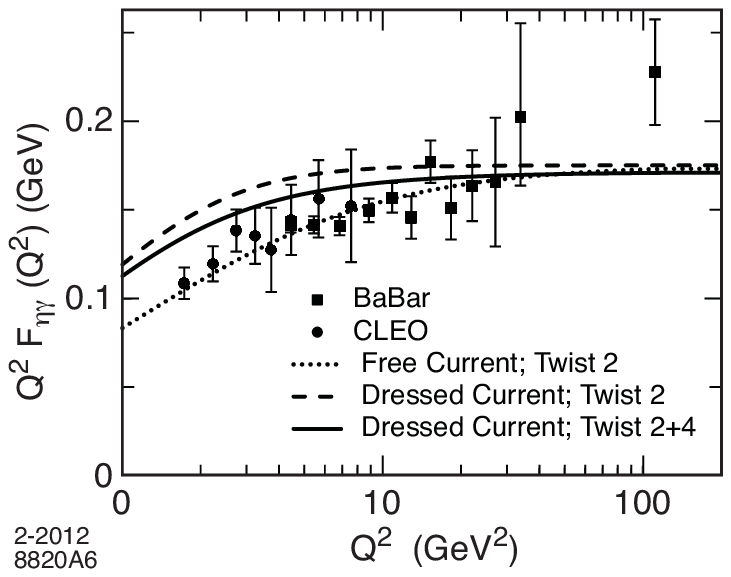}
\includegraphics[width=8.2cm]{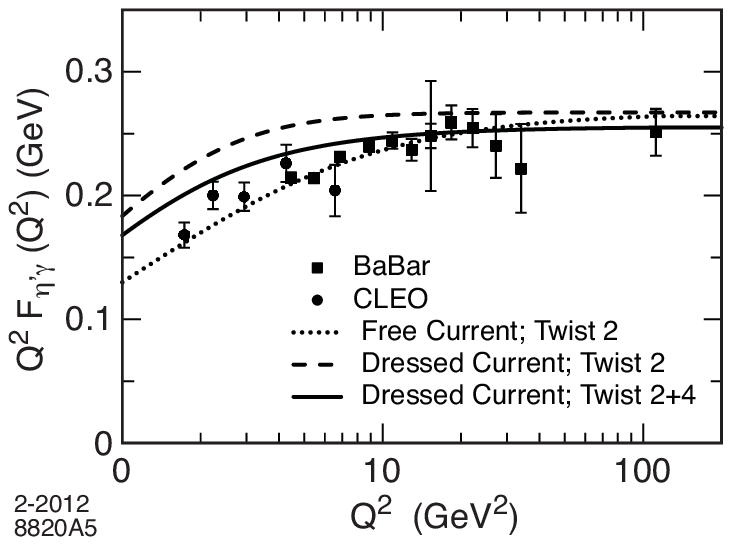}
\caption{\small The $\gamma \gamma^* \rightarrow \eta$ transition form factor $Q^2 F_{\eta \gamma}(Q^2)$  (left). The dotted curve is the asymptotic result.
The dashed and solid curves include the effects of using a confined EM current for twist-two plus twist-two plus twist-four respectively. 
Same  for the $\gamma \gamma^* \rightarrow \eta^\prime$ transition form factor
 $Q^2 F_{\eta^\prime \gamma}(Q^2)$ (right). The data are from \cite{Aubert:2009mc, CELLO,  CLEO}. }\label{fig:EtaTFF_Q2Feta}
\end{center}
\end{figure}

The  $\eta$ and $\eta^\prime$ mesons result from the mixing of
the neutral states $\eta_8$ and $\eta_1$ of the SU(3)$_F$ quark model.
The TFFs for the  $\eta$ and $\eta^\prime$ mesons have the same expression as  the pion transition form factor,
except for an overall multiplying factor $c_P=1, \, \frac{1}{\sqrt{3}}$, and $\frac{2\sqrt{2}}{\sqrt{3}}$ for the $\pi^0$, $\eta_8$ and $\eta_1$, respectively~\cite{Brodsky:2011xx}. The results for the $\eta$ and $\eta^\prime$ transitions form factors are shown in
Fig. \ref{fig:EtaTFF_Q2Feta}. The calculations agree very well with available experimental data over a large range of $Q^2$.
The rapid growth of the large $Q^2$ data for the pion-photon transition form factor reported by the BaBar Collaboration is difficult to explain within the current framework of QCD.
The analysis presented here thus indicates the importance of additional measurements of the  pion-photon transition form factor at large $Q^2$.

\section{Baryons in light-front holography \label{NucleonSpec}}

The study of the excitation spectrum of baryons is one of the most challenging aspects of particle physics. In fact, dedicated experimental programs are in place  to determine the spectrum  of nucleon excitations and its internal structure.
Important computational efforts in 
lattice QCD aim to the reliable extraction of the excited nucleon eigenstates. 
Lattice calculations of the ground state light hadron masses agree with experimental values within 5\%~\cite{Dudek:2011zz}. However, the
excitation spectrum of the nucleon represents a formidable challenge to lattice QCD due to the enormous computational complexity required beyond the leading ground state configuration~\cite{Edwards:2011jj}.  Moreover, a large basis of interpolating operators is required  since excited nucleon states are classified according to  irreducible representations of the lattice, not the total angular momentum.

 As we shall discuss below, the analytical exploration of the baryon spectrum and nucleon form factors, using light-front gauge/gravity duality ideas, leads, in contrast,  to simple formulas and rules which describe quite well the systematics of the established light-baryon resonances and elastic and transition nucleon form factors, which can be tested against new experimental findings. The gauge/gravity duality can give us important insights into the strongly coupled dynamics of nucleons using simple analytical methods. 
 
We can extended the holographic ideas to spin-$\half$ hadrons by considering the propagation of spin-$\half$ Dirac modes in AdS space~\cite{Polchinski:2002jw}. The action for a Dirac field in AdS$_{d+1}$ is 
\begin{equation}  \label{SF}
S_F = \int d^d x \, dz  \sqrt{g} \,   \left(\frac{i}{2} \bar \Psi e^M_A \Gamma^A D_M \Psi - \frac{i}{2}  ( D_M \bar \Psi) e^M_A \Gamma^A  \Psi
 - \mu \bar \Psi \Psi \right),
\end{equation}
where  $\sqrt{g} = \left(\frac{R}{z}\right)^{d+1}$ and $e^M_A$ is the inverse vielbein, $e^M_A = \left(\frac{z}{R}\right) \delta^M_A$. The covariant derivative of the spinor field is $D_M = \partial_M  - \frac{i}{2} \omega_M^{AB} \Sigma_{AB}$ where $\Sigma_{A B}$ are
the generators of the Lorentz group in the spinor representation, $\Sigma_{A B} = \frac{i}{4}  \left[\Gamma_A, \Gamma_B\right]$, and the tangent space Dirac matrices obey the usual anti-commutation relation $\left\{\Gamma^A, \Gamma^B\right\} = \eta^{A B}$.
For $d$ even we can choose the set of gamma matrices
$\Gamma_A = \left(\Gamma_\mu, \Gamma_z\right)$ with $\Gamma_z = - \Gamma^z 
=  \Gamma_0 \Gamma_1 \cdots \Gamma_{d-1}$. For $d=4$  we have $\Gamma_A = \left(\gamma_\mu, - i \gamma_5\right)$,  where $\gamma_\mu$ and  $\gamma_5$ are the usual 4-dimensional Dirac matrices with $\gamma_5 = i \gamma_0 \gamma_1 \gamma_2 \gamma_3$ and $\gamma_5^2=+1$.
The spin connection in AdS is $w_M^{A B} = \left(\eta^{A z} \delta_M^B - \eta^{B z} \delta^A_M\right)/z$, thus the equation of motion  $ \left(i  e^M_A \Gamma^A D_M - \mu\right) \Psi = 0$ leads to the Dirac equation in AdS space
\begin{equation} \label{AdSDEq}
\left[ i \left( z \eta^{M N} \Gamma_M \partial_N + \frac{d}{2} \Gamma_z \right) - \mu R \right] \Psi =0,
\end{equation}
where the $d + 1$ dimensional mass $\mu$  is a priory an arbitrary parameter.~\footnote{The spinor action (\ref{SF}) is often complemented by an additional surface term  in the
UV boundary~\cite{Henningson:1998cd} $\lim_{\epsilon \to 0} \int d^d x \sqrt{g_\epsilon} \bar \Psi \Psi$ where $g_\epsilon$ is the metric induced in the boundary surface by the metric $g$ of AdS$_{d+1}$.
The additional term is required to preserve the $O(d+1,1)$ isometry  group of  AdS$_{d+1}$ and to compute a two-point correlation function in the conformal boundary  theory~\cite{Mueck:1998iz}. The equation of motion (\ref{AdSDEq}) is not modified by the surface term.}

One can also take as starting 
point the construction of light-front wave equations in physical space-time for baryons by studying  the LF transformation properties of spin 1/2 states~\cite{deTeramond:2010we}. The light-front wave equation describing baryons is a matrix eigenvalue equation $D_{LF} \vert \psi \rangle = \mathcal{M} \vert \psi \rangle$ with $H_{LF} = D_{LF}^2$. In a $2 \times 2$ spinor  component representation
\begin{eqnarray} \label{LFDEq}  \nonumber
  \frac{d}{d\zeta} \psi_+ + \frac{\nu+\half}{\zeta}\psi_+ &=& \mathcal{M} \psi_- , \\
- \frac{d}{d\zeta} \psi_-  + \frac{\nu+\half}{\zeta}\psi_-  &=& \mathcal{M} \psi_+ .
\end{eqnarray}
As shown below, we can identify $\nu$ with the orbital angular momentum $L$: $\nu = L+1$.

Upon the substitution $z \to \zeta$ and
\begin{equation} \label{psiPsi}
\Psi(x,z) = e^{-i P \cdot  x}  z^2 \psi(z) u({P}),
\end{equation}
in  (\ref{AdSDEq}) we recover for $d=4$ its LF expression (\ref{LFDEq}), provided that $ \vert \mu R \vert = \nu + \half $. The baryon invariant mass is 
$P_\mu P^\mu = \mathcal{M}^2$ and the spinor $u({P})$ is a four-dimensional spinor which obeys  the Dirac equation  
$(\Pslash - \mathcal{M}) u({P}) = 0$.
Thus the eigenvalue equation $H_{LF} \psi_\pm = \mathcal{M}^2 \psi_\pm$ for the upper and lower components leads to the wave equation
\begin{equation} \label{LFWEB}
\left(-\frac{d^2}{d\zeta^2}
- \frac{1 - 4 \nu^2}{4\zeta^2}  \right) \psi_+(\zeta) = \mathcal{M}^2 \psi_+(\zeta),
\end{equation}
and
\begin{equation} \label{LFWEB}
\left(-\frac{d^2}{d\zeta^2}
- \frac{1 - 4(\nu + 1)^2}{4\zeta^2}  \right) \psi_-(\zeta) = \mathcal{M}^2 \psi_-(\zeta),
\end{equation}
with solutions
\begin{equation}
\psi_+ \sim \sqrt \zeta  J_\nu(\zeta \mathcal{M}),  \hspace{20pt} \psi_- \sim \sqrt \zeta  J_{\nu+1}(\zeta \mathcal{M}) .
\end{equation}

The solution of the spin-$\threehalf$ Rarita-Schwinger equation for the field $\Psi_M$ in AdS space is more involved, but considerable simplification occurs in the $\Psi_z = 0$ gauge for physical polarization along Minkowski coordinates $\Psi_\mu$, where it becomes similar to the spin-$\half$ solution~\cite{Volovich:1998tj, Matlock:1999fy}.

\subsection{A hard-wall model for baryons \label{HWM}}

The hermiticity of the LF Dirac operator $D_{LF}$ in the eigenvalue equation  $D_{LF} \vert \psi \rangle = \mathcal{M} \vert \psi \rangle$  implies that the surface term $\psi_+^*(\zeta) \psi_-(\zeta) - \psi_-^*(\zeta) \psi_+(\zeta)$ should vanish at the boundary. Thus in a truncated space holographic model, the light front modes $\psi_+$ or $\psi_-$ should vanish at the boundary $\zeta = 0$ and $\zeta = \zeta_0$. This condition fixes the boundary conditions and determine the baryon spectrum in the truncated hard-wall model. A similar surface term arises when one computes  the equation of motion from the action (\ref{SF}). In fact,  integrating by parts (\ref{SF}) and using the equation of motion we find
\begin{equation}
S_F= - \lim_{\epsilon \to 0} \int \frac{d^dx}{2 z^d} \Big ( \bar \Psi_+ \Psi_- - \bar \Psi_- \Psi_+ \Big) \Big \vert_\epsilon^{z_0},
\end{equation}
where  $\Psi_\pm = \half \left(1 \pm \gamma_5 \right) \Psi$, and $R$ has units $R=1$.
The baryon mass spectrum thus follows from the  LF
 ``bag'' boundary conditions $\psi_\pm\left(\zeta_0\right) = 0$  or the AdS  boundary conditions  $\Psi_\pm\left(z_0\right) = 0$  at the IR value, $ z_0  = 1/\Lambda_{\rm QCD}$,
where the LF  invariant impact variable $\zeta$ (\ref{zetan}) is identified with the AdS  holographic coordinate $z$,  $z \to \zeta$. We find
 \begin{equation}
 \mathcal{M}^+ = \beta_{\nu,k} \, \Lambda_{\rm QCD},  \hspace{20pt}
 \mathcal{M}^- = \beta_{\nu+1,k} \, \Lambda_{\rm QCD},
 \end{equation}
 with a scale-independent  mass ratio determined by the zeros of Bessel functions $\beta_{\nu, k}$.

 In the usual AdS/CFT correspondence the baryon is an $SU(N_C)$ singlet bound state of $N_C$ quarks in the large $N_C$ limit. Since  there are no quarks in this theory, quarks are introduced as external sources at the AdS asymptotic boundary~\cite{Witten:1998xy, Gross:1998gk}.  The baryon is constructed  as an $N_C$ baryon vertex located in the interior of AdS. In this top-down string approach baryons are usually described as solitons
 or Skyrmion-like objects~\cite{Hong:2007kx, Hata:2007mb}.   In contrast, the bottom-up light-front holographic approach described here is based on the precise mapping of AdS expressions to light-front QCD.
 Consequently, we construct baryons corresponding to $N_C=3$ not $N_C \to  \infty$.  The corresponding
 interpolating operator for an $N_C = 3$ physical baryon
$ \mathcal{O}_{3 + L} =  \psi D_{\{\ell_1} \dots
 D_{\ell_q } \psi D_{\ell_{q+1}} \dots
 D_{\ell_m\}} \psi$,  $L = \sum_{i=1}^m \ell_i$, is a 
 twist-3,  dimension $9/2 + L$ with scaling behavior given by its
 twist-dimension $3 + L$.  We thus require $\nu = L+1$ to match the short distance scaling behavior.   One can interpret $L$ as the maximal value of $|L^z|$ in a given LF Fock state.

 In the case of massless quarks, the nucleon eigenstate ($u_\pm = \half \left(1 \pm \gamma_5 \right) u$)
 \begin{eqnarray}  \nonumber
 \psi(\zeta)  &=& \psi_+(\zeta) u_+ + \psi_-(\zeta) u_- \\
 &=& C \sqrt \zeta \left(J_\nu(\zeta \mathcal{M}) u_+ +  J_{\nu+1}(\zeta \mathcal{M}) u_- \right) ,
 \end{eqnarray}
has components $\psi_+$ and $\psi_-$ with different orbital angular momentum, $L^z = 0$ and $L^z = +1$,
combined with spin components $S^z = +1/2$ and $S^z =  - 1/2$ respectively, but  with equal probability~\footnote{For the truncated-space 
model,  (\ref{pmP}) follows from the identity
$\int_0^1 x dx \left[J_\alpha^2(x \beta) - J_{\alpha+1}^2(x \beta) \right] = J_\alpha(\beta) J_{\alpha+1}(\beta)/\beta$,
independently of the component wavefunction chosen to fix the boundary conditions at $\zeta = \zeta_0$.}
\begin{equation} \label{pmP}
\int d \zeta \vert \psi_+(\zeta) \vert^2 =  \int d \zeta \vert \psi_-(\zeta) \vert^2,
\end{equation} 
a manifestation of the chiral invariance of the theory for massless quarks.
Thus in light-front holography,  the spin of the proton is carried by the quark orbital angular momentum: $J^z=  \langle L^z\rangle=\pm 1/2$ since $\langle\sum S^z_q \rangle= 0$~\cite{Brodsky:2011zj}, and not by its gluons.

An important feature of  bound-state relativistic theories  is that hadron eigenstates have in general Fock components with different $L$ components. In the holographic example discussed above,  the proton has $S$ and P components with equal probability. In the case of QED, the ground state $1S$ state of the Dirac-Coulomb equation has both $L=0$ and $L=1$ components.   By convention, in both light-front QCD and QED, one labels the eigenstate with its minimum value of $L$.  For example, the symbol $L$ in the light-front AdS/QCD spectral  prediction for mesons (\ref{M2SFMLS}) refers to the {\it minimum } $L$  (which also corresponds to the leading twist) and $S$ is the total internal spin of the hadron.

 \begin{table}[htdp]
 \caption{\small Classification of confirmed baryons listed by the PDG~\cite{Nakamura:2010zzi}. The labels $L$,  $S$ and  $n$ refer to the internal  orbital angular momentum, internal spin and radial quantum number respectively. The even-parity baryons correspond to the $\bf 56$ multiplet of $SU(6)$ and the odd-parity to the ${\bf 70}$.}
  \begin{center} 
 {\begin{tabular}{@{}cccc@{}}
 \hline\hline \vspace{0pt}
 $ L$ & $S$ &   $n$ &   Baryon State
 \vspace{2pt}
 \\[0.2ex]
 \hline
 \multicolumn{4}{c}{}\\[-2.5ex]
 0  & $\half$ &   0  & $N{\half^+}(940)$\\[0.0ex]
 0  & $\half$ &   1  & $N{\half^+}(1440)$\\[0.0ex]
 0  & $\half$ &   2  & $N{\half^+}(1710)$\\[0.0ex]
 0 &  $\threehalf$&    0  &$\Delta{\threehalf^+}(1232)$\\[0.0ex]
 0 & $\threehalf$&    1  &$\Delta{\threehalf^+}(1600)$\\[0.0ex]
 1 & $\half$ &  0  & $N{\half^-}(1535)~~ N{\threehalf^-}(1520)$ \\[0.0ex]
 1 & $\threehalf$ &   0  & $N{\half^-}(1650)~~ N{\threehalf^-}(1700)~~N{\fivehalf^-}(1675)$\\[0.0ex]
 1 & $\half$ &  0  &$\Delta{\half^-}(1620)~~ \Delta{\threehalf^-}(1700)$ \\[0.0ex]
 2  & $\half$  &  0  &$N{\threehalf^+}(1720)~~ N{\fivehalf^+}(1680)$ \\[0.0ex]
 2  & $\half$ &   1  &$~~~~~~~~~~~~~~~ ~~ N{\fivehalf^+}(1900)$ \\[0.0ex]
 2  & $\threehalf$ & 0  & $\Delta{\half^+}(1910)~~ \Delta{\threehalf^+}(1920)
                     ~~ \Delta{\fivehalf^+}(1905)~~\Delta{\sevenhalf^+}(1950)$\\[0.0ex]
 3 & $\half$ &   0  & $N{\fivehalf^-}~~ ~~~~~~ N{\sevenhalf^-}$ \\[0.0ex]
 3 & $\threehalf$ &  0  &   $N{\threehalf^-}~~~~~~ ~~ N{\fivehalf^-}~~~~~~ ~~
                     N{\sevenhalf^-}(2190)~~ N{\ninehalf^-}(2250)$\\[0.0ex]
 3  & $\half$ &  0   & $\Delta{\fivehalf^-}~~~~~ ~~ \Delta{\sevenhalf^-}$ \\[0.0ex]
 4 & $\half$ &   0  & $N{\sevenhalf^+}~~~~~~ ~~ N{\ninehalf^+}(2220)$ \\[0.0ex]
 4 & $\threehalf $ &   0  & $\Delta{\fivehalf^+}~~~~~~ ~~ \Delta{\sevenhalf^+} ~~~~~~ ~~
                       \Delta{\ninehalf^+}~~~~~~ ~~\Delta{\elevenhalf^+}(2420)$\\[0.0ex]
 5 & $\half$ &   0  &$N{\ninehalf^-}~~~~~~ ~~ N{\elevenhalf^-}~~~~~~~$ \\[0.0ex]
 5 & $\threehalf$ &  0  & $N{\sevenhalf^-}~~~~~~~ ~~ N{\ninehalf^-}~~~~~~~
      N{\elevenhalf^-}(2600) ~~N{\thirteenhalf^-} $\\[1.0ex]
 \hline\hline
 \end{tabular}}
 \end{center}
 \label{baryons}
 \end{table}

We list in Table \ref{baryons} the confirmed (3-star and 4-star) baryon states from the updated Particle Data Group~\cite{Nakamura:2010zzi}.~\footnote{A recent exploration of the properties of baryon resonances derived from a multichannel partial wave analysis~\cite{Anisovich:2011fc} report additional resonances not included in the Review of Particle Properties~\cite{Nakamura:2010zzi}.}
To determine the internal spin, internal orbital angular momentum and radial quantum number assignment of the $N$ and $\Delta$ excitation spectrum from the total angular momentum-parity PDG assignment, it is  convenient to use the conventional $SU(6) \supset SU(3)_{flavor} \times SU(2)_{spin}$ multiplet structure, but other model choices are also possible~\cite{Klempt:2009pi}. \footnote{In particular the $\Delta\fivehalf^-(1930)$ state (not shown in Table \ref{baryons}) has been given the non-$SU(6)$ assignment $S = 3/2$, $L =1$, $n=1$  in Ref. ~\cite{Klempt:2009pi}. This assignment will be further discussed in the section below.}

 \begin{figure}[h]
\centering
\includegraphics[angle=0,width=8.1cm]{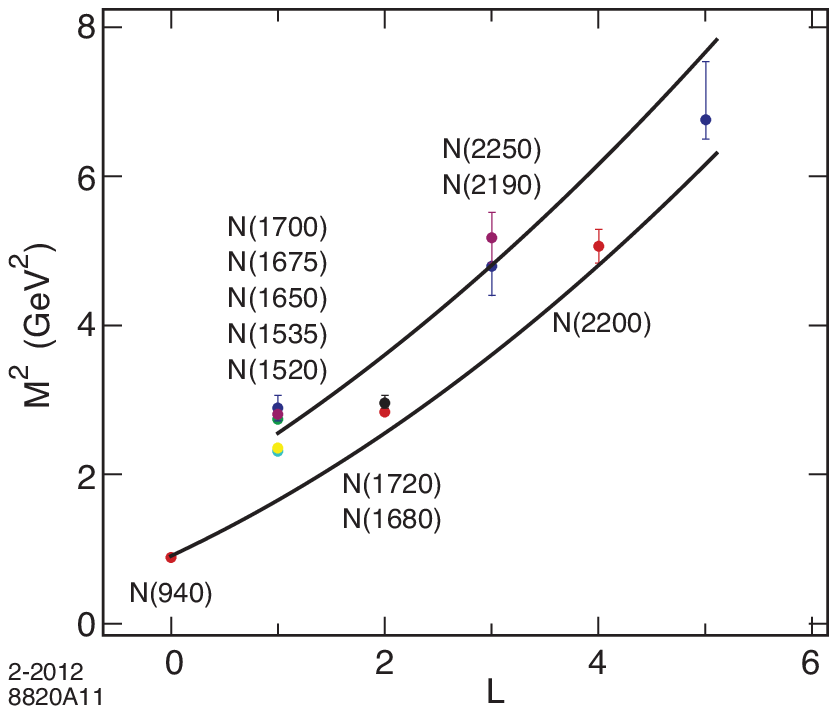} \hspace{0pt}
\includegraphics[angle=0,width=8.0cm]{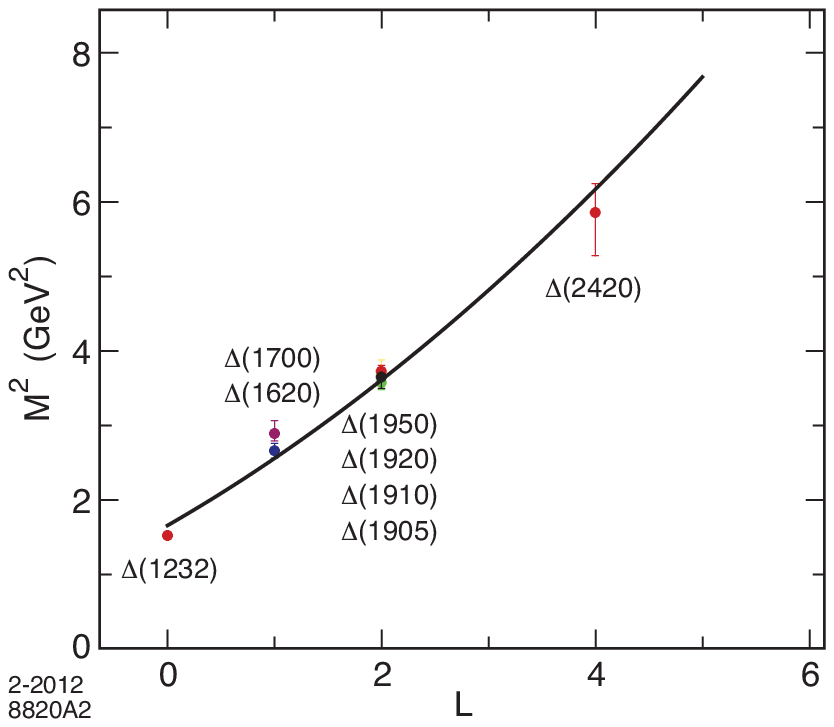}
\caption{\small Light baryon orbital spectrum  ($n=0$) for $\Lambda_{\rm QCD} =  0.25$ GeV. Predictions for the nucleons (left)  and for the $\Delta$ trajectories (right).}
\label{baryonsHWM}
\end{figure}

We show in Fig. \ref{baryonsHWM} the model predictions for the orbital excitation spectrum of baryons which follows from the boundary conditions $ \psi_\pm\left(\zeta = 1/ \Lambda_{\rm QCD}\right) = 0$ in a truncated-space model in the infrared region~\cite{deTeramond:2005su}.~\footnote{The results shown here  in Fig. \ref{baryonsHWM} give better results for the lower mass baryons as compared with Ref. \cite{deTeramond:2005su} where naive conformal dimensions were used instead.} The figure  shows the predicted orbital spectrum of the nucleon and  $\Delta$ orbital resonances  for $n=0$. The only parameter is the value of $\Lambda_{QCD}$ which we take as 0.25 GeV. Orbital excitations are approximately aligned along  two trajectories corresponding to even and odd parity states, with exception of the $\Delta{\half^-}(1620)$ and  $\Delta{\threehalf^-}(1700)$ states which are in the same trajectory. The spectrum shows a clustering of states with the same orbital $L$, consistent with a strongly suppressed spin-orbit force. This remarkable prediction for the baryons  is not a peculiarity of the hard-wall model, but 
is an important property of light-front holographic models.

In the quark-diquark model of Jaffe and Wilczek~\cite{Wilczek:2004im}, nucleon states with $S =1/2$ in 
Fig. \ref{baryonsHWM} (a)  correspond to ``good" diquarks, $S = 3/2$ nucleons and all the $\Delta$ states in Fig. \ref{baryonsHWM} (b) to ``bad" diquarks, with exception of the $\Delta(1930)$ which does not follow the simple $3q$ quark-diquark pattern. As for the case for mesons discussed in Sec. \ref{hardwallmesons}, the hard-wall model predicts $\mathcal{M} \sim 2n + L$, in contrast to the usual Regge behavior $\mathcal{M}^2 \sim n + L$ found in experiment~\cite{Klempt:2007cp}.  The radial modes are also not well described in the truncated-space model.  For example, the first AdS radial state has a mass 1.85 GeV, which is thus difficult to identify with the Roper $N(1440)$ resonance.
This problem is not present in the soft wall model for baryons discussed below.

\subsection{A soft-wall model for baryons \label{SWB}}

For fermion fields in AdS one cannot break conformality with the introduction of a dilaton in the action since it can be rotated away leaving the action conformally
 invariant.~\footnote{This remarkable property was first pointed out in Ref.~\cite{Kirsch:2006he}, and later derived independently in Ref.~\cite{wuhan:2009}.}
As a result, one must introduce an effective confining potential $V(z)$ in the action of a Dirac field propagating
in AdS$_{d+1}$  space to break the conformal invariance of the theory and generate a baryon spectrum
\begin{equation}  \label{SFV}
S_F = \int d^d x \, dz  \sqrt{g} \,   \left(\frac{i}{2} \bar \Psi e^M_A \Gamma^A D_M \Psi - \frac{i}{2}  ( D_M \bar \Psi) e^M_A \Gamma^A  \Psi
 - \mu \bar \Psi \Psi - V(z)  \bar \Psi \Psi \right).
\end{equation}
The variation of the action (\ref{SFV}) leads to the Dirac equation in  AdS
\begin{equation} \label{AdSDVEq}
\left[ i \left( z \eta^{M N} \Gamma_M \partial_N + \frac{d}{2} \Gamma_z \right) - \mu R - R V(z)\right] \Psi =0.
\end{equation}

As in the case for the hard wall model described in the previous section, the corresponding light-front wave equation in physical space-time  follows from identifying the transverse LF coordinate $\zeta$ with the AdS holographic variable $z$, $z \to \zeta$, and the  substitution (\ref{psiPsi}) in (\ref{AdSDVEq}). For d = 4 we find the 
matrix eigenvalue equation in the $2 \times 2$ spinor component representation
\begin{eqnarray} \label{LFDEqV}  \nonumber
  \frac{d}{d\zeta} \psi_+ + \frac{\nu+\half}{\zeta}\psi_+  + U(\zeta) \psi_+ &=& \mathcal{M} \psi_- , \\
- \frac{d}{d\zeta} \psi_-  + \frac{\nu+\half}{\zeta}\psi_-  +  U(\zeta) \psi_-&=& \mathcal{M} \psi_+ ,
\end{eqnarray}
where $U(\zeta) = \frac{R}{\zeta} \, V(\zeta)$ is the effective confining potential in the light-front Dirac equation.

Instead of choosing a dilaton profile to reproduce linear Regge behavior, as described in Sec. \ref{softwallmesons} for the case of mesons, we choose the confining interaction $V$ in (\ref{SFV}) to reproduce  linear Regge trajectories for the baryon mass spectrum $\mathcal{M}^2$. This  ``soft-wall'' model  for baryons in a higher dimensional AdS space, has also a LF analogue; it corresponds to a Dirac equation in physical space-time  in presence of an effective linear  confining potential $U$ defined at equal LF time. For the potential $U = \kappa^2 \zeta$ equation (\ref{LFDEqV}) 
 is equivalent to the system of second order equations
\begin{equation} \label{LFWEB}
\left(-\frac{d^2}{d\zeta^2}
- \frac{1 - 4 \nu^2}{4\zeta^2} + \kappa^4 \zeta^2+ 2 (\nu + 1)  \kappa^2 \right) \psi_+(\zeta) = \mathcal{M}^2 \psi_+(\zeta),
\end{equation}
and
\begin{equation} \label{LFWEB}
\left(-\frac{d^2}{d\zeta^2}
- \frac{1 - 4(\nu + 1)^2}{4\zeta^2} + \kappa^4 \zeta^2+ 2  \nu \kappa^2 \right) \psi_-(\zeta) = \mathcal{M}^2 \psi_-(\zeta).
\end{equation}
As a consequence, when one squares the Dirac Equation with $U(\zeta)$, one generates a Klein-Gordon equation with the potential $\kappa^4 z^2$.  This is consistent with the same confining potential which appears in the meson equations.
 The LF equation 
$H_{LF} \psi_\pm = \mathcal{M}^2 \psi_\pm$
 has thus the two-component solution
\begin{equation}
\psi_+(\zeta) \sim \zeta^{\frac{1}{2} + \nu} e^{-\kappa^2 \zeta^2/2}
  L_n^\nu(\kappa^2 \zeta^2) ,\hspace{20pt}
\psi_-(\zeta) \sim  \zeta^{\frac{3}{2} + \nu} e^{-\kappa^2 \zeta^2/2}
 L_n^{\nu+1}(\kappa^2 \zeta^2), 
\end{equation}
 with equal probability for the properly normalized components.  
 The eigenvalues are
\begin{equation} \label{M2nu}
\mathcal{M}^2 = 4 \kappa^2 (n + \nu + 1),
\end{equation}
identical for both plus and minus eigenfunctions.   
Note that, as expected, the potential $\kappa^4 \zeta^2$ in the second order equation matches the soft-wall potential for mesons discussed in Sec. \ref{softwallmesons}. However, in contrast to the case for  mesons, the dilaton modification of the action gives  little guidance for finding  an effective potential for baryons, since the dilaton can be scaled away by a field redefinition. Consequently the overall energy scale  is left unspecified for the baryons~\cite{deTeramond:2010we}. The remarkable regularities observed in the nucleon spectrum and the analytical properties of the AdS/LF equations allows us, nonetheless, to built precise rules to describe the observed baryon spectrum and make predictions for, as  yet undiscovered, new baryon excited states.

\begin{figure}[h]
\centering
\includegraphics[angle=0,width=8.1cm]{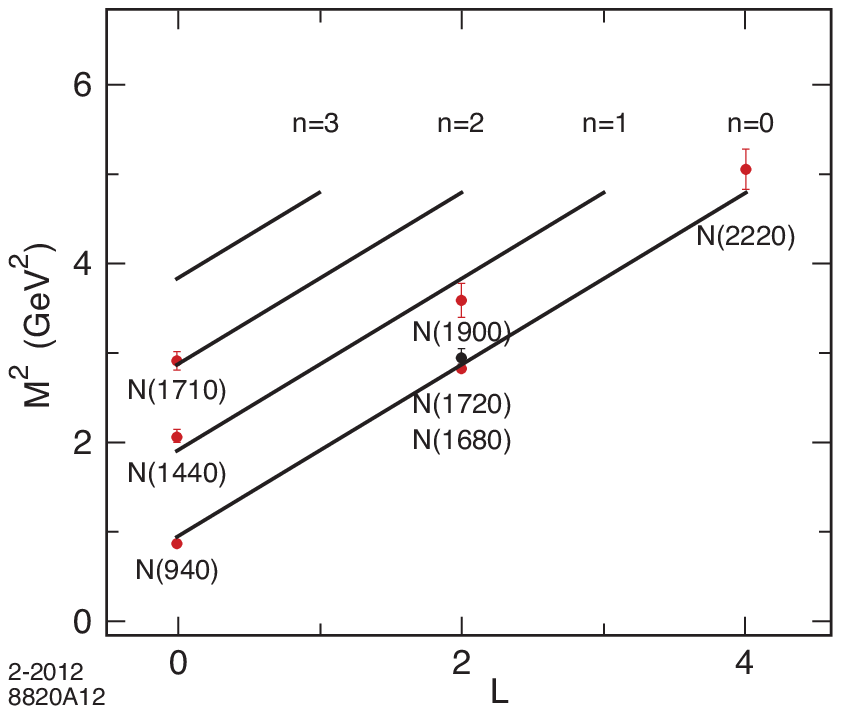} \hspace{0pt}
\includegraphics[angle=0,width=8.1cm]{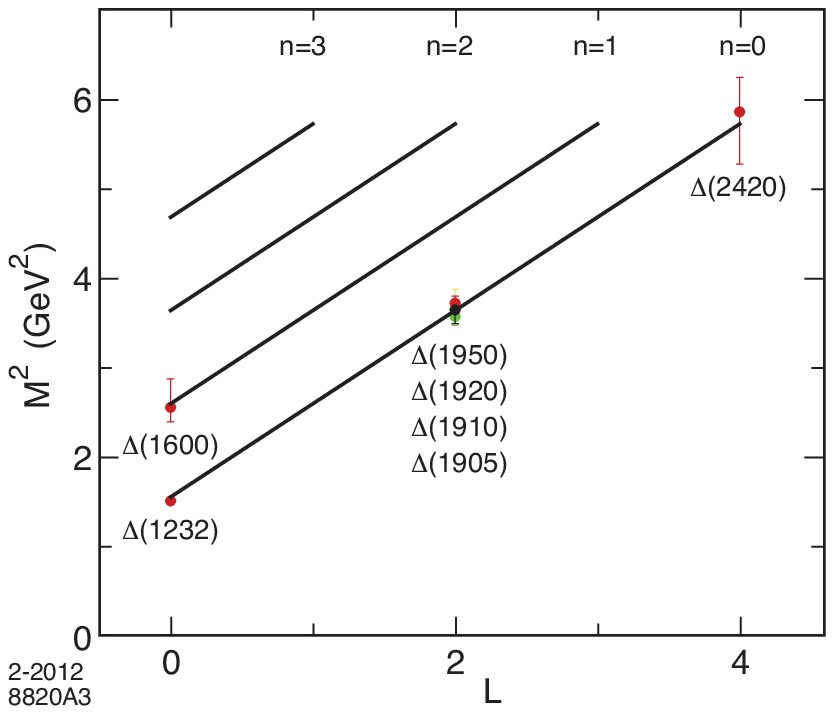}
\caption{\small Orbital and radial baryon excitations for the positive-parity Regge trajectories for  the  $N$ (left) and $\Delta$ (right)  families for $\kappa= 0.49-0.51$ GeV.}
\label{BaryonsSWpnL}
\end{figure}

Before computing the baryon spectrum we must fix the overall mass scale and the parameter $\nu$.  Since our starting point for finding the bound state equation of motion for baryons is the light-front method, we shall require the mass scale to be identical for mesons and baryons while maintaining chiral symmetry for the pion~\cite{deTeramond:2010we} in the LF Hamiltonian equations. In practice, these constraints require a subtraction of $- 4 \kappa^2$ from (\ref{M2nu}).~\footnote{This subtraction to the mass scale may be understood as the displacement required to describe nucleons with  $N_C = 3$ as a composite system with leading twist  $3+L$;  {\it  i.e.}, a  quark-diquark bound state with a twist-2 composite diquark rather than an elementary twist-1 diquark.}

As is the case for the truncated-space 
model, the value of $\nu$ is determined by the short distance scaling behavior, $\nu = L +1$.  Higher-spin fermionic modes
 $\Psi _{\mu_1 \cdots \mu_{J-1/2}}$, $J > 1/2$, with all of its polarization indices along the $3+1$ coordinates follow by shifting dimensions for the fields as shown for the case of mesons in Ref.~\cite{deTeramond:2009xk}~\footnote{The detailed study of higher fermionic spin wave equations in modified AdS spaces is based on our collaboration with Hans Guenter Dosch~\cite{BDdT:2012}. See also the discussion in Ref.~\cite{Gutsche:2011vb}.}.
Therefore, as in the meson sector,  the increase  in the 
mass $\mathcal{M}^2$ for baryonic states for increased radial and orbital quantum numbers is
$\Delta n = 4 \kappa^2$, $\Delta L = 4 \kappa^2$ and $\Delta S = 2 \kappa^2,$ 
relative to the lowest ground state,  the proton; {\it i.e.}, the slope of the spectroscopic trajectories in $n$ and $L$ are identical.
Thus for the positive-parity nucleon sector  
\begin{equation} 
\mathcal{M}^{2 \,(+)}_{n, L, S} =  4 \kappa^2 \left(n + L + \frac{S}{2} + \frac{3}{4} \right),  \label{M2p} 
\end{equation}
where the internal spin $S = \half$  or $\threehalf$.

The resulting predictions for the spectroscopy of positive-parity light baryons  are shown in Fig. \ref{BaryonsSWpnL}.
Only confirmed PDG~\cite{Nakamura:2010zzi} states are shown. The Roper state $N(1440)$ and the $N(1710)$ are well accounted for in this model as the first and second radial states of the proton. Likewise, the $\Delta(1660)$ corresponds to the first radial state of the $\Delta(1232)$ as shown in in Fig. \ref{BaryonsSWpnL}.
The model is  successful in explaining the parity degeneracy observed in the light baryon spectrum, such as the $L\! =\!2$, $N(1680)\!-\!N(1720)$ degenerate pair and the $L=2$, $\Delta(1905), $ $ \Delta(1910),$ $\Delta(1920), $ $ \Delta(1950)$ states which are degenerate within error bars. 
The parity degeneracy of baryons shown in Fig. \ref{BaryonsSWpnL}   is also a property of the 
hard-wall model described in the previous section, but in that case the radial states are not well described~\cite{deTeramond:2005su}.

In order to have a comprehensive description of the baryon spectrum, we need to extend (\ref{M2p}) 
to the negative-parity baryon sector.  In the case of the hard-wall model, this was realized by choosing the boundary conditions for the plus or minus components of the AdS  wave function $\Psi^\pm$.  In practice, this amounts to allowing the negative-parity spin baryons to have a larger spatial extent, a point also raised in~\cite{Klempt:2010cf}. In the soft-wall model there are no boundary conditions to set in the infrared since the wave function vanishes exponentially for large values of $z$. We note, however, that setting boundary conditions on the wave functions, as done in Sec.  \ref{HWM}, is equivalent to choosing the branch $\nu = \mu R - \half$ for the negative-parity spin-$\half$ baryons and $\nu = \mu R + \half$ for the positive parity spin-$\threehalf$ baryons.  This gives  a factor $4 \kappa^2$ between the  lower-lying and the
higher-lying nucleon trajectories as illustrated in Fig. \ref{NucleonsSWpm}, where we compare the lower nucleon trajectory  corresponding to the  $J = L + S$ spin-$\half$ positive-parity nucleon family with the upper nucleon trajectory corresponding to the $J = L + S -1$ spin-$\threehalf$ negative-parity nucleons. As is clearly shown in the figure, the gap is precisely the factor $4 \kappa^2$.

\begin{figure}[h]
\centering
\includegraphics[angle=0,width=8.6cm]{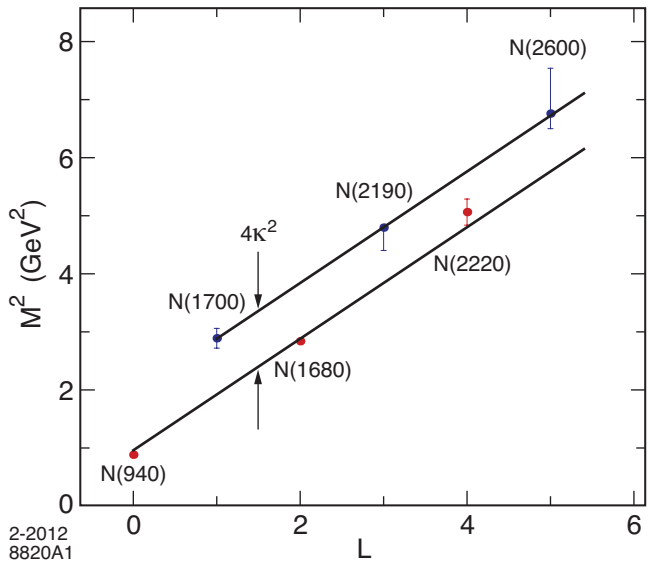} 
\caption{\small Spectrum gap between the negative-parity spin-$\threehalf$ nucleons and the spin-$\half$ positive-parity nucleon families for $\kappa= 0.49$ GeV.}
\label{NucleonsSWpm}
\end{figure}

If we apply the same spin-change rule previously discussed  for the positive-parity nucleons, we would expect that the trajectory for the family of spin-$\half$ negative-parity nucleons is  lower by the factor $2 \kappa^2$ compared to the spin-$\threehalf$ minus-parity nucleons according to the spin-change rule previously discussed. Thus the formula for the negative-parity baryons
\begin{equation} 
\mathcal{M}^{2 \,(-)}_{n, L, S} =  4 \kappa^2 \left(n + L + \frac{S}{2} + \frac{5}{4} \right),  \label{M2m} 
\end{equation}
where  $S = \half$  or $\threehalf$.
It is important to recall that our formulas for the baryon spectrum are the result of an analytic inference, rather than formally derived.

\begin{figure}[h]
\centering
\includegraphics[angle=0,width=8.15cm]{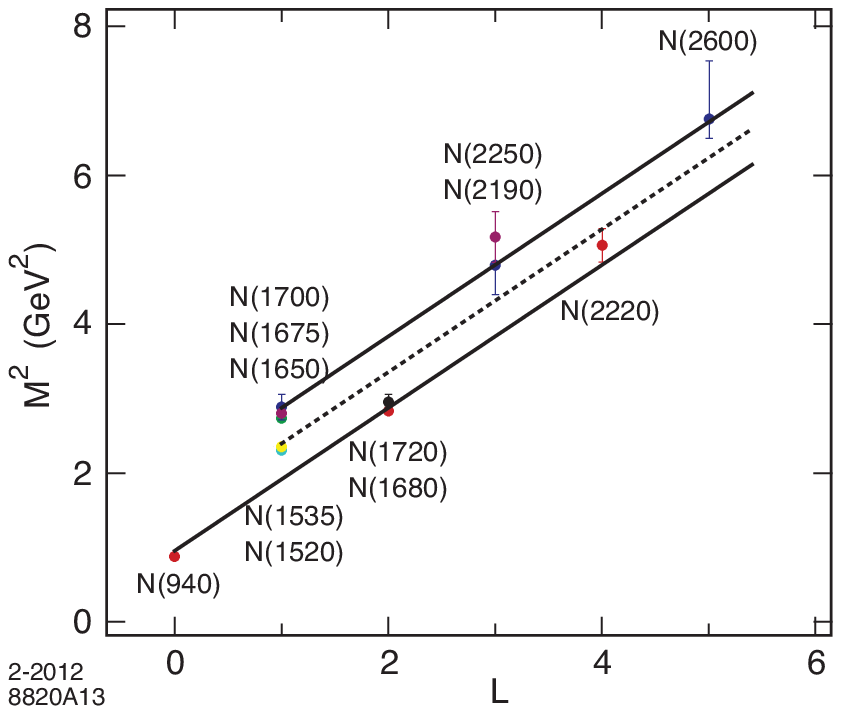} 
\includegraphics[angle=0,width=8.05cm]{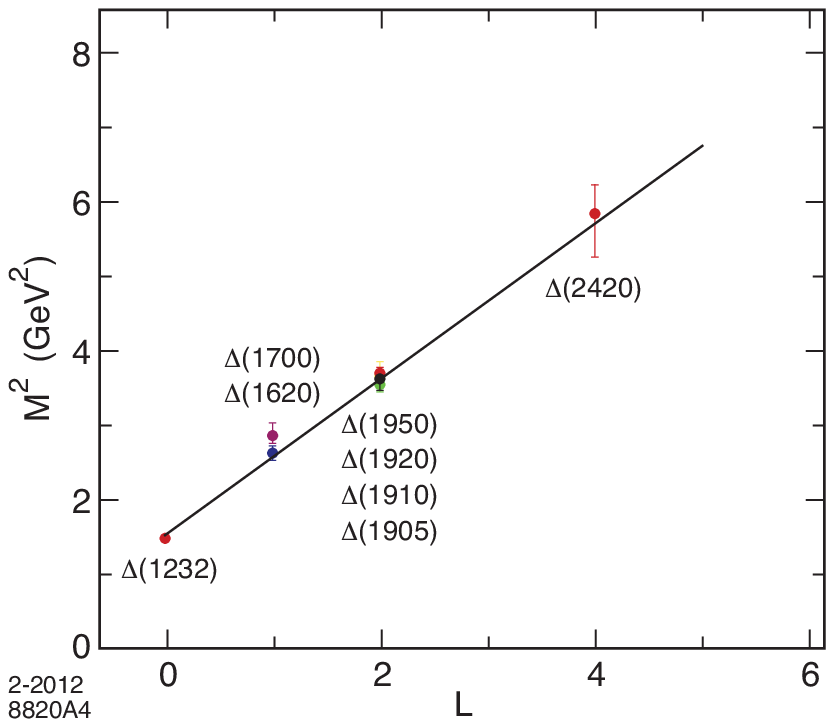}
\caption{\small Baryon orbital trajectories for  the  $N$  (left) and $\Delta$ families (right) for $n=0$ and $\kappa= 0.49-0.51$ GeV. The lower  and upper  nucleon trajectories (left) correspond respectively to the   the spin-$\half$, positive-parity, and  to the spin-$\threehalf$, negative-parity,  families.
 The middle dotted trajectory (left) corresponds to spin-$\half$ negative-parity nucleons. Plus and minus-parity states for the $\Delta$ states (right) are in the same Regge trajectory.}
\label{BaryonsSWpm}
\end{figure}

The full baryon orbital excitation spectrum listed in Table \ref{baryons} for $n=0$ is shown in Fig. \ref{BaryonsSWpm}.
We note that  $\mathcal{M}^{2 \,(+)}_{n, L, S = \frac{3}{2}} = \mathcal{M}^{2 \,(-)}_{n, L, S =  \frac{1}{2}}$ and consequently the positive and negative-parity $\Delta$ states lie in the same trajectory, consistent with the experimental results. Only the confirmed PDG~\cite{Nakamura:2010zzi} states listed in Table \ref{baryons} are shown. Our results for the $\Delta$ states agree with those of Ref.~\cite{Forkel:2007cm}. ``Chiral partners" as the $N(1535)$ and the $N(940)$ with different orbital angular momentum are  non-degenerate from the onset. Using (\ref{M2p}) and (\ref{M2m}) we find the relation
\begin{equation}
\frac{\mathcal{M}_{N(1535)}}{\mathcal{M}_{N(940)}}= \sqrt{\fivehalf},
\end{equation}
which is consistent with experiment to a good accuracy.  One can in fact also build the entire negative-parity excitation spectrum starting from the proton partner, the $J=1/2$ negative-parity nucleon state $N(1535)$, using the same rules {\it e.g.},
an increase in mass $\mathcal{M}^2$ of $4 \kappa^2$  for a unit change in the radial quantum number, $4 \kappa^2$ for a change in one unit in the orbital quantum number and $2 \kappa^2$ for a change of one unit of spin relative to the lowest negative-parity state, the $N(1535)$.

With the exception of the $\Delta(1930)$ state (which is not included in  Table \ref{baryons}), all the confirmed baryon excitations  are well described by formulas (\ref{M2p}) and and (\ref{M2m}).  If we follow the non-$SU(6)$ quantum number assignment for the $\Delta(1930)$ given in Ref.~\cite{Klempt:2009pi}, namely  $S = 3/2$, $L =1$, $n=1$  we find from (\ref{M2m}) the value $\mathcal{M}_{\Delta(1930)} = 4 \kappa \simeq 2$ GeV, consistent with the experimental result 1.96 GeV~\cite{Nakamura:2010zzi}. Expected results from new experiments are important to find out if new baryonic excitations follow the simple pattern described by Eqs. (\ref{M2p}) and (\ref{M2m}).

An important feature of light-front holography is that it predicts a similar multiplicity of states for mesons and baryons,  consistent with what is 
observed experimentally~\cite{Klempt:2007cp}. This remarkable property could have a simple explanation in the cluster decomposition of the
holographic variable, which labels a system of partons as an active quark plus a system of $n-1$ spectators. From this perspective, a baryon with $n=3$ looks in light-front holography as a quark--scalar-diquark system. It is also interesting to notice that in the hard wall model the proton mass is entirely due to the kinetic energy of the light quarks, whereas in the soft-wall model described here, half of the invariant mass squared $\mathcal{M}^2$ of the proton is due to the kinetic energy of the partons and half is due  to the confinement potential.

\section{Nucleon form factors}

Proton and neutron electromagnetic form factors are among the most basic observables of the nucleon, and thus central for our
understanding the nucleon's structure and dynamics. In general two form factors are required to describe the elastic scattering of 
electrons by spin-$\half$ nucleons, the Dirac and  Pauli form factors, $F_1$ and  $F_2$
\begin{equation} \label{NFF}
\langle P' \vert J^\mu(0) \vert P \rangle = u(P') \left[ \gamma^\mu F_1(q^2) +  \frac{i \sigma^{\mu \nu} q^\nu}{2 \mathcal{M}} F_2(q^2)\right] u({P}),
\end{equation}
where $q = P' - P$.
In the light-front formalism one can identify
the Dirac and Pauli form factors from the LF spin-conserving and spin-flip current matrix elements of the $J^+$ current~\cite{Brodsky:1980zm}.

On the higher dimensional gravity side the spin-non-flip amplitude for the  EM transition  corresponds to
the  non-local coupling of an external EM field $A^M(x,z)$  propagating in AdS with a  fermionic mode
$\Psi_P(x,z)$, given by the left-hand side of the equation below 
 \begin{multline} \label{FF}
 \int d^4x \, dz \,  \sqrt{g}  \,  \bar\Psi_{P'}(x,z)
 \,  e_M^A  \, \Gamma_A \, A^M(x,z) \Psi_{P}(x,z) \\  \sim 
 (2 \pi)^4 \delta^4 \left( P'  \! - P - q\right) \epsilon_\mu u(P') \gamma^\mu F_1(q^2) u({P}),
 \end{multline} 
 where $e^A_M = \left(\frac{R}{z}\right) \delta_M^A$ is the vielbein with curved space indices  $M, N = 1, \cdots 5$ and tangent indices
 $A, B = 1, \cdots, 5$. The expression on the right-hand side  represents the Dirac EM form factor in physical space-time. It is the EM matrix element  (\ref{NFF}) of the local quark current  $J^\mu = e_q \bar q \gamma^\mu q$  with local coupling to the constituents. In this case one can also show  that a precise mapping of the $J^+$ elements  can be carried out at fixed LF time, providing an exact correspondence between the holographic variable $z$ and the LF impact variable $\zeta$ in ordinary space-time with the 
result~\cite{Brodsky:2008pg}
 \begin{equation} \label{pmFFAdS}
G_\pm(Q^2)  =  g_\pm R^4 \int \frac{dz}{z^4} \, V(Q^2, z)  \, \Psi^2_\pm(z), 
\end{equation}
for the components $\Psi_+$ and $\Psi_-$ with angular momentum $L^z = 0$ and $L^z = +1$ respectively.
The effective charges $g_+$ and $g_-$ are determined from the spin-flavor structure of the theory.

A precise mapping for the Pauli form factor using light-front holographic methods has not been carried out. To study the spin-flip nucleon form factor $F_2$ using holographic methods, Abidin and Carlson~\cite{Abidin:2009hr} 
 propose to introduce a non-minimal electromagnetic coupling with the  `anomalous' gauge invariant term 
\begin{equation} \label{F2AdS}
\int d^4x~dz~\sqrt{g}  ~ \bar\Psi
 \,  e_M^A\,  e_N^B \left[\Gamma_A, \Gamma_B\right] F^{M N}\Psi,
 \end{equation}
 in the five-dimensional action, since the structure of (\ref{FF}) can only account for $F_1$.  Although this is a practical  
 avenue, the overall strength of the new term has to be fixed by the static quantities and thus some predictivity is lost.

Light-front holographic QCD methods have also been used to obtain generalized parton distributions (GPDs) of the nucleon in the  zero skewness limit
in Refs.  \cite{Vega:2010ns} and \cite{Vega:2012iz}   for the soft and hard-wall models respectively.  GPDs are nonperturbative, and thus holographic methods are well suited to explore their analytical structure.~\footnote{See also the discussion in Ref. \cite{Nishio:2011xa}.} In the sections below we discuss elastic and transition nucleon form factors using light-front holographic ideas.~\footnote{A study of the EM nucleon to $\Delta$ transition form factors has been carried out in the framework of the Sakai and Sugimoto model in Ref.~\cite{Grigoryan:2009pp}.}~\footnote{LF holographic methods can also be used to study the flavor separation of the elastic nucleon form factors which have been determined recently up to $Q^2 = 3.4 ~{\rm GeV}^2$~\cite{Cates:2011pz}. This will be described elsewhere. See also Ref.~\cite{Rohrmoser:2011tw}.}

\subsection{Computing  nucleon elastic form factors in light-front holographic QCD}

In order to compute the individual features of the proton and neutron form factors  one needs to incorporate the spin-flavor structure of the nucleons,  properties  which are absent in models of the gauge/gravity correspondence.
The spin-isospin symmetry can be readily included in AdS/QCD by weighting the different Fock-state components  by the charges and spin-projections of the quark constituents; e.g., as given by the $SU(6)$  spin-flavor symmetry. 
We label by $N_{q \uparrow}$  and $N_{q \downarrow}$ the probability to find the constituent $q$ in a nucleon with spin up or down respectively. For the  $SU(6)$ wave function we have 
\begin{equation}
N_{u \uparrow} = \frac{5}{3}, \hspace{20pt}  N_{u \downarrow} = \frac{1}{3}, \hspace{20pt}  
N_{d \uparrow} = \frac{1}{3},  \hspace{20pt}  N_{d \downarrow} = \frac{2}{3},
\end{equation}
for the proton and
\begin{equation}
N_{u \uparrow} = \frac{1}{3}, \hspace{20pt}  N_{u \downarrow} = \frac{2}{3}, \hspace{20pt}  
N_{d \uparrow} = \frac{5}{3},  \hspace{20pt}  N_{d \downarrow} = \frac{1}{3},
\end{equation}
for the neutron.  The effective charges $g_+$ and $g_-$ in (\ref{pmFFAdS})  are computed by the sum of the charges of the struck quark composed by the corresponding probability for the $L^z = 0$ and $L^z = + 1$ components $\Psi_+$ and $\Psi_-$ respectively. We find $g^+_p = 1$, $g^-_p = 0$, $g_+^n = - \frac{1}{3}$ and $g_-^n = \frac{1}{3}$. The nucleon Dirac form factors in the $SU(6)$ limit are thus given by
\begin{eqnarray}
F_1^p(Q^2) &\!=\!& R^4 \int \frac{dz}{z^4} \, V(Q^2, z)  \, \Psi_+^2(z), \\
F_1^n(Q^2) &\!=\!& - \frac{1}{3}R^4 \int \frac{dz}{z^4} \,   V(Q^2, z) \left[  \Psi_+^2(z) - \Psi_-^2(z)  \right],
\label{pnF1AdS}
\end{eqnarray}
where $F_1^p(0) = 1$ and  $F_1^n(0) = 0$.

\begin{figure}[h]
\begin{center}
 \includegraphics[width=8.0cm]{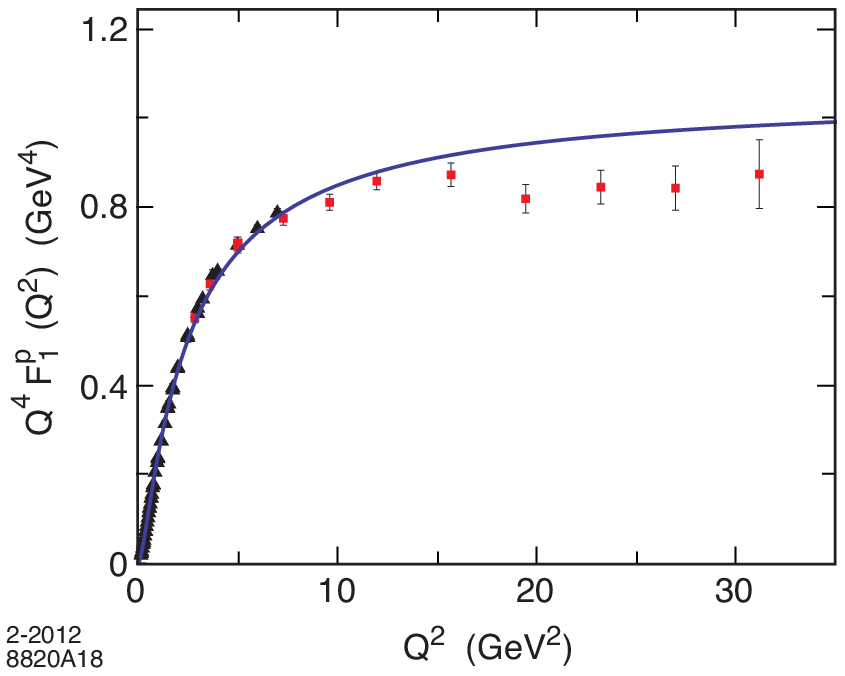}   \hspace{0pt}
\includegraphics[width=8.0cm]{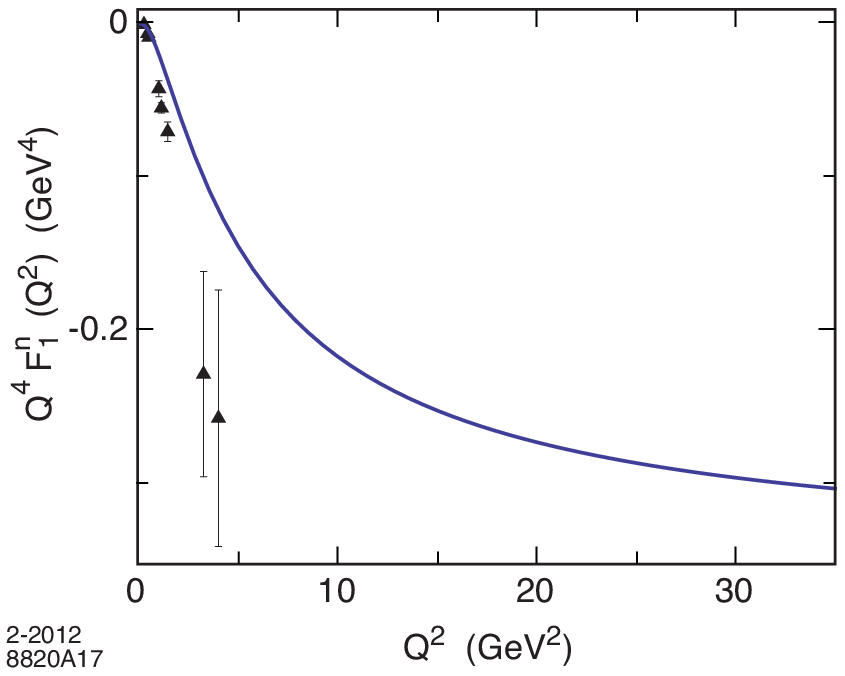}
 \caption{\small Predictions for  $Q^4 F_1^p(Q^2)$ (left) and   $Q^4 F_1^n(Q^2)$ (right) in the
soft wall model. Data compilation  from Diehl~\cite{Diehl:2005wq}.}
\label{fig:nucleonFF1}
\end{center}
\end{figure}

In the soft-wall model the plus and minus  components of the twist-3 nucleon wave function are
\begin{equation} \label{AdSNWF}
\Psi_+(z) = \frac{ \sqrt{2} \kappa^2}{R^2} z^{7/2} e^{- \kappa^2 z^2/2}, \hspace{20pt}
\Psi_-(z) =                \frac{\kappa^3}{R^2} z^{9/2} e^{- \kappa^2 z^2/2} ,
\end{equation}
and $V(Q^2, z)$ is given by (\ref{eq:Vkappa}). The results for $F_1^{p,n}$ follow from the analytic form (\ref{Ftau}) for any twist $\tau$.
We find
\begin{equation} \label{protonF1p}
F_1^p(Q^2) =  F_+(Q^2),
 \end{equation}
 and
 \begin{equation} \label{neutronF1n}
 F_1^n(Q^2) = -\frac{1}{3} \left(F_+(Q^2) - F_-(Q^2)\right),
 \end{equation}
where we have, for convenience,  defined  the twist-2 and twist-3 form factors
\begin{equation} \label{Fp}
F_+(Q^2) =  \frac{1}{{\Big(1 + \frac{Q^2}{\mathcal{M}^2_\rho} \Big) }
 \Big(1 + \frac{Q^2}{\mathcal{M}^2_{\rho'}}  \Big) },
 \end{equation}
and
  \begin{equation} \label{Fm}
 F_-(Q^2) =  
  \frac{1}{\Big(1 + \frac{Q^2}{\mathcal{M}^2_\rho} \Big) 
 \Big(1 + \frac{Q^2}{\mathcal{M}^2_{\rho'}}  \Big)
       \Big(1  + \frac{Q^2}{\mathcal{M}^2_{\rho^{''}}} \Big)} .
\end{equation}
As discussed in Sec.  \ref{EFFDC}, the multiple pole structure in (\ref{Fp}) and (\ref{Fm}) is derived from the dressed EM current propagating in AdS space.

 The results for $Q^4 F_1^p(Q^2)$ and $Q^4 F_1^n(Q^2)$  are shown in
Fig. \ref{fig:nucleonFF1}. To compare with physical data we have shifted the poles in expression (\ref{Ftau}) to their physical values located at $M^2 = 4 \kappa^2(n + 1/2)$ 
following the  discussion in Sec. \ref{caveats}.  The value $\kappa = 0.545$ GeV  is determined from the $\rho$ mass.

\begin{figure}[h]
\begin{center}
 \includegraphics[width=8.0cm]{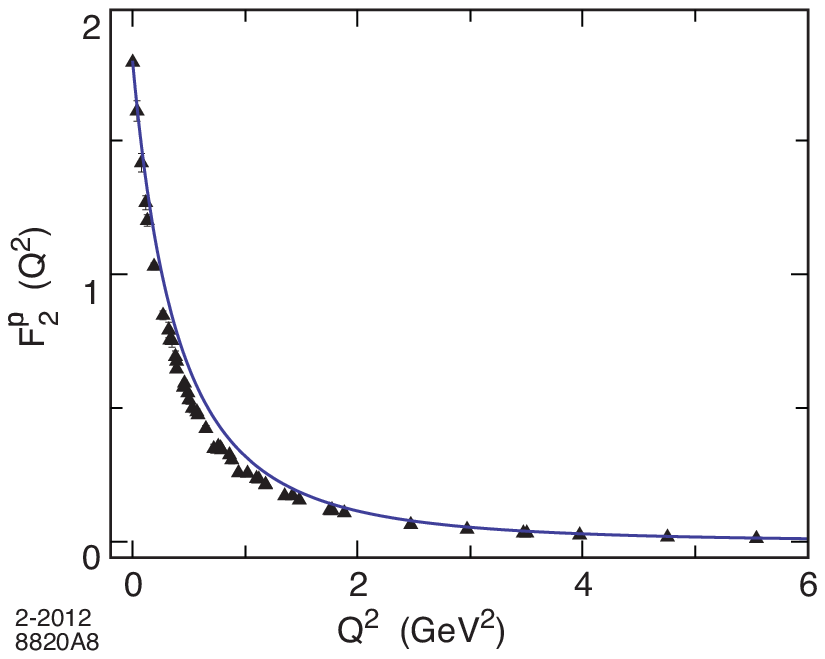}   \hspace{0pt}
\includegraphics[width=8.0cm]{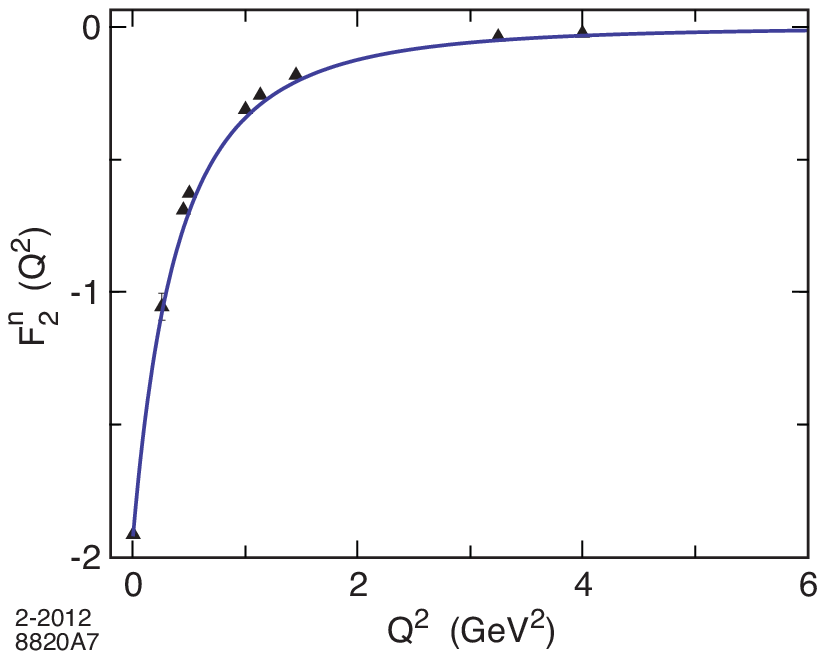}
 \caption{\small Predictions for  $F_2^p(Q^2)$ (left)  and  $F_2^n(Q^2)$ (right) in the
soft wall model. Data compilation  from Diehl~\cite{Diehl:2005wq}.}
\label{fig:nucleonFF2}
\end{center}
\end{figure}

The  expression for the elastic nucleon form factor $F_2^{p,n}$  follows  from (\ref{NFF}) and (\ref{F2AdS}). \begin{equation} \label{F2}
F_2^{p,n}(Q^2) \sim \int \frac{d z}{z^3} \Psi_+(z) V(Q^2,z) \Psi_-(z).
\end{equation}
Using the twist-3 and twist-4 AdS soft-wall wavefunctions $\Psi_+$ and $\Psi_-$ (\ref{AdSNWF})
we find
\begin{equation}
 F_2^{p,n}(Q^2) =  
  {\chi_{p,n}} F_-(Q^2),
\end{equation}
where the amplitude (\ref{F2}) has been normalized to the static quantities $\chi_p$ and $\chi_n$ and $F_-(Q^2)$ is given by (\ref{Fm}). The experimental values $\chi_p = 1.793$ and $\chi_n = -1.913$ are consistent with the $SU(6)$ prediction~\cite{Beg:1964nm} $\mu_P/\mu_N = -3/2$. In fact
$(\mu_P/\mu_N)_{\rm exp} = - 1.46$ where $\mu_P = 1 + \chi_p$ and $\mu_N = \chi_n$.
The results for $F_2^p(Q^2)$ and $F_2^n(Q^2)$ for  $\kappa = 0.545$ GeV are shown in
Fig. \ref{fig:nucleonFF2}. 

We compute the charge and magnetic root-mean-square (rms) radius from the usual electric and magnetic nucleon form factors
\begin{equation}
G_E(q^2) = F_1(q^2) + \frac{q^2}{4 \mathcal{M}^2} F_2(q^2)
\end{equation}
and
\begin{equation}
G_M(q^2) = F_1(q^2) + F_2(q^2).
\end{equation}
Using the definition
\begin{equation}
\langle r^2 \rangle = -  \frac{6}{F(0)} \frac{d F(Q^2)}{d Q^2} \Big|_{Q^2 =0},
\end{equation}
we find the values $\sqrt{\langle r_E \rangle_p} = 0. 802 ~{\rm fm}$, $\sqrt{\langle r^2_M \rangle_p} = 0. 758 ~{\rm fm}$,
 $\langle r^2_E \rangle_n  = -0.10 ~{\rm fm^2}$ and $\sqrt{\langle r^2_M \rangle_n} = 0.768 ~{\rm fm}$, compared with the
 experimental values   $\sqrt{\langle r_E \rangle_p} = (0.877 \pm 0.007) ~{\rm fm}$, 
 $\sqrt{\langle r^2_M \rangle_p} = (0.777 \pm 0.016) ~{\rm fm}$,
 $\langle r^2_E \rangle_n  = (- 0.1161 \pm 0.0022) ~{\rm fm^2}$ and 
 $\sqrt{\langle r^2_M \rangle_n} = (0.862 \pm 0.009) ~{\rm fm}$ from electron-proton scattering experiments~\cite{Nakamura:2010zzi}.~\footnote{The neutron charge radius is defined by $\langle r_E^2 \rangle_n = -  6 \frac{d G_E(Q^2)}{d Q^2} \Big|_{Q^2 =0}$. }
 The muonic hydrogen measurement gives  $\sqrt{\langle r_E \rangle_p} = 0.84184(67)~{\rm fm}$ from  Lamb-shift measurements~\cite{Pohl:2010zz}.~\footnote{Other soft and hard-wall model predictions of the
 nucleon  rms radius are given in Refs. \cite{Abidin:2009hr, Vega:2010ns, Vega:2012iz}.}

Chiral effective theory predicts that the slopes are singular for zero pion mass. For example, the slope of the Pauli form factor of the proton at $q^2=0$ computed by Beg and Zepeda diverges as $1/ m_\pi$~\cite{Beg:1973sc}. ÊÊThis comes from the simple triangle diagram $\gamma^* \to \pi^+ \pi^- \to p \bar p.$ ÊOne can also argue from dispersion theory that the singular behavior of the form factors as a function of the pion mass comes from the two-pion cut.  Lattice theory computations of nucleon form factors require in fact the strong dependence at small pion mass to extrapolate the predictions to the physical pion mass~\cite{Collins:2011mk}. Ê The two-pion calculation~\cite{Beg:1973sc} is a Born computation which probably does not exhibit vector dominance. To make a reliable computation in the hadronic basis of intermediate states
one evidently has to include an infinite number of states. On the other hand, chiral divergences do not appear in AdS/QCD when we use the dressed current since, as shown is Sec. \ref{EFFDC}, the holographic analysis with a dressed EM current in AdS generates instead a nonperturbative multi-vector meson pole structure.\footnote{In the limit of a free propagating current in AdS, we obtain logarithmic divergent results: $\langle r_p^2\rangle_{F_1} = 3 \ln\left(\frac {4 \kappa^2}{Q^2}\right)\Big\vert_{Q^2 \to 0}$ and $\langle r_p^2\rangle_{F_2} = \frac{9}{2} \ln\left(\frac {4 \kappa^2}{Q^2}\right)\Big\vert_{Q^2 \to 0}$.}

\subsection{Computing nucleon transition form factors in light-front holographic QCD}

As an illustrative example we consider in this section  the 
form factor for the $\gamma^* p \to N(1440) P_{11}$ transition measured recently at JLab.   We shall weight  the different Fock-state components by the charges and spin-projections of the quark constituents using the $SU(6)$  spin-flavor symmetry as in the previous section.  The expression for the spin non-flip proton form factors for the transition $n,L \to n' L$ is~\cite{Brodsky:2008pg} 
\begin{equation} \label{F1}
{F_1^p}_{n, L \to n', L}(Q^2)  =    R^4 \int \frac{dz}{z^4} \, \Psi_+^{n' \!, \,L}(z) V(Q^2,z)  \Psi_+^{n, \, L}(z),
\end{equation}
where we have factored out the plane wave dependence of the AdS fields
\begin{equation} \label{Psip}
\Psi_+(z) = \frac{\kappa^{2+L}}{R^2}  \sqrt{\frac{2 n!}{(n+L+1)!}} \,z^{7/2+L} L_n^{L+1}\!\left(\kappa^2 z^2\right) 
e^{-\kappa^2 z^2/2}.
\end{equation}
The orthonormality of the Laguerre polynomials in (\ref{Psip}) implies that the nucleon form factor at $Q^2 = 0$ is one if $n = n'$ and zero otherwise.  Using the integral
representation of the bulk-to-boundary propagator $V(Q^2,z)$ given by (\ref{Vx}) we find the twist-3  spin non-flip  transition form factor
 \begin{equation} \label{RoperF1}
 {F_1^p}_{N \to N^*}(Q^2) = \frac{\sqrt{2}}{3} \frac{\frac{Q^2}{\mathcal{M}^2_\rho}}{\Big(1 + \frac{Q^2}{\mathcal{M}^2_\rho} \Big) 
 \Big(1 + \frac{Q^2}{\mathcal{M}^2_{\rho'}}  \Big)
       \Big(1  + \frac{Q^2}{\mathcal{M}^2_{\rho^{''}}} \Big)}.
\end{equation}

 \begin{figure}[!]
 \begin{center}
\includegraphics[angle=0,width=8.0cm]{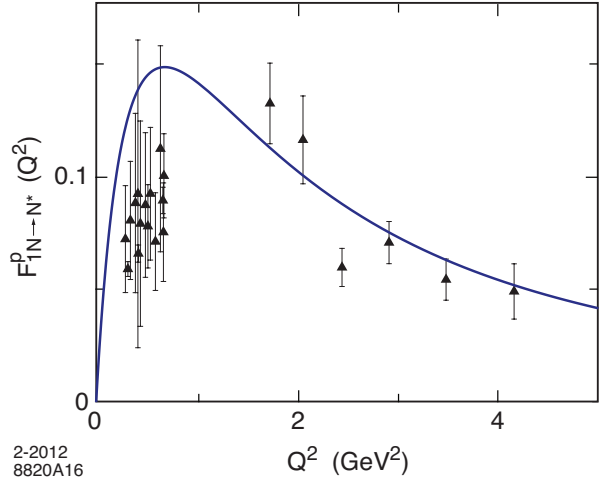}
\caption{\small Proton transition form factor  ${F_1^p}_{N\to N^*}(Q^2)$ to the first radial excited state. Data from JLAB~\cite{Aznauryan:2009mx}.}
\label{pFFs}
\end{center}
\end{figure}

The result (\ref{RoperF1}),
 compared with available data in Fig.~\ref{pFFs}, correspond to the valence approximation.   The transition form factor
 (\ref{RoperF1}) is expressed in terms of  the mass of the $\rho$ vector meson and its first two radial excited states, with no additional parameters. The results  in Fig.~\ref{pFFs}  are in good agreement with experimental data. The transition  to the  $N(1440) P_{11}$ state  corresponds to  the first radial excitation of the three-quark  ground state of the nucleon. In fact,  the Roper resonance $N(1440)P_{11}$ and the $N(1710)P_{11}$ are well accounted in the light-front holographic framework as the first and second radial states of the nucleon family as shown in Sec. \ref{SWB} (See Fig.  \ref{BaryonsSWpnL}). It is certainly worth to extend the simple computations described here and perform a systematic study of the different transition form factors measured at JLab. This study will help to discriminate among models and compare with the new results expected from  the JLab 12 GeV Upgrade Project,  in particular at photon
virtualities $Q^2>5  ~{\rm GeV}^2$, which correspond to the experimental
coverage of the CLAS12 detector at JLab~\cite{Aznauryan:2011qj}.

\section{Higher Fock components in light-front holographic QCD}

The light-front Hamiltonian eigenvalue equation (\ref{LFH}) is a matrix in Fock space which represents an infinite number of coupled integral equations for the Fock components $\psi_n = \langle n \vert \psi \rangle$. The resulting potential in quantum field theory can be considered as an instantaneous four-point effective interaction 
in LF time, similar to the instantaneous gluon exchange in the light-cone gauge $A^+ = 0$,
which leads
to $q q \to qq$, $q \bar q \to q \bar q$, $ q \to q q \bar q$ and $\bar q \to \bar q q \bar q$ as in QCD(1+1).  
Higher Fock states can have any number of extra $q \bar q$ pairs, but surprisingly no dynamical gluons.
Thus in holographic QCD, gluons are absent in the confinement potential.\footnote{This result
 is consistent with the flux-tube interpretation of QCD~\cite{Isgur:1984bm}  where soft gluons interact so strongly that they are sublimated into a color confinement potential for quarks. The absence of constituent glue in hadronic physics has been invoked also in  Ref. \cite{Klempt:2004ih}, where the role of the confining potential is attributed to an instanton induced interaction.}
This unusual property of AdS/QCD may explain the dominance of quark interchange~\cite{Gunion:1972qi}
 over quark annihilation or gluon exchange contributions in large angle elastic scattering~\cite{Baller:1988tj}.~\footnote{In Ref. \cite{Brodsky:2011pw} we discuss a number of experimental results in hadron physics which support this picture.}

In order to illustrate the relevance of higher Fock states and the absence of dynamical gluons at the hadronic scale, we will discuss  a simple semi-phenomenological model of the elastic form factor of the pion where we include the first two components in a Fock expansion of the pion wave function
$\vert \pi \rangle  = \psi_{q \bar q /\pi} \vert q \bar q  \rangle_{\tau=2}
+  \psi_{q \bar q q \bar q} \vert q \bar q  q \bar q  \rangle_{\tau=4} + \cdots$ ,
where the $J^{PC} = 0^{- +}$ twist-two and twist-4 states $\vert q \bar q \rangle$  and  $\vert q \bar q q \bar q  \rangle$ are created by the interpolating operators
$\bar q \gamma^+ \gamma_5  q$ and $ \bar q \gamma^+ \gamma_5  q  \bar q q$ respectively.

 \begin{figure}[h]
\centering
\includegraphics[width=7.45cm]{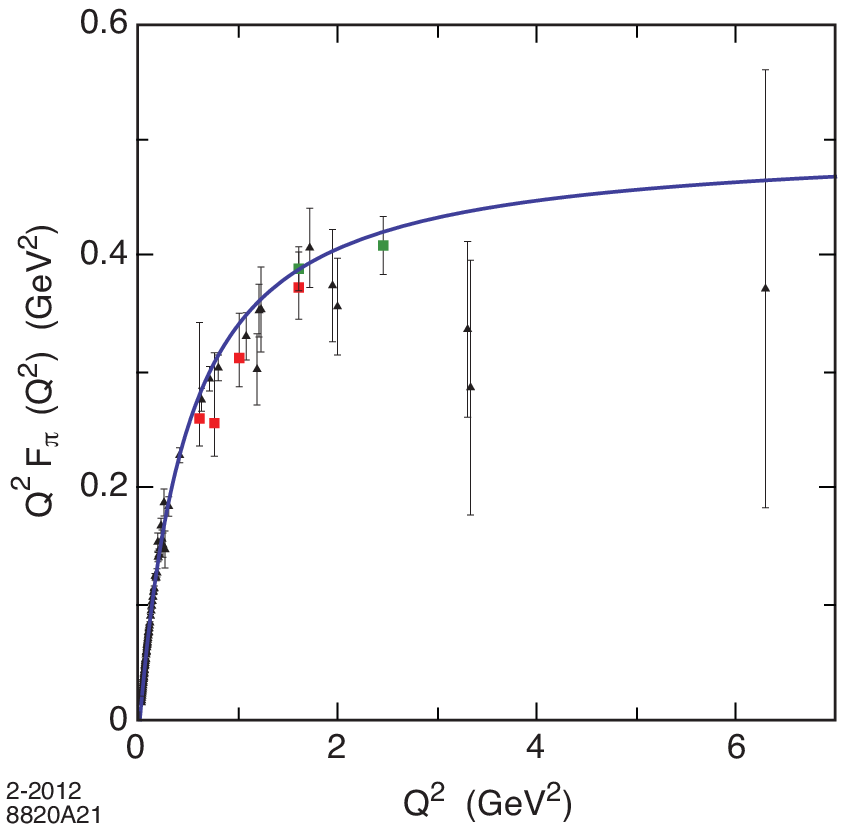} \hspace{0pt}
\includegraphics[width=8.2cm]{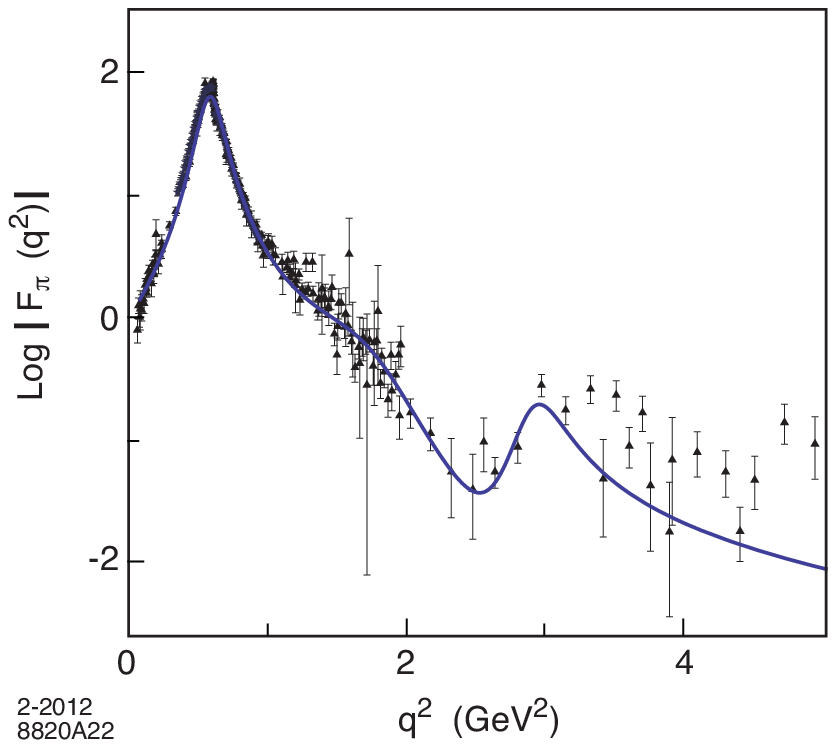}
\caption{\small Structure of the space-like (left)  and time-like (right) pion  form factor in light-front holography for a truncation of the pion wave function up to twist four.
Triangles are the data compilation  from Baldini  {\it et al.}~\cite{Baldini:1998qn},  squares are JLAB data~\cite{Tadevosyan:2007yd}. }
\label{pionFFhfs}
\end{figure}

Since the charge form factor is a diagonal operator, the final expression for the form factor corresponding to the truncation up to twist four is the sum of two terms, a monopole and a three-pole term.
In the strongly coupled semiclassical gauge/gravity limit hadrons have zero widths and are stable. One can nonetheless modify the formula (\ref{Ftau}) by introducing 
a finite width:
$q^2 \to q^2 + \sqrt 2 i \mathcal{M} \Gamma$.  We choose the values $\Gamma_\rho =  140$ MeV,   $\Gamma_{\rho'} =  360$ MeV and  $\Gamma_{\rho''} =  120$ MeV.  The results for the pion form factor with twist two and four Fock states  are shown in Fig. \ref{pionFFhfs}. The results correspond to $P_{q \bar q q \bar q}$ = 13 \%, the admixture of the
$\vert q \bar q q \bar q  \rangle$ state. The value of $P_{q \bar q q \bar q}$ (and the widths) are input in the model. The value of $\kappa$ is determined from the $\rho$ mass and the masses of the radial excitations follow from setting the poles at their physical locations, $\mathcal{M}^2 \to 4 \kappa^2(n + 1/2)$, as discussed in Sec. \ref{caveats}.
The time-like structure of the pion form factor displays a rich pole structure with constructive and destructive interfering phases;  this
is incompatible with the admixture of the twist-three 
state $\vert q \bar q g \rangle$ containing a dynamical gluon since the interference in this case is opposite in sign.

\section{Conclusions \label{conclusions}}

As we have shown,  the exact light-front Hamiltonian $H_{LF} \vert \psi \rangle = \mathcal{M}^2 \vert \psi \rangle$ for QCD can be systematically reduced to a relativistic frame-independent semiclassical wave equation~\cite{deTeramond:2008ht}
\begin{equation} \label{LFWE2}
\left(-\frac{d^2}{d\zeta^2}
- \frac{1 - 4L^2}{4\zeta^2} + U(\zeta) \right)
\phi(\zeta) = \mathcal{M}^2 \phi(\zeta),
\end{equation}
for the  valence Fock state of mesons. The unmodified AdS equations
correspond to the kinetic energy terms of  the massless constituent quarks with  relative orbital angular momentum $L= L^z$. The effective potential $U(\zeta)$ corresponds to the color-confining potential 
and follows from the  truncation of AdS space, in a modified effective AdS action, and light-front holography. The variable $\zeta$ is the invariant separation of the constituents. This frame-independent light-front wave equation is comparable in simplicity to Schr\"odinger theory in atomic physics which is formulated at equal instant time. We have also derived an analogous light-front Dirac equation for holographic QCD which describes light-quark baryons with finite color $N_C=3.$

Remarkably, these light-front  equations are equivalent to the equations of motion in a  higher dimensional warped  space  asymptotic to AdS space. The mapping of the gravity theory to the boundary quantum field theory, quantized at fixed  light-front time, thus gives a precise relation between holographic wave functions and the light-front wave functions which describe the internal structure of the hadrons and their electromagnetic couplings.   This mapping provides the basis for a profound connection between physical QCD quantized in the light-front and the physics of hadronic modes in a higher dimensional AdS space.
However, the derivation of the effective color-confining  potential $U(\zeta)$ directly from QCD, remains an open question.

Despite some limitations of AdS/QCD~\cite{Csaki:2008dt}, the light-front holographic  approach to  the gauge/gravity duality, {\it Light-Front Holography},  has already provided significant physical  insight into the strongly-coupled nature and internal structure of hadrons; in fact, it is one of the few tools available.  As we have seen, the resulting model provides a simple and successful framework for describing  nonperturbative  hadron dynamics: the systematics of the excitation spectrum of hadrons: the mass eigenspectrum,  observed multiplicities and degeneracies. It also provides powerful new analytical tools for computing hadronic transition amplitudes, incorporating the scaling behavior and the transition from the hard-scattering perturbative domain, where quark and gluons are the relevant degrees of freedom, to the long range confining hadronic region.

The dressed current in AdS includes the nonperturbative pole structure. Consequenly, the  approach incorporates both the long-range confining hadronic domain and the constituent conformal short-distance  quark particle  limit in a single framework. The results display  a simple analytical structure  which allows us to explore  dynamical properties in Minkowski space-time; in many cases these studies are not amenable to Euclidean  lattice gauge theory computations.   In particular, the excitation dynamics of nucleon resonances encoded in the nucleon transition form factors  can provide fundamental insight into the strong-coupling dynamics of QCD. New theoretical  tools are thus of primary interest for the interpretation of the results expected at the new mass scale and kinematic regions accessible to the JLab 12 GeV Upgrade Project.

The semiclassical approximation to light-front QCD 
described in this article is expected to break down at short distances
where gluons become dynamical degrees of freedom and hard gluon exchange and quantum corrections become important. 
One can systematically improve the semiclassical approximation, for example,  by introducing nonzero quark masses and short-range Coulomb-like gluonic
corrections,  thus extending the predictions of the model to the dynamics and spectra of heavy and heavy-light quark systems.  The model can also be improved by applying Lippmann-Schwinger methods to systematically improve the light-front Hamiltonian of the semiclassical holographic approximation.
One can also use the holographic LFWFs  as basis functions for diagonalizing the full light-front  QCD Hamiltonian\cite{Vary:2009gt} as well as  the input boundary functions to study the evolution of structure functions and distribution amplitudes at a low energy scale.

\section*{Acknowledgements}

\noindent Invited lectures presented by GdT at the Niccol\`o Cabeo International School of Hadronic Physics, Ferrara, Italy, May 2011. GdT is grateful to the organizers and especially to Paola Ferretti Dalpiaz for her outstanding hospitality. We thank E. Klempt, V. E. Lyubovitskij and S. D. Glazek for helpful comments. We are  grateful to  F.-G. Cao, A. Deur, H. G. Dosch and J. Erlich  for collaborations. This research was supported by the Department of Energy  contract DE--AC02--76SF00515.

\newpage

\appendix
\appendixpage

\section{AdS boundary conditions and interpolating operators \label{interop}}

The formal statement of the duality between a gravity theory on $(d+1)$-dimensional 
Anti-de Sitter $AdS_{d+1}$ space and the strong coupling limit of a conformal
field theory (CFT)  on the $d$-dimensional asymptotic boundary of $AdS_{d+1}$
at $z=0$ is expressed in terms of the $d+1$ partition function for
a field $\Phi(x ,z)$ propagating in the bulk
\begin{equation}
Z_{grav}[\Phi] = e^{i S_{eff}[\Phi]} =
\int \mathcal{D}[\Phi] e^{ i S[\Phi]},
\label{eq:Zgrav}
\end{equation}
where $S_{eff}$ is the effective action of the $AdS_{d+1}$ theory,
and the $d$-dimensional generating functional of correlation functions
of the conformal field theory in presence of an external source $\Phi_0(x^\mu)$ 
\begin{equation}
  Z_{CFT}[\Phi_0] =  e^{ i W_{CFT}[\Phi_0]}
  = \left< \exp\left(i \int d^dx \, \Phi_0(x) \mathcal{O}(x)\right) \right>.
\label{eq:ZCFT}
\end{equation}
The functional $W_{CFT}$ is the generator of connected 
Green's functions of the boundary theory and $\mathcal{O}$ is a QCD local
interpolating operator. 

According to the AdS/CFT correspondence, 
to every operator in the conformal field theory  there corresponds an AdS field. We use the isometries of AdS space to map the scaling dimensions of the local interpolating operators defined at the AdS boundary into the modes propagating inside AdS space.
The precise relation of the gravity theory on AdS space
to the conformal field theory at its boundary 
is~\cite{Gubser:1998bc}
\begin{equation}
 Z_{grav}\big[\Phi(x,z) \big \vert_{z = 0} = \Phi_0(x) \big]
 = Z_{CFT}\left[\Phi_0\right],
\label{eq:grav-CFT}
\end{equation}
where the partition function (\ref{eq:Zgrav}) on $AdS_{d+1}$ is integrated 
over all possible configurations
$\Phi$ in the bulk which approach its boundary value $\Phi_0$.
If we neglect the contributions from quantum fluctuations 
to the gravity partition function, then the generator $W_{CFT}$ of
connected Green's functions of the four-dimensional gauge theory
(\ref{eq:ZCFT}) is precisely equal to the classical (on-shell) gravity action
(\ref{eq:Zgrav})
\begin{equation}
W_{CFT}\left[\phi_0\right] =
S_{eff}\big[\Phi(x,z) \big\vert_{z=0} = \Phi_0(x)\big]_{\rm on-shell},
\end{equation}
evaluated in terms of the classical solution to the bulk equation of motion.
This defines  the semiclassical approximation to the conformal field theory.
In the bottom-up phenomenological approach, the effective action
 in the bulk is  usually modified for large values of $z$ to incorporate confinement and is truncated at the quadratic level.

In the limit $z \to 0$, the independent solutions behave as
\begin{equation} \label{eq:Phiz0}
\Phi(x , z) \to z^\tau \,\Phi_+(x) + z^{d - \tau} \,\Phi_-(x),
\end{equation}
where $\tau$ is the scaling dimension.
The non-normalizable solution $\Phi_-$ has the leading boundary behavior
and is the boundary value of the bulk
field $\Phi$ which couples to a QCD gauge invariant operator 
$\mathcal{O}$ in the $z \to 0$ asymptotic boundary, thus $\Phi_- = \Phi_0$.
The normalizable solution $\Phi_+$ is the response
function and corresponds to the physical
states~\cite{Balasubramanian:1998sn}. 
The  interpolating
operators $\mathcal{O}$ of the boundary conformal theory are constructed
from local gauge-invariant products of quark and gluon fields and
their covariant derivatives, taken at the same point in
four-dimensional space-time in the $x^2 \to 0$ limit. 
According to (\ref{eq:ZCFT}) the scaling dimensions of $\mathcal{O}$ are matched to the conformal scaling behavior of
the AdS fields in the limit $z \to 0$ and are thus encoded into
the propagation of the modes inside AdS space. 

Integrating by parts, and using the equation of motion for the field in AdS, the bulk
contribution to the action vanishes, and one is left with a non-vanishing surface
term in the ultraviolet boundary
\begin{equation} 
S = R^{d-1} \lim_{z \to 0} 
\int d^d x \, \frac{1}{z^{d-1}} \, \Phi \partial_z \Phi, 
\label{eq:SUV}
\end{equation}
which can be identified with the boundary  QFT functional $W_{CFT}$. 
Substituting the leading dependence (\ref{eq:Phiz0}) of $\Phi$ near $z =0$  in the ultraviolet
surface action (\ref{eq:SUV}) and using the functional relation
\begin{equation}
\frac{\delta W_{CFT} }{\delta \Phi_0} = \frac{\delta S_{\rm eff}}{\delta\Phi_0},
\end{equation}
one finds that 
$\Phi_+(x)$ is related to the expectation values of $\mathcal O$
in the presence of the source $\Phi_0$~\cite{Balasubramanian:1998sn}
\begin{equation} \label{eq:DimPhi}
\left\langle 0 \vert {\mathcal O}(x) \vert 0 \right\rangle_{\Phi_0}
\sim \Phi_+(x).
\end{equation}
The exact relation depends on the normalization of the fields 
chosen~\cite{Klebanov:1999tb}. The field $\Phi_+$ thus acts as a
classical field, and it is the boundary limit of the normalizable string
solution which propagates in the bulk.

\end{document}